\documentclass[aps, prl, amsmath, twocolumn, 10pt, superscriptaddress]{revtex4-2}

\usepackage{bbold}
\usepackage{threeparttable}
\usepackage{booktabs}
\usepackage{graphicx}
\usepackage{braket}
\usepackage{upgreek}
\usepackage[psdextra, colorlinks, citecolor=blue]{hyperref}
\usepackage{hyperref}

\usepackage{xcolor}

\newcommand{\SMrefs}[1]{}

\begin{document}

\title{Tailoring Germanium Heterostructures for Quantum Devices with Machine Learning}

\author{Patrick Del Vecchio}
\email{p.delvecchio@tudelft.nl}
\affiliation{QuTech and Kavli Institute of Nanoscience, Delft University of Technology, Delft, Netherlands}
\author{Kevin Rossi}
\affiliation{Department of Materials Science and Engineering, Delft University of Technology, Delft, Netherlands}
\affiliation{Climate Safety and Security Centre, TU Delft The Hague Campus, Delft University of Technology, 2594 AC, The Hague, The Netherlands}
\author{Giordano Scappucci}
\affiliation{QuTech and Kavli Institute of Nanoscience, Delft University of Technology, Delft, Netherlands}
\author{Stefano Bosco}
\email{s.bosco@tudelft.nl}
\affiliation{QuTech and Kavli Institute of Nanoscience, Delft University of Technology, Delft, Netherlands}

\begin{abstract}
Germanium (Ge) quantum wells are emerging as versatile platforms for quantum devices, supporting high-quality spin qubits and integration with superconducting leads. These applications benefit from strong intrinsic spin-orbit interaction (SOI), enabling efficient electrical control and engineering of spin degrees of freedom. The most advanced Ge/SiGe heterostructures to date, based on compressively strained Ge channels within strain-relaxed silicon-germanium (SiGe) barriers, exhibit weak SOI due to the  heavy-hole character of the wave function, posing challenges for spin-based quantum devices and requiring complex device designs for fast qubit manipulation. In this work, we demonstrate that concrete heterostructure modifications can overcome these limitations, enhancing SOI by up to three orders of magnitude. Specifically, we propose to enrich unstrained Ge channels by localized, strained silicon spikes. Leveraging a multi-objective Bayesian optimization, we optimize the spike profile to maximize SOI, while ensuring compatibility with current epitaxial growth processes and robustness against realistic variations of growth parameters. Our heterostructure substantially enhances device performance, yielding up to two orders of magnitude higher quantum-dot spin qubit quality factors than state-of-the-art materials. We also predict GHz-scale spin splittings for hybrid superconducting Andreev spin qubits. These novel Ge heterostructures with engineered Si concentration profiles can open pathways to scalable quantum and spintronic applications.
\end{abstract}

\maketitle

\footnotetext[1]{Details on the $k\cdot p$ model, the perturbative framework, the ASQ spin splitting derivation, the quantum dot calculations and the machine learning methodological details are provided in the Supplemental Material}

\paragraph{Introduction.--}

Germanium (Ge) quantum wells are rapidly becoming leading platforms for spin-based quantum technologies~\cite{Loss1998,Scappucci2020,Burkard2023}. They support high-coherence spin qubit processors~\cite{Hendrickx2020N,Hendrickx2021,John2025,Zhang2025} and provide natural hosts for quantum simulations~\cite{Jirovec2025arxiv,Farina2025}. Their compatibility with superconductivity has also sparked interest for hybrid superconducting-semiconducting devices~\cite{Jakob2025,Lakic2025,Aggarwal2021,Hinderling2024,Johannsen2026,Luethi2023,Laubscher2024,MichelPino2025,Babkin2025,Adelsberger2023,Tosato2023,Steele2025,Fabris2026,Borovkov2026,Sagi2024,Strohbeen2023,Kiyooka2025}, where Ge offers distinct advantages as a clean material~\cite{Lodari2019,Tosato2022} that can be isotopically purified and made largely insensitive to residual nuclear spins~\cite{Moutanabbir2024,Daoust2026,Fischer2008,Bosco2021PRL}.

A key ingredient to fully unlock the potential of Ge heterostructures for coherent spin physics is their intrinsic Rashba  spin–orbit interaction (SOI)~\cite{Winkler2003,Moriya2014,Morrison2014,Hassan2017}, which provides direct electrical control of the spin state~\cite{Hardy2019,Hendrickx2020NC,Zhang2025,vanRiggelen2022,Wang2024,Hendrickx2021,Jirovec2021,Wang2022,Liu2023,Lawrie2023,Stehouwer2025,Jirovec2025arxiv,SaezMollejo2025,John2025,Jirovec2025PRA,Ivlev2025,Bulaev2007}, and enables tuneable spin response in quantum dots~\cite{Bosco2021PRB,Valvo2025}, potentially compensating for disorder and variability that fundamentally limits large-scale integration~\cite{Martinez2026,Martinez2022,Stehouwer2023,Sangwan2025}.  In hybrid superconducting-semiconducting architectures, large intrinsic SOI ensures strong coupling to superconducting resonators~\cite{DePalma2024,Bosco2022PRL,Janik2025,Yu2023}, efficient manipulation of superconducting (Andreev) spin qubits~\cite{Padurariu2010,PitaVidal2025,Hays2021,Hoffman2024}, and topological phases of matter~\cite{PitaVidal2023,Bargerbos2022,VanLoo2026,Bordin2025,tenHaaf2025,Nayak2008,Sau2010,Lutchyn2010,Sarma2015,Luethi2023,Schrade2017,Rainis2013}. Yet, state-of-the-art Ge heterostructures, based on compressively-strained channels ($\varepsilon$-Ge) sandwiched by strain-relaxed silicon-germanium (SiGe) barriers, show negligible intrinsic SOI due to  large epitaxial strain fields, posing challenges for efficient spin manipulation.

\begin{figure}[t]
    \centering
    \includegraphics[width=1.0\linewidth]{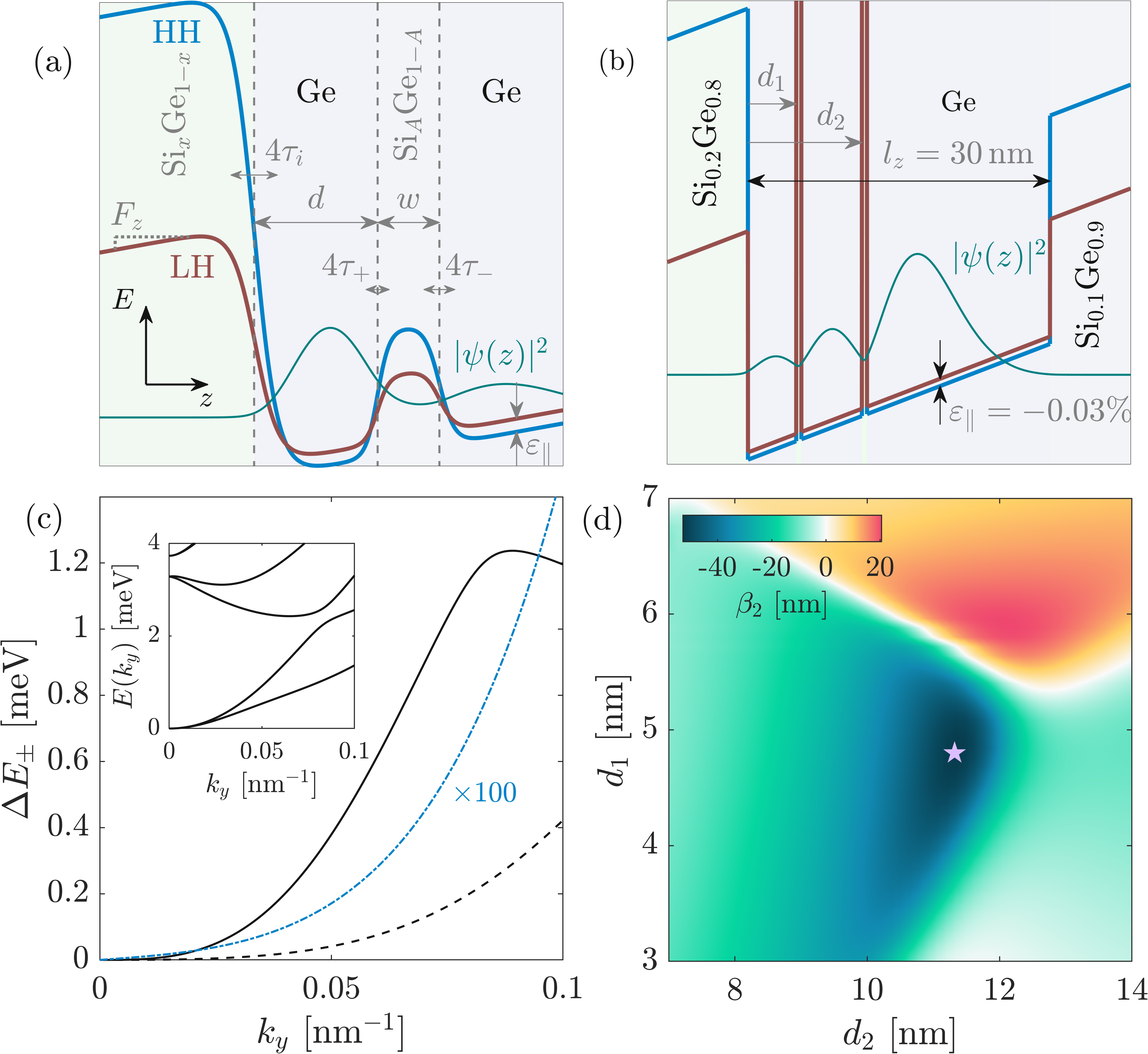}
    \caption{\textit{Ge+ heterostructures.}
    (a)-(b) Energy band alignment against position $z$ of Ge+ with (a) Si bump and (b) Si double spikes for a typical set of parameters. Optimization parameters are displayed in light gray. (c) SOI energy $E_\mathrm{so}$ of the HH ground state at $k_x=0$ against $k_y$ (black solid lines) for the manually optimized double spikes (see Table \ref{tab:params} for the list of parameters). For comparison, we show $E_\mathrm{so}$ in unstrained Ge with black dashed lines and in a $16\,\mathrm{nm}$-wide $\varepsilon$-Ge quantum well with blue dashed-dotted lines (rescaled by 2 orders of magnitude). Insets show the energy dispersion $E(k_y)$ of the optimal Ge+ heterostructure.}
    \label{fig:inputs}
\end{figure}

In this work, we introduce a strategy to consistently enhance SOI by engineering Si concentration gradients in unstrained Ge channels (Ge$+$). Our proposal builds on recent advances in the growth of unstrained Ge channels with top strained SiGe barriers ($\varepsilon$-SiGe)~\cite{Costa2026}  and  boosts SOI by incorporating (a) a smooth localized Si bump  and (b) two sharp Si spikes, see Fig.~\ref{fig:inputs}. We adopt multi-objective Bayesian optimization \cite{gardner2018gpytorch} to  improve the bump and spikes design, identifying profiles that maximize SOI while remaining stable against growth imperfections. The Si bumps and spikes enhance SOI by a factor $\sim 2$ and $\sim 15$ compared to unstrained Ge, respectively. This SOI is three orders of magnitude larger than state-of-the-art $\varepsilon$-Ge. Importantly, Ge+ is fully compatible with current material growth techniques and mirrors strategies successfully applied in electron-based systems~\cite{McJunkin2022,Woods2023,Woods2024}.  By comparing semiconducting and superconducting spin qubits across different heterostructures, we find that Ge+ devices outperform the state-of-the-art by orders of magnitude, demonstrating that engineering Si concentration spikes in Ge provides an efficient tool for coherent spin control.


\paragraph{SOI in Ge heterostructures.--}
Hole nanostructures are generally well-described by a $6$-band $k\cdot p$ Hamiltonian $H_{k\cdot p}$  that includes heavy, light, and split-off holes (HH, LH, SOH, respectively). This Hamiltonian includes the mixing of different hole species originating from finite momentum (Luttinger-Kohn Hamiltonian~\cite{Luttinger1955,Winkler2003,Voon2009,Eissfeller2011}), from epitaxial biaxial strain resulting from  lattice constant mismatch between SiGe and Ge (Bir-Pikus Hamiltonian~\cite{Bir1974}), and the effects of Si concentration gradients along the growth direction modeled by the virtual crystal approximation, see the supplemental material (SM) for more details~\cite{Note1}.

In Ge/$\varepsilon$-SiGe planar heterostructures the ground state has an HH character and its long wavelength  in-plane dynamics is nicely captured by a $2\times 2$ low-energy Hamiltonian~\cite{Note1}

\begin{equation}\label{HHeff}
    \mathcal{H}_\mathrm{eff} = \frac{\hbar^2}{2m_0}\gamma k^2 -\frac{i\hbar^2}{2m_0}\left(\beta_2k_+^3 -\beta_3k^2k_- \right)\sigma_-+ \text{h.c.} \ .
\end{equation}

We introduced the in-plane crystal momentum $\mathbf{k}=(k_x,k_y)$, with $k_\pm = k_x\pm ik_y=k e^{\pm i \chi}$ and $k=|\textbf{k}|$, the in-plane effective mass $m_0/\gamma$, and the free electron mass $m_0$. Importantly, when spatial inversion symmetry is broken along the growth direction $z$, e.g. from gate electric fields or asymmetric Si profiles~\cite{Winkler2003,Moriya2014}, the coupling between HH and LH generates an intrinsic cubic in momentum Rashba SOI that is parametrized by the two lengths $\beta_{2,3}$. Here, $\sigma_\pm = (\sigma_x\pm i\sigma_y)/2$, where the Pauli matrices $\sigma_i$ act on the groundstate pseudospin. Analytical formulas for the parameters of $\mathcal{H}_\mathrm{eff}$ are provided in the SM~\cite{Note1}.

More generally, we quantify  SOI by the  spin-splitting energy $E_\mathrm{so}(\textbf{k}) \equiv E_+(\textbf{k}) - E_-(\textbf{k})$ at zero magnetic field, with  $E_{\sigma=\pm}(\mathbf{k})$ being the energy eigenstate of pseudo-spin $\sigma=\pm1$ satisfying $E_+(\mathbf{k}) = E_-(-\mathbf{k})$.
From Eq.~\eqref{HHeff}, one finds the long-wavelength relation

\begin{equation}\label{Eso}
    E_\mathrm{so}(\mathbf{k}) = \frac{\hbar^2}{m_0}k^3\sqrt{\beta_2^2+\beta_3^2-2\beta_2\beta_3\cos 4\chi} + \mathcal{O}(k^5) \ .
\end{equation}

While  Eq.~\eqref{Eso} (and Eq.~\eqref{HHeff}) provides the correct spin-splitting at low momentum, as $k$ increases, HH and LH become increasingly mixed and $\mathcal{O}(k^5)$ corrections to $E_\mathrm{so}$ becomes significant. For this reason,  we also evaluate $E_\mathrm{so}$ including $\mathcal{O}(k^5)$ corrections by solving the full $6$-band $k\cdot p$ Hamiltonian $H_{k\cdot p}$~\cite{Eissfeller2011,Eissfeller2012} via the envelope function approximation~\cite{Bastard1975,Note1}.

\paragraph{Engineering SOI by Si concentration gradients.--}
The composition of the heterostructure along the growth direction can substantially modify the intrinsic SOI. In particular, by locally enriching the Ge channel with Si, we can efficiently engineer $E_\mathrm{so}$. To understand the origin of this effect, we consider the explicit expression of the SOI lengths $\beta_{2,3}$ in Eq.~\eqref{HHeff}. Keeping only $2$\textsuperscript{nd} order perturbative terms, neglecting the SOH and focusing only on $\beta_2$ (typically $\beta_3/\beta_2=(\gamma_3-\gamma_2)/(\gamma_3+\gamma_2)\approx 0.14$ in Ge), we find

\begin{equation}\label{b2}
    \beta_2 \approx \frac{\sqrt{3}\hbar^2}{m_0}\sum_j{\frac{\mu_j}{\Delta_j}\int{dz\,f_1^h\left[2\gamma_3f_j^{\ell\prime}+\left(\gamma'_3 - \kappa'\right)f_j^\ell\right]}} \ .
\end{equation}

\noindent We introduce $\mu_j=(\sqrt{3}/2)\int{dz\, f_1^h(z)[\gamma_2(z)+\gamma_3(z)]f_j^\ell(z)}$ with $\gamma_{2,3}(z)$ being position-dependent Luttinger parameters and $\kappa(z)$ being the position-dependent hole $g$-factor parameter. Also, $\Delta_j\equiv E_j^\mathrm{LH}-E_1^\mathrm{HH}$ is the HH-LH energy splitting between the HH ground state and the $j$-th LH subband and $f^{h,\ell}(z)$ are the envelope functions of the HHs and LHs, respectively. 

The first term in Eq.~\eqref{b2}, which depends on the derivative $f_j^{\ell\prime}(z)$, is the well-known contribution to $\beta_2$ coming from  broken spatial inversion symmetry~\cite{Winkler2003,Wang2022}. The second term arises from the spatial variation of the material constants, including abrupt changes of composition at the interfaces between neighboring materials, and engineered modulations of composition in the channel. This is the origin of the enhanced SOI in our Ge+ heterostructures.

\paragraph{Machine learning-optimized Si bump.--}
We first maximize  SOI in Ge+ channels comprising a single smooth Si-poor bump placed a few nm below the top $\varepsilon$-SiGe barrier, see Fig.~\ref{fig:inputs}(a). We optimize $9$ structural parameters: the bump thickness $w$ and amplitude $A$ (directly connected to its Si content), the barrier-bump distance $d$, the Si content  $x$ in the $\varepsilon$-Si$_{x}$Ge$_{1-x}$ barrier, the residual epitaxial in-plane strain $\varepsilon_\parallel$ in Ge, the applied electric field  $F_z$, the interface broadening of the top barrier $4\tau_i$, and the  broadening of the upper (lower) side of bump $4\tau_+$ ($4\tau_-$). The Si concentration profile $x(z)$ at each interface is modeled by the logistic function $x(z) = x_\mathrm{T} + (x_\mathrm{B}-x_\mathrm{T})\left(1+e^{-z/(4\tau)}\right)^{-1}$, where $x_\mathrm{T}$ ($x_\mathrm{B}$) is the concentration away from the top (bottom) of the interface. 

We adopt multi-objective Bayesian optimization~\cite{gardner2018gpytorch,Note1} to chart the Pareto front of structural parameters that maximize $\beta_2$ and simultaneously minimize instabilities from electric-field fluctuations by maximizing $\log (|\partial^2\beta_2/\partial F_z^2|^{-1})$. The exploration is successful in efficiently identifying numerous optimal solutions (see Fig.~S13 in \cite{Note1}). For all identified candidates, we adjust $F_z$ to ensure $\partial \beta_2/\partial F_z = 0$ and keep the heterostructure with the highest $\beta_2$. The specifications of the optimized Ge$+$ are shown in the first column of Table \ref{tab:params}. The SOI enhancement is $\sim 40$-fold compared to unstrained Ge and $4$ orders of magnitude larger compared to $\varepsilon$-Ge. However, it has a relatively small $\Delta_1$ gap, which prevents $E_\mathrm{so}(\mathbf{k})$ from increasing at the prescribed $k^3$-rate past a HH-LH anti-crossing near $k\sim 0.03\,\mathrm{nm}^{-1}$ (see Fig. S2~\cite{Note1}) and limits its value to $E_\mathrm{so}\sim 0.4\,\mathrm{meV}$ at larger momenta. This limitation partially offsets the SOI enhancement provided by the bump, motivating a trade-off between large SOI and large HH-LH gap.

\paragraph{Ge+ with Si spikes.--}
To increase $\Delta_1$ while maintaining large SOI, we consider a different structure comprising two sharp Si-rich spikes ($50\,\%$ Si) a few monolayers thick ($\sim 0.5\,\mathrm{nm}$) located at distances $d_1$ and $d_2$ respectively, below the top $\varepsilon$-Si$_{0.2}$Ge$_{0.8}$ barrier, see Fig. \ref{fig:inputs}(b). Importantly, integrating such Si spikes is experimentally feasible with mainstream chemical vapor deposition processes by exploiting self-saturating growth of Si on Ge at low temperature~\cite{scappucci_buried-channel_2023}. We restrict ourselves to two spikes to limit challenges in fabrication, sources of alloy disorder, and strain-relaxation hotspots in the Ge channel. We also note the inclusion of a bottom $\varepsilon$-Si$_{0.1}$Ge$_{0.9}$ barrier a distance $l_z$ below the top barrier to enhance $\Delta_j$.

We first approach spiked Ge+ by manually optimizing $|\beta_2|$ against $d_{1,2}$, while keeping fixed the remaining parameters: $F_z= 1.5\,\mathrm{mV}/\mathrm{nm}$, $4\tau_i = 0$, and $l_z = 30\,\mathrm{nm}$. We include a baseline residual strain $\varepsilon_\parallel = -0.03\%$ in Ge. Using these parameters, $|\beta_2(d_1,d_2)|$ is maximal at $d_1 = 4.8\,\mathrm{nm}$ and $d_2 = 11.3\,\mathrm{nm}$ (see Table \ref{tab:params}, 2\textsuperscript{nd} column). The corresponding SOI $\beta_2 = -51.5\,\mathrm{nm}$ is $15$-times greater than unstrained Ge and $3$ orders of magnitude larger than $\varepsilon$-Ge at equal $F_z$. A larger HH-LH gap $\Delta_1\approx3.28\,\mathrm{meV}$ enables a spin splitting $E_\mathrm{so}(\mathbf{k})$ up to $\sim 1.2\,\mathrm{meV}$ before the anti-crossing, see Fig. \ref{fig:inputs}(c).

\begin{table}[t]
    \centering
    \begin{threeparttable}
    \caption{Structural parameters and SOI coefficients for the optimized bumped Ge$+$ and the two spiked Ge$+$ heterostructures.}
    \begin{tabular}{l l l l l}
        \toprule
         & & Bump & Spikes (man.) & Spikes (ML) \\
        \midrule
        $F_z$ & [mV/nm] & $1.288$ & $1.5$\tnote{*} & $0.9448$ \\
        $d$ & [nm] & $8.07$ & $(4.8,11.3)$ & $(6.36,14.40)$ \\ 
        $w$ & [nm] & $1.70$ & $0.5$\tnote{*} & $0.5$\tnote{*} \\
        $A$ & [\%] & $8.8$ & $50$\tnote{*} & $50$\tnote{*} \\
        $x$ & [\%] & $28.9$ & $20$\tnote{*} & $20$\tnote{*} \\
        $\varepsilon_\parallel$ & [\%] & $-9.78\cdot 10^{-4}$ & $-0.03$\tnote{*} & $-9.20\cdot 10^{-3}$ \\
        $4\tau_i$ & [nm] & $3.01$ & $0$\tnote{*} & $1.582$ \\
        $4\tau_+$ & [nm] & $3.70$ & $(0,0)$\tnote{*} & $(0,0)$\tnote{*} \\
        $4\tau_-$ & [nm] & $3.77$ & $(0,0)$\tnote{*} & $(0,0)$\tnote{*} \\
        \midrule
        $\Delta_1$ & [meV] & $0.62$ & $3.28$ & $1.62$ \\
        $\beta_2^{\mathrm{Ge}+}$ & [nm] & $-186.2$ & $-51.5$ & $-75.9$ \\
        \midrule
        $\beta_2^\mathrm{Ge}$ & [nm] & $4.543$\tnote{$\dagger$} & $3.41$\tnote{$\dagger$} & $9.43$\tnote{$\dagger$} \\
        $\beta_2^{\varepsilon\text{-Ge}}$ & [nm] & $0.021$\tnote{$\ddagger$} & $0.0145$\tnote{$\ddagger$} & $0.0238$\tnote{$\ddagger$} \\
        \bottomrule
    \end{tabular}
    \label{tab:params}
    \begin{tablenotes}
        \item [*] Kept constant in optimization. \\
        \item [$\dagger$] Bump/spike-less Ge at corresponding $F_z$ ($A=0\%$). \\
        \item [$\ddagger$] $16\,\mathrm{nm}$ $\varepsilon$-Ge well between unstrained Si$_{0.2}$Ge$_{0.8}$ barriers at corresponding $F_z$.
    \end{tablenotes}
    \end{threeparttable}
\end{table}

Importantly, Ge+ with Si-spikes is robust against various heterostructure parameters. Fig. \ref{fig:inputs}(d) shows $\beta_2(d_1,d_2)$ at $F_z = 1.5\,\mathrm{mV}/\mathrm{nm}$, with the optimal solution marked by a purple star. The magnitude $|\beta_2|$ remains $>40\,\mathrm{nm}$ for variations of $\pm 0.5\,\mathrm{nm}$ in $(d_1,d_2)$-space around the optimal solution. Interestingly, there are distances $(d_1,d_2)$ where $\beta_2$ vanishes [white regions in Fig. \ref{fig:inputs}(d)]. As expected, larger $|\beta_2|$ regions coincide with smaller $\Delta_1$ regions, and larger residual strains increase $\Delta_1$ and decrease $\beta_2$ [see Fig. S3(a)-(b)~\cite{Note1}]. Finally, spiked Ge+ is also robust against variations of the well width $l_z$ when $l_z>25\,\mathrm{nm}$ [see Fig. S3(c)~\cite{Note1}].


\paragraph{Machine learning-optimized Si spikes.--}
We search for potentially higher SOI spiked Ge+ systems by performing  again a multi-objective Bayesian optimization over five of the aforementioned parameters: $d_{1,2}$, $\varepsilon_\parallel$, $F_{z}$, and $4\tau_i$, and define the objectives to optimize $\beta_2$ and $\log (|\partial\beta_2/\partial F_z|^{-1})$. We take $l_z\to\infty$ for simplicity. 
The optimization considered a relatively narrow range of parameters, close to the manually obtained solution. From the candidates sampled during the optimization (see
Fig. S14 in \cite{Note1}), we then filter and focus on heterostructures showing an extremum against $F_z$ with large $\beta_2$. The specifications of the optimal spiked Ge+ found with machine learning are displayed in column $3$ of Table \ref{tab:params}. Compared to the manually optimized system, notably, it shows a larger $|\beta_2|$ and a smaller $\Delta_1$ at lower residual strain $|\varepsilon_\parallel|$.

Similar to bumped Ge+, the smaller $\Delta_1$ energy partially offsets the SOI enhancements from the spikes, and leads to a maximal Rashba splitting energy of only $0.69\,\mathrm{meV}$ at the anti-crossing with the first LH level near $k \approx 0.055\,\mathrm{nm}^{-1}$. A larger Rashba splitting may be obtained by increasing the magnitude of $\varepsilon_\parallel$, which would directly increase $\Delta_1$, but at the cost of a smaller $\beta_2$. We also note that machine learning converged to larger $d_{1,2}$ and smaller $F_z$ compared to the manually optimized system, which suggests the existence of a family of optimal solutions in the $(F_z,d_1,d_2)$ parameter space that trade large $\beta_2$ for small $\Delta_1$. We confirm this with one last series of manual optimizations on the distances $(d_1,d_2)$ that maximize $\beta_2$ for different $F_z$'s, while keeping fixed $\varepsilon_\parallel = -0.03\%$ and $\tau_i = 0$ for simplicity (see Fig. S4 and S5 in \cite{Note1}). The heterostructure in the $2$\textsuperscript{nd} column of Table \ref{tab:params} is thus a compromise between enhancing $\beta_2$ and $1/\Delta_1$, given that smaller $\Delta_1$ saturates $E_\mathrm{so}(\mathbf{k})$ to smaller values. We focus on this system in the remaining of this work.

\begin{figure}[t]
    \centering
    \includegraphics[width=1.0\linewidth]{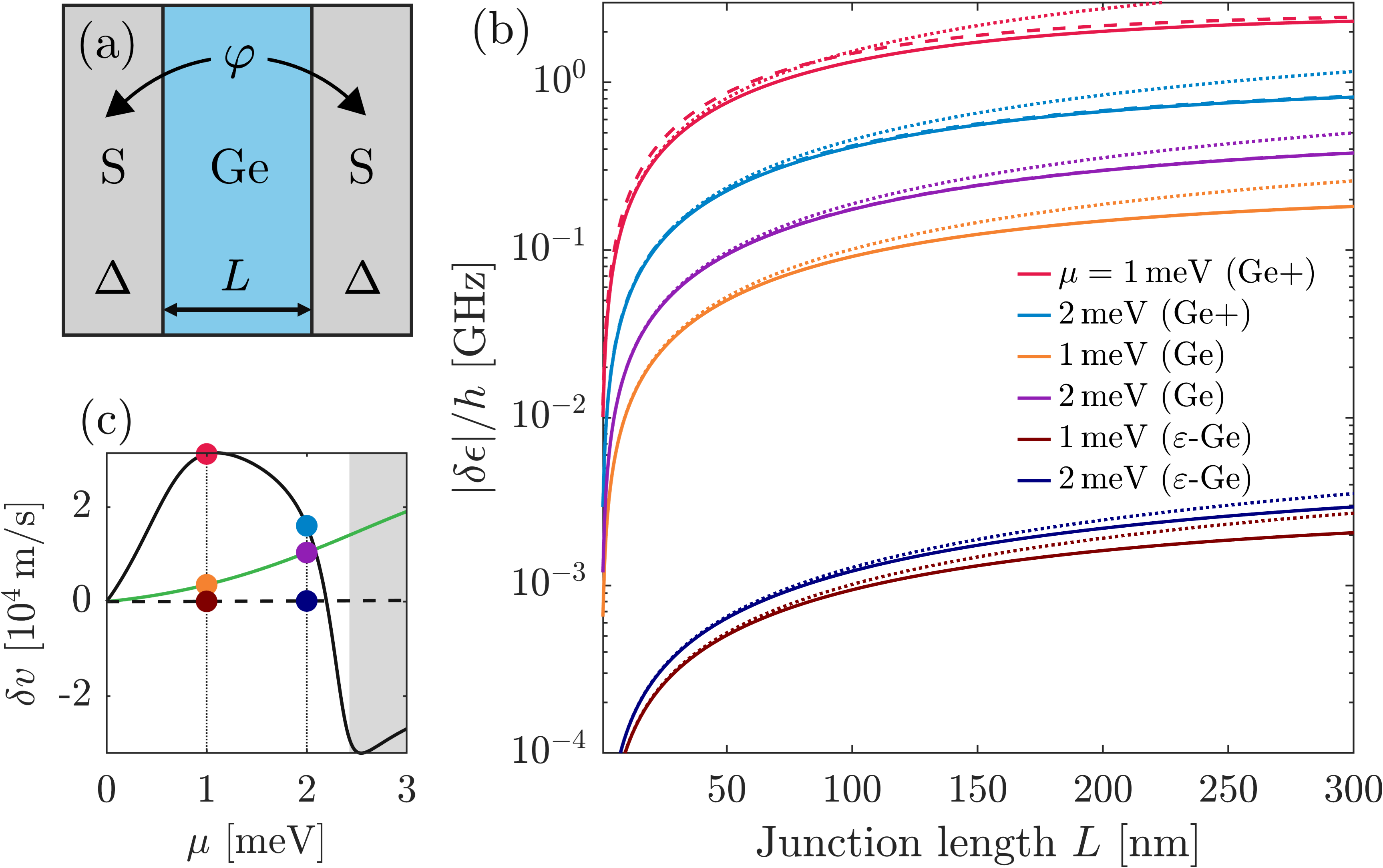}
    \caption{\textit{Ge+ Andreev spin qubits.} (a) Schematic of a Josephson junction with a semiconducting channel width $L$, phase difference $\varphi$, and superconducting gap $\Delta$. (b) $\delta\epsilon/h$ at $\varphi=\pi/2$ against the junction length $L$. Solid lines are the exact computation of $\delta\epsilon$, dashed (dotted)lines correspond to the approximate Eq.~\eqref{deL} (Eq. \eqref{deS}). (c) Fermi velocity difference $\delta v$ against $\mu$ for the optimal Ge+ spiked system (black line) and for the reference Ge and $\varepsilon$-Ge (green and dashed lines). $\delta v\approx 0.026\cdot10^4\,\mathrm{m}/\mathrm{s}$ for $\varepsilon$-Ge at $\mu = 3\,\mathrm{meV}$. The shadowed region indicates the starting point of the 2\textsuperscript{nd} well subband in Ge+.}
    \label{fig:asq}
\end{figure}

\paragraph{Ge+ for hybrid systems.--}
To demonstrate the advantage of Ge+, we analyze the hybrid superconducting-semiconducting system sketched in Fig. \ref{fig:asq}(a). Specifically, we consider an Andreev spin qubit (ASQ)~\cite{Bargerbos2023,Hays2021,Kurilovich2021} confined in a Josephson junction with a semiconducting channel of length $L$ and two superconducting regions with superconducting gap $\Delta$ and phase difference $\varphi$. We note that a recent analysis of Ge ASQ~\cite{Coppini2026} shows that unstrained Ge already significantly outperforms $\varepsilon$-Ge providing larger spin splittings $\delta\epsilon$ due to larger SOI. 

We assume the chemical potential $\mu\gg\Delta$, such that the Hamiltonian within the three regions can be linearized in momentum close to $\mu$. We neglect quantization effects arising from the width $W$ of the junction by assuming $W\to\infty$, i.e., $k=0$ along that direction~\cite{Coppini2026}. The cubic SOI in Ge gives rise to two different Fermi velocities $v_\pm=v \pm\delta v/2$ for the spins providing a spin-split bound state at no magnetic field. Larger spin splittings, approaching the GHz regime, can enable efficient microwave control and readout of the superconducting spin qubit. Here, we evaluate $v_\pm$ from the full diagonalization of $H_{k\cdot p}$.
The spin-splitting energy is~\cite{Note1},

\begin{equation}\label{deL}
    \delta\epsilon=\epsilon_--\epsilon_+ = -\frac{\epsilon/v}{1+E_L/\sqrt{\Delta^2-\epsilon^2}}\delta v+\mathcal{O}\left(\delta v^3\right) \ ,
\end{equation}

\noindent where $E_L = \hbar v/L$, and $\epsilon$ is the $\varphi$-dependent energy of the bound state at $\delta v=0$. Notably, $|\delta v|\approx(4\beta_2/\gamma)\mu/\hbar$ to lowest order of the chemical potential~\cite{Note1}. In short junctions, i.e. $L\ll\hbar v/\Delta$, the spin splitting is 

\begin{equation}\label{deS}
    \delta\epsilon \approx \frac{\Delta^2}{2E_L}\left[\sin\varphi + \frac{\Delta}{2E_L}\left(\cos\frac{\varphi}{2} + 3\cos\frac{3\varphi}{2}\right)\right]\frac{\delta v}{v} \ .
\end{equation}

The spin splitting of the lowest energy bound state is shown in Fig. \ref{fig:asq}(b) against $L$ at fixed $\mu$ using $\Delta=70\,\upmu\mathrm{eV}$~\cite{Lakic2025}. Ge+ show consistently spin splittings above $100\,\mathrm{MHz}$, entering also the GHz regime. In contrast to Ge and $\varepsilon$-Ge, the behavior of $\delta\epsilon$ against $\mu$ is not-monotonic for Ge+ as $\delta v$ is also a non-monotonic function of $\mu$, see Fig. \ref{fig:asq}(c). In Ge+, the anti-crossing with the LH subband leads to $\delta v$ reaching a maximum near $\mu = 1\,\mathrm{meV}$ before decreasing at larger $\mu$. This in turn leads to a smaller $\delta\epsilon$ at $\mu=2\,\mathrm{meV}$ compared to $\mu=1\,\mathrm{meV}$. Despite the anti-crossing, $\delta\epsilon$ approaches $\sim 2\,\mathrm{GHz}$ at $L\sim 200\,\mathrm{nm}$, compared to $\delta\epsilon\approx1.6\,\mathrm{MHz}$ for $\varepsilon$-Ge.

\begin{figure}
    \centering
    \includegraphics[width=\linewidth]{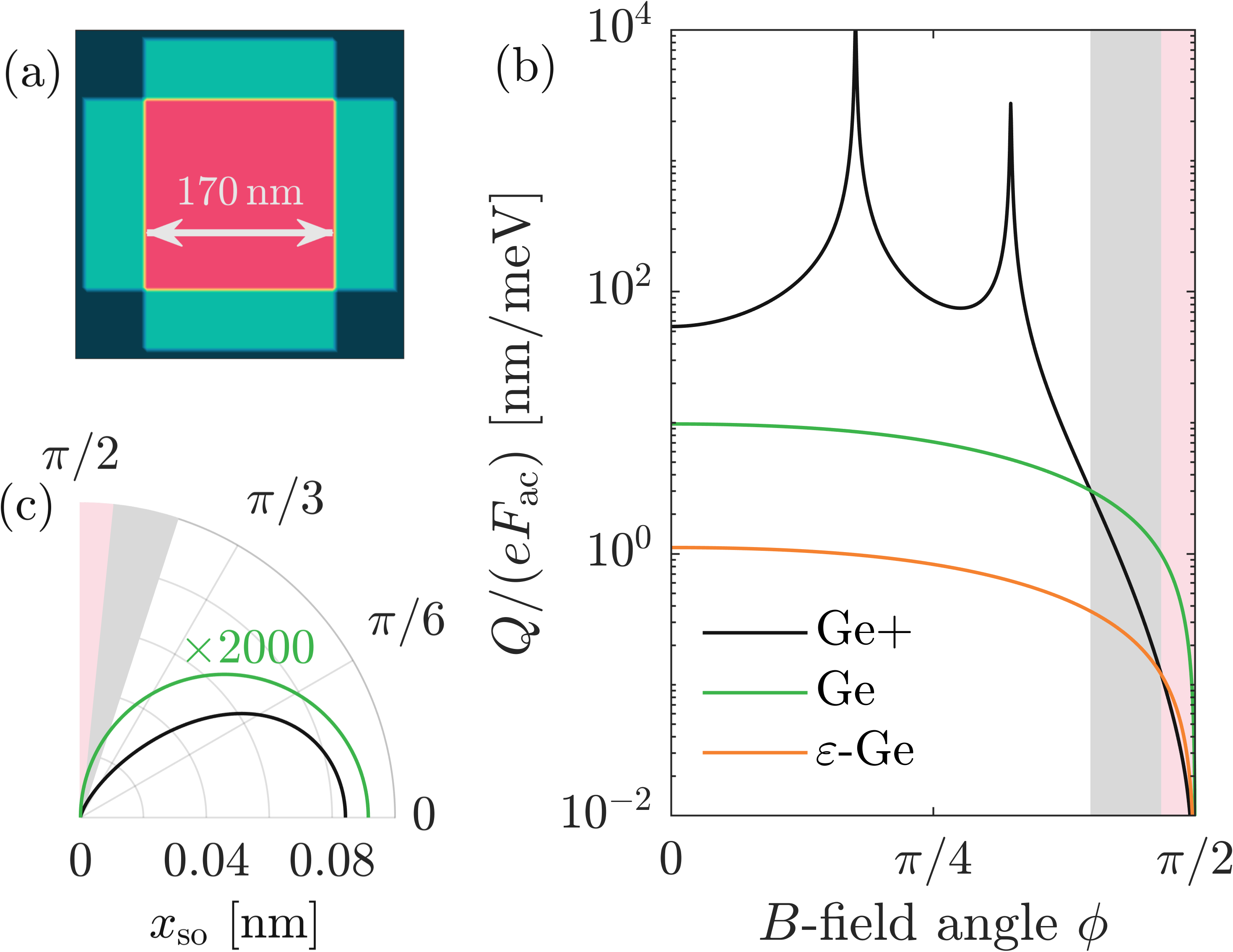}
    \caption{\textit{Ge+ dot spin qubits.} (a) Top view of the gates that define the quantum dot. The color palette indicates the applied voltage. (b) Qubit $Q$-factor rescaled by the driving field amplitude $eF_\mathrm{ac}$ against the $B$-field in-plane angle $\phi$. $B = 0.1\,\mathrm{T}$. (c) Spin-dipole moment $x_\mathrm{so}$  against $\phi$ using the same color legend as in panel (b).}
    \label{fig:qubits}
\end{figure}

\paragraph{Ge+ for spin qubits.--}
Interestingly, the enhanced SOI in Ge+ improves the performance of gate-defined spin qubits by enabling faster qubit operation without a significant increase in decoherence. We consider a single hole confined in the electrostatic potential underneath a plunger gate and four surrounding identical barrier gates, see Fig.~\ref{fig:qubits}(a). The gate dimensions and voltages are similar to those used in experiments~\cite{John2025,Tosato2025}, and are chosen to produce an electric field $F_z\sim 1.5\,\mathrm{mV}/\mathrm{nm}$ at the dot. The spin qubit is encoded in the spin doublet of the lowest-energy orbital in the quantum dot, is spin-split by an in-plane magnetic field at angle $\phi$ with respect to the $x$ axis, and is driven electrically by an oscillating electric field $\tilde{\mathbf{F}}(t)\parallel\mathbf{e}_x$ resonant with the qubit Larmor frequency~\cite{Bulaev2007}. The qubit dephases through charge noise  modeled by small fluctuations of gate voltages~\cite{Piot2022}. Assuming $1/f$ charge noise, we estimate the dephasing time as $T_2^*\sim\hbar/\sqrt{\sum_i{\varepsilon_i^2}}$~\cite{Bassi2026,Piot2022,SaezMollejo2025,Hendrickx2024}, where $\varepsilon_i$ is the variation of the qubit Larmor frequency caused by the five uncorrelated gate voltage fluctuations $\delta_i = 1\,\mathrm{mV}$. The energy variations $\varepsilon_i$ are computed by expanding small changes of the electrostatic potential in the vicinity of the center of the dot, and using the qubit orbital wavefunctions simulated from the full $k\cdot p$ Hamiltonian of the quantum dot~\cite{Note1,Terrazos2021,Wang2021}. The Rabi frequency of the qubit is proportional to the spin-dipole moment $x_\mathrm{so}$~\cite{Efros2006}, which is the transverse component of the $x$ position operator in the quantum dot ground-state subspace for a given $B$-field configuration~\cite{Bosco2021PRB,Bosco2022PRL}. Here, we focus on the effects of Si gradients in the virtual crystal approximation, but we remark that in realistic devices, atomistic disorder at the interfaces~\cite{Szmulowicz2004}, gate-induced shear strains~\cite{AbadilloUriel2023}, and quantum dots deformations~\cite{Bosco2021PRB}, could further modify our estimations.

We use the quality factor $Q = (T_2^*x_\mathrm{so}/\hbar)eF_\mathrm{ac}$ of the qubit to meaningfully compare the performance of the heterostructures at equivalent resonant driving field amplitudes $F_\mathrm{ac}$. In Fig. \ref{fig:qubits}(b) we compare $Q$ for different heterostructures. While the significant boost in SOI of Ge+ could cause larger dephasing~\cite{Note1}, the $Q$-factor is substantially enhanced for a large portion of $B$-field in-plane angles compared to unstrained and $\varepsilon$-Ge. At angles $45^\circ\pm13.32^\circ$, $T_2^*$ is boosted due to vanishing longitudinal components of the operators $z$ and $x^2+y^2$, providing a sweet spot for charge noise~\cite{Bassi2026,Piot2022,Hendrickx2024}. Beyond $72^\circ$ (gray shading) and $84.15^\circ$ (red shading) away from the driving axis, Ge+ becomes less advantageous than Ge and $\varepsilon$-Ge, respectively, as $x_\mathrm{so}$ becomes too small to compensate the reduction of $T_2^*$, see Fig. \ref{fig:qubits}(c). Remarkably, Fig. \ref{fig:qubits}(b) suggests that unstrained Ge already exhibits an $\sim 8$-fold $Q$-factor improvement over $\varepsilon$-Ge for all $B$-field in-plane angles.

\paragraph{Conclusion.--}
We propose an enriched Ge+ heterostructure that incorporates Si concentration gradients to significantly enhance SOI. We identify multiple routes to engineer Si gradients - implemented via a single Si-poor bump and two Si-rich spikes - and leverage multi-objective Bayesian optimization to chart the Pareto front of heterostructures display optimal trade-offs between SOI maximization, and its robustness against parameter variations. We further show that devices built on ML-optimized Ge+ have the potential to substantially outperform state-of-the-art superconducting and semiconducting spin-qubit devices, opening new paths toward Ge-based quantum technologies.

\paragraph{Acknowledgment.--}
 We thank members of the Bosco, Scappucci, Rimbach-Russ, and Veldhorst groups for valuable discussions. We acknowledge that this work was carried out in parallel with Ref~\cite{Coppini2026} on strain engineering of Ge Andreev spin qubit. We are grateful to the authors - V. Coppini, A. Manesco, A. Akhmerov, V. Fatemi, and B. Van  Heck-  for their insightful discussions and collaboration.  This research was supported by the EU through the H2024 QLSI2 project,  by the Army Research Office under Award Number: W911NF-23-1-0110, by NCCR Spin (grant number 225153), and by The Netherlands Ministry of Defense under Awards No. R23/009. The views, conclusions, and recommendations contained in this document are those of the authors and are not necessarily endorsed nor should they be interpreted as representing the official policies, either expressed or implied, of The Netherlands Ministry of Defense, of the Army Research Office or the U.S. Government. The U.S. Government and The Netherlands Ministry of Defense is authorized to reproduce and distribute reprints for Government purposes notwithstanding any copyright notation herein.

\paragraph{Data availability statement.--}
The code is openly available in our repository: \url{https://github.com/kevinrossitud/gesi_mobo}.

%

\end{document}


\title{\texorpdfstring{Supplemental Material:\\
Tailoring Germanium Heterostructures for Quantum Devices with Machine Learning}{}
}

\author{Patrick Del Vecchio}
\email{p.delvecchio@tudelft.nl}
\affiliation{QuTech and Kavli Institute of Nanoscience, Delft University of Technology, Delft, Netherlands}
\author{Kevin Rossi}
\affiliation{Department of Materials Science and Engineering, Delft University of Technology, Delft, Netherlands}
\affiliation{Climate Safety and Security Centre, TU Delft The Hague Campus, Delft University of Technology, 2594 AC, The Hague, The Netherlands}
\author{Giordano Scappucci}
\affiliation{QuTech and Kavli Institute of Nanoscience, Delft University of Technology, Delft, Netherlands}
\author{Stefano Bosco}
\email{s.bosco@tudelft.nl}
\affiliation{QuTech and Kavli Institute of Nanoscience, Delft University of Technology, Delft, Netherlands}

\begin{abstract}
    In this Supplemental Material, we describe the $k\cdot p$ formalism (Section \ref{sec:kp}) and the method by which it is diagonalized for heterostructures (Section \ref{sec:kp_}). In Section \ref{sec:pert}, we present the perturbative framework used to calculate spin-orbit parameters and effective Hamiltonians. Sections \ref{sec:bump} and \ref{sec:spikes} contain additional results on the Si-bump and Si-spikes systems, respectively, including SOI spin splittings, and the SOI strength $\beta_2$ against various heterostructure parameters. In Section \ref{sec:asq} we present our framework leading to the Andreev bound state energy formula, and the spin splittings for arbitrary junction length and in the short junction limit. This section ends with a few expansions of various powers of Fermi velocities in terms of the chemical potential and band structure parameters. Then, the methodology for diagonalizing the quantum dot Hamiltonian and calculating $T_2^*$ is outlined in Section \ref{sec:qd}. Finally, the machine-learning optimization method is described in Section \ref{sec:ml-method}.
\end{abstract}

\maketitle

\section{\texorpdfstring{General 6-band $k\cdot p$ Hamiltonian}{}}\label{sec:kp}
The electronic band structure of the heterostructures are evaluated from $6$-band $k\cdot p$ theory. We consider a two-dimensional hole gas (2DHG) under the influence of a gate electric field $\mathbf{F} = F_z\mathbf{e}_z$ applied along the $z$ direction, which is perpendicular to the 2DHG plane, an externally applied uniform magnetic field

\begin{equation}
    \mathbf{B} = B\left(\mathbf{e}_x\sin\theta\cos\phi+\mathbf{e}_y\sin\theta\sin\phi+\mathbf{e}_z\cos\theta\right),
\end{equation}

\noindent and an in-plane confinement represented by an effective parabolic potential 

\begin{equation}
    V_\parallel(x,y) = -\alpha_0\left(\frac{x^2}{l_x^4} + \frac{y^2}{l_y^4}\right),
\end{equation}

\noindent characterized by the lengths $l_x$ and $l_y$ and where $\alpha_0 = \hbar^2/(2m_0)$. The minus sign in the potential comes from our electron-energy convention (the valence band has negative effective mass). To agree with literature, the energy dispersion provided in the main text are reverted in their sign, i.e. positive effective mass. 

The total Hamiltonian can be written as a sum of multiple contributions: a kinetic part $H_k(\mathbf{K})$, a spin-orbit part $H_\mathrm{so}$, an epitaxial strain part $H_\varepsilon$ and a potential energy part $V(\mathbf{r})$. We write the mechanical wavevector $\mathbf{K}=\mathbf{k}+e\mathbf{A}/\hbar$, where $\bra{\mathbf{r}}\mathbf{k} = -i\nabla\bra{\mathbf{r}}$ and $\mathbf{A}$ is the vector potential for the externally applied magnetic field $\mathbf{B} = \nabla\times\mathbf{A}$. The components of $\mathbf{K}$ obey $\mathbf{K}\times\mathbf{K} = e\mathbf{B}/(i\hbar)$. We first write the Hamiltonian in the basis of the bulk Bloch states at the $\Gamma$-point in the cartesian representation:

\begin{equation}
    \mathcal{B}_X = \left\{\ket{p_x+},\ket{p_y+},\ket{p_z+},\ket{p_x-},\ket{p_y-},\ket{p_z-}\right\}.
\end{equation}

The Hamiltonian is

\begin{equation}
    H_{k\cdot p}(\mathbf{K}) = H_k(\mathbf{K}) + H_\mathrm{so} + H_\varepsilon + V(\mathbf{r}),
\end{equation}

\noindent where $V(\mathbf{r}) = \mathcal{E}_{\Gamma_5^+}(z) + eF_zz+V_\parallel(x,y)$, where $\mathcal{E}_{\Gamma_5^+}$ is the valence band edge energy at the $\Gamma$-point {\it without} spin-orbit interaction. The kinetic term is

\begin{equation}
    H_k(\mathbf{K}) = \mathbb{1}_{2\times 2}\otimes H^k_{vv}(\mathbf{K})+H_B^k(\mathbf{K})\otimes\mathbb{1}_{3\times 3} + H_q(\mathbf{K}),
\end{equation}

\noindent with

\begin{align}
    H_{vv}^k(\mathbf{K}) = \begin{bmatrix}
        K_xLK_x+K_yMK_y+K_zMK_z & K_xN_+K_y+K_yN_-K_x & K_xN_+K_z+K_zN_-K_x \\
        K_yN_+K_x+K_xN_-K_y & K_xMK_x+K_yLK_y+K_zMK_z & K_yN_+K_z+K_zN_-K_y \\
        K_zN_+K_x+K_xN_-K_z & K_zN_+K_y+K_yN_-K_z & K_xMK_x+K_yMK_y+K_zLK_z
    \end{bmatrix},
\end{align}

\noindent where $L$, $M$ and $N_\pm$ are related to the Luttinger parameters $\gamma_{1,2,3}$ and $\kappa$:

\begin{equation}
    \begin{bmatrix}
        L \\
        M \\
        N_+ + \alpha_0 \\
        N_- - \alpha_0
    \end{bmatrix} = -\alpha_0\begin{bmatrix}
        1 & 4 & 0 & 0 \\
        1 & -2 & 0 & 0 \\
        0 & 0 & 3 & 3 \\
        0 & 0 & 3 & -3
    \end{bmatrix}\begin{bmatrix}
        \gamma_1 \\
        \gamma_2 \\
        \gamma_3 \\
        \kappa
    \end{bmatrix},
\end{equation}

\noindent with

\begin{equation}
    H_B^k(\mathbf{K}) = \frac{i\alpha_0g_0}{2}\sum_{\alpha\beta\gamma}{\epsilon_{\alpha\beta\gamma}K_\alpha K_\beta\sigma_\gamma},
\end{equation}

\noindent with $g_0\approx 2$ and $\sigma_\alpha$ the Pauli matrices. The $H_q$ term is

\begin{equation}
    H_q(\mathbf{K}) = U_{JX}^\dagger\left[\left(-2i\alpha_0\sum_{\alpha\beta\gamma}{\epsilon_{\alpha\beta\gamma}K_\alpha qK_\beta J_\gamma^3}\right)\oplus\mathbb{0}_{2\times 2}\right]U_{JX},
\end{equation}

\noindent where $q$ is the cubic-in-$J$ correction to the hole $g$-factor, $J_\alpha$ are the unitless spin $3/2$ matrices ($[J_\alpha,J_\beta] = i\epsilon_{\alpha\beta\gamma}J_\gamma$ and basis ordering $m_j=3/2,1/2,-1/2,-3/2$) and where the unitary $U_{JX}$ is defined further below. The spin-orbit term is proportional to the bulk split-off gap $\Delta_0$ and given by

\begin{equation}
    H_\mathrm{so} = \frac{\Delta_0}{3}\sum_\alpha{\sigma_\alpha\otimes\lambda_\alpha},
\end{equation}

\noindent where

\begin{align}
    \lambda_x &= \begin{bmatrix}
        0 & 0 & 0 \\
        0 & 0 & -i \\
        0 & i & 0
    \end{bmatrix}, &
    \lambda_y &= \begin{bmatrix}
        0 & 0 & i \\
        0 & 0 & 0 \\
        -i & 0 & 0
    \end{bmatrix}, &
    \lambda_z &= \begin{bmatrix}
        0 & -i & 0 \\
        i & 0 & 0 \\
        0 & 0 & 0
    \end{bmatrix}.
\end{align}

\noindent  $H_\mathrm{so}$ has eigenvalues $0$ ($\times 2$), $-2\Delta_0/3$ ($\times 2$) and $+\Delta_0/3$ ($\times 4$). Finally, the strain term is $H_\varepsilon = \mathbb{1}_{2\times 2}\otimes H_{vv}^\varepsilon$, where (assuming a diagonal strain tensor)

\begin{equation}
    H_{vv}^\varepsilon = \begin{bmatrix}
        l\varepsilon_{xx} + m\varepsilon_{yy} + m\varepsilon_{zz} & 0 & 0 \\
        0 & m\varepsilon_{xx} + l\varepsilon_{yy} + m\varepsilon_{zz} & 0 \\
        0 & 0 & m\varepsilon_{xx} + m\varepsilon_{yy} + l\varepsilon_{zz} \\
    \end{bmatrix},
\end{equation}

\noindent where $\varepsilon_{ij}$ are the components of the strain tensor and $l$ and $m$ are related to the $a_v$ and $b$ deformation potential constants:

\begin{equation}
    \begin{bmatrix}
        l \\
        m
    \end{bmatrix} = \begin{bmatrix}
        1 & 2 \\
        1 & -1
    \end{bmatrix}\begin{bmatrix}
        a_v \\
        b
    \end{bmatrix}.
\end{equation}

Note that the ordering between material constants (e.g. $\gamma_1$, $a_v$, $\mathcal{E}_{\Gamma_5^+}$, etc \dots) and wavevector operator components $K_i$ must be preserved as indicated. This is because (a) different components of $K_i$ do not mutually commute (c.f. $\mathbf{K}\times\mathbf{K} = e\mathbf{B}/(i\hbar)$) and improper ordering would lead to incorrect Zeeman splittings and (b) the $k_z$ operator does not commute with position-dependent material parameters. In fact, if $\gamma\ket{z} = \ket{z}\gamma(z)$ where $\gamma(z)$ is the value of $\gamma$ at position $z$, then for any envelope function $\braket{z|\psi} = \psi(z)$: $\bra{z}[\gamma,k_z]\ket{\psi} = i\gamma^\prime(z)\psi(z)$ where $\gamma^\prime(z) = \frac{d}{dz}\gamma(z)$.

The $k\cdot p$ Hamiltonian is instead often written in a basis that diagonalizes $H_\mathrm{so}$. A somewhat conventional basis is the following~\cite{Winkler2003}:

\begin{equation}
    \mathcal{B}_J = \left\{\Ket{\frac{3}{2},\frac{3}{2}},\Ket{\frac{3}{2},\frac{1}{2}},\Ket{\frac{3}{2},-\frac{1}{2}},\Ket{\frac{3}{2},-\frac{3}{2}},\Ket{\frac{1}{2},\frac{1}{2}},\Ket{\frac{1}{2},-\frac{1}{2}}\right\},
\end{equation}

\noindent where

\begin{align}
    \Ket{\frac{3}{2},\frac{3}{2}}\equiv\Ket{\mathrm{HH}+} &= -\frac{1}{\sqrt{2}}\left(\ket{p_x+}+i\ket{p_y+}\right), \\
    \Ket{\frac{3}{2},\frac{1}{2}}\equiv\Ket{\mathrm{LH}+} &= \frac{1}{\sqrt{6}}\left(2\ket{p_z+}-\ket{p_x-}-i\ket{p_y-}\right), \\
    \Ket{\frac{3}{2},-\frac{1}{2}}\equiv\Ket{\mathrm{LH}-} &= \frac{1}{\sqrt{6}}\left(\ket{p_x+}-i\ket{p_y+}+2\ket{p_z-}\right), \\
    \Ket{\frac{3}{2},-\frac{3}{2}}\equiv\Ket{\mathrm{HH}-} &= \frac{1}{\sqrt{2}}\left(\ket{p_x-}-i\ket{p_y-}\right), \\
    \Ket{\frac{1}{2},\frac{1}{2}}\equiv\Ket{\mathrm{SO}+} &= -\frac{1}{\sqrt{3}}\left(\ket{p_z+}+\ket{p_x-}+i\ket{p_y-}\right), \\
    \Ket{\frac{1}{2},-\frac{1}{2}}\equiv\Ket{\mathrm{SO}-} &= -\frac{1}{\sqrt{3}}\left(\ket{p_x+}-i\ket{p_y+}-\ket{p_z-}\right).
\end{align}

\noindent The unitary $U_{JX}$ connecting the two bases is therefore

\begin{equation}
    U_{JX} = \begin{bmatrix}
        -s_2 & is_2 & 0 & 0 & 0 & 0 \\
        0 & 0 & s_{23} & -s_6 & is_6 & 0 \\
        s_6 & is_6 & 0 & 0 & 0 & s_{23} \\
        0 & 0 & 0 & s_2 & is_2 & 0 \\
        0 & 0 & -s_3 & -s_3 & is_3 & 0 \\
        -s_3 & -is_3 & 0 & 0 & 0 & s_3 \\
    \end{bmatrix},
\end{equation}

\noindent with $s_2=1/\sqrt{2}$, $s_3 = 1/\sqrt{3}$, $s_6 = 1/\sqrt{6}$ and $s_{23} = \sqrt{2/3}$.

\section{\texorpdfstring{First diagonalization of the $k\cdot p$ matrix at $B=0$ and $\mathbf{k}_\parallel = 0$}{}}\label{sec:kp_}
The $6$-band $k\cdot p$ matrix is first written down at $k_x = k_y = 0$ (and $B=0$) and $V_\parallel(x,y)=0$, whose eigenstates provide an orthonormal basis on which the full Hamiltonian at $\mathbf{K} \neq 0$ is projected. In a basis similar to $\mathcal{B}_J$ but with a different ordering:

\begin{equation}
    \mathcal{B}_\mathcal{J} = \left\{\ket{\mathrm{LH}+},\ket{\mathrm{SO}+},\ket{\mathrm{HH}+},\ket{\mathrm{LH}-},\ket{\mathrm{SO}-},\ket{\mathrm{HH}-}\right\},
\end{equation}

\noindent the Hamiltonian $H_0^\mathrm{qw}\equiv H_{k\cdot p}(B=0;k_x=k_y=0)$ takes the following form:

\begin{align}\label{H0}
  H_0^\mathrm{qw} &= \begin{bmatrix}
    H_{\sigma=+} & 0 \\
    0 & H_{\sigma=-}
  \end{bmatrix} \\
  H_\sigma &= H_\sigma^k + H_\sigma^\varepsilon + V_z,
\end{align}

\noindent where

\begin{align}
  H_\sigma^k &= \alpha_0\begin{bmatrix}
  -k_z(\gamma_1+2\gamma_2)k_z & 2\sqrt{2}\sigma k_z\gamma_2k_z & 0 \\
  & -k_z\gamma_1k_z & 0 \\
  \dagger & & -k_z\gamma_-k_z
  \end{bmatrix}, \\
  H_\sigma^\varepsilon &= a_v\mathrm{Tr}\,\varepsilon + b\cdot\delta\varepsilon\begin{bmatrix}
  -1 & \sqrt{2}\sigma & 0 \\
  & 0 & 0 \\
  \dagger & & 1
  \end{bmatrix}, \\
  V_z &= \mathcal{E}_{\Gamma_5^+} + \frac{\Delta_0}{3} + eF_zz - \Delta_0\begin{bmatrix}
  0 & 0 & 0 \\
  & 1 & 0 \\
  \dagger & & 0
  \end{bmatrix}.
\end{align}

Here, $\gamma_\pm = \gamma_1 + (1\pm 3)\gamma_2$, $\delta\varepsilon = \varepsilon_{xx} - \varepsilon_{zz}$. We assume that all layers are pseudomorphic to each other, i.e., the in-plane lattice constant does not change along the growth direction and $\varepsilon_{xx}=\varepsilon_{yy}$. The out-of-plane component of the strain tensor is evaluated with $\varepsilon_{zz} = -(2c_{12}/c_{11})\varepsilon_{xx}$, where $c_{11}$ and $c_{12}$ are material-specific elastic constants~\cite{Madelung1991}. The energy band offsets between Ge and SiGe and the deformation potential constant $b$ are obtained by linearly interpolating the values reported in Ref.~\cite{VandeWalle1986}. The Luttinger parameters $\gamma_{1,2,3}$ and $g$-factor parameter $\kappa$ are taken from Ref.~\cite{Winkler2003}. The three $\gamma_{1,2,3}$ and $\kappa$ are interpolated in the full composition range of SiGe according to the method outlined in Ref.~\cite{Winkler1996}. We stress again that the ordering between $k_z$ and material parameters must be preserved.

The block-diagonal structure of Eq.~\eqref{H0} suggests that its eigenstates are of two types ($2\times 2 = 4$ including pseudospin $\sigma$):

\begin{subequations}\label{types}
  \begin{align}
    \ket{\h{l};\sigma} &= \Ket{\frac{3}{2}, \frac{3\sigma}{2}}\ket{f^h_l}, \\
    \ket{\e{j};\sigma} &= \Ket{\frac{3}{2}, \frac{\sigma}{2}}\ket{f^\ell_j} + \sigma\Ket{\frac{1}{2}, \frac{\sigma}{2}}\ket{f^s_j},
  \end{align}
\end{subequations}

\noindent where the first ($\h{}$ states) consists of pure HH subbands with associated envelope functions $\braket{z|f_l^h}$ and spin-independent energies $E_l^\h{}$, and where the second ($\e{}$ states) consists of superpositions of pure LH and pure split-off holes with envelope functions $\braket{z|f^\ell_j}$ and $\braket{z|f^s_j}$, respectively, with spin-independent energies $E_j^\e{}$. Here, $l$ and $j$ are subband indices for $\h{}$ and $\e{}$ subbands, respectively. To compute the envelopes and energies, we start from the spin-up ($\sigma=+$) block $H_+$ in Eq.~\eqref{H0}, which we diagonalize for different heterostructure compositions by the finite difference method using a mesh-spacing $\delta z = 0.01\,\mathrm{nm}$. The eigenstates of $H_-$ are the time-reversed conjugates of the eigenstates of $H_+$.

Using the eigenstates of $H_0^\mathrm{qw}$ as the new basis for the Hilbert space, the full 2DHG Hamiltonian $H_{k\cdot p}$ for $\mathbf{K}_\parallel\neq 0$ takes the following form (bold symbols indicate that we are in the eigenbasis of $H_0^\mathrm{qw}$):

\begin{equation}\label{Hk}
\begin{split}
    \mathbf{H}_{k\cdot p}(\mathbf{K}_\parallel) = \mathbf{E}_0^\mathrm{qw} &+ \alpha_0\left[\mathbf{M}_\gamma K_\parallel^2 + \frac{\cos\theta}{2l_B^2}\mathbf{M}_g + \frac{\sin\theta}{2l_B^2}\left(\frac{\sin\theta}{2l_B^2}\mathbf{N}_\gamma r_\phi - \tilde{\mathbf{N}}_\gamma\right)r_\phi\right] \\
    &+ \alpha_0\left[i\mathbf{M}_1 K_- + \mathbf{M}_2K_-^2 + \frac{\sin\theta}{2l_B^2}\left(e^{-i\phi}\mathbf{N}_g + i\mathbf{N}_1^+r_\phi K_- + i\mathbf{N}_1^-K_-r_\phi\right) + \text{h.c.}\right] \\
    &+ V_\parallel(x,y),
\end{split}
\end{equation}

\noindent where $K_\pm = K_x\pm iK_y$, $K_\parallel^2 = K_x^2 + K_y^2 = \{K_-,K_+\}/2$, $\{A,B\} = AB+BA$, $r_\phi = x\sin\phi-y\cos\phi$ and $l_B=\sqrt{\hbar/(eB)}$. The gauge for the in-plane part of the $B$-field is set as $A_z = \mathbf{e}_z\cdot\mathbf{A}= -Br_\phi\sin\theta$, while the gauge for the out-of-plane part of the field is left arbitrary for now, provided it does not include any $z$ operator. Regardless of the gauge the in-plane components of the wavenumber satisfy the commutation relation $[K_-,K_+]/2 = \cos\theta/l_B^2$. The Hamiltonian in Eq. \eqref{Hk} implicitly depends on strain and on the SOI strength $\Delta_0$ through the envelopes $\ket{f_j^{\ell,s,h}}$ and the energies $E_j^\tau$ ($\tau=\{\h{},\e{}\}$). Those terms do not explicitly appear in \eqref{Hk} because the $k\cdot p$ Hamiltonian does not contain terms such as $\Delta_0k_\pm$ or $\delta\varepsilon\cdot k_\pm$. It also integrates out the $z$ dimension through the matrices $\mathbf{M}$ and $\mathbf{N}$. We have

\begin{equation}
  \mathbf{E}_0^\mathrm{qw} = \begin{bmatrix}
    \mathbf{E}^\h{} & 0 & 0 & 0 \\
    0 & \mathbf{E}^\e{} & 0 & 0 \\
    0 & 0 & \mathbf{E}^\e{} & 0 \\
    0 & 0 & 0 & \mathbf{E}^\h{}
    \end{bmatrix},
\end{equation}

\noindent with $\mathbf{E}^\tau = \mathrm{diag}\{E_1^\tau,E_2^\tau,\dots\}$ ($\tau = \{\h{},\e{}\}$) are the energies of $H_0^\mathrm{qw}$. The $\mathbf{M}$-matrices are

\begin{align}
  \mathbf{M}_\gamma &= \begin{bmatrix}
    \boldsymbol{\Gamma}_\parallel^\h{} & 0 & 0 & 0 \\
    0 & \boldsymbol{\Gamma}_\parallel^\e{} & 0 & 0 \\
    0 & 0 & \boldsymbol{\Gamma}_\parallel^\e{} & 0 \\
    0 & 0 & 0 & \boldsymbol{\Gamma}_\parallel^\h{}
  \end{bmatrix}, & \mathbf{M}_1 &= \begin{bmatrix}
    0 & \mathbf{T}^\times & 0 & 0 \\
    0 & 0 & \mathbf{T}^\e{} & 0 \\
    0 & 0 & 0 & \mathbf{T}^{\times\dagger} \\
    \mathbf{T}^\h{} & 0 & 0 & 0
  \end{bmatrix}, \\
  \mathbf{M}_g &= \begin{bmatrix}
      \mathbf{G}_\perp^\h{} & 0 & 0 & 0 \\
      0 & \mathbf{G}_\perp^\e{} & 0 & 0 \\
      0 & 0 & -\mathbf{G}_\perp^\e{} & 0 \\
      0 & 0 & 0 & -\mathbf{G}_\perp^\h{}
  \end{bmatrix}, & \mathbf{M}_2 &= \begin{bmatrix}
    0 & 0 & \boldsymbol{\upmu} & 0 \\
    0 & 0 & 0 & \boldsymbol{\upmu}^\dagger \\
    \boldsymbol{\updelta}^\dagger & 0 & 0 & 0 \\
    0 & \boldsymbol{\updelta} & 0 & 0
  \end{bmatrix}.
\end{align}

The non-zero blocks of these matrices are all explicitly expanded in terms of the eigenstates of $H_0^\mathrm{qw}$:

\begin{subequations}
  \begin{align}
    \Gamma_{\parallel l,l^\prime}^\h{} &= -\braket{f^h_l|\gamma_1 + \gamma_2|f^h_{l^\prime}}, \\
    \Gamma_{\parallel j,j^\prime}^\e{} &= -\braket{f^\perp_j|\gamma_-|f^\perp_{j^\prime}} - \braket{f^\circ_j|\gamma_1 + \gamma_2|f^\circ_{j^\prime}}, \\
    G_{\perp l,l^\prime}^\h{} &= -\bra{f_l^h}6\kappa+\frac{27q}{2}\ket{f_{l^\prime}^h} \\
    G_{\perp j,j^\prime}^\e{} &= -6\bra{f_j^\circ}\kappa\ket{f_{j^\prime}^\circ} + 2\left(\braket{f_j^\perp|f_{j^\prime}^\perp} - 2\braket{f_j^\circ|f_{j^\prime}^\circ}\right) - \frac{1}{2}\bra{f_j^\ell}q\ket{f_{j^\prime}^\ell} \\
    T_{l,j}^\times &= -\frac{i}{\sqrt{2}}\bra{f^h_l}\left(3u_+\ket{f^\perp_j} + \frac{7\sqrt{6}}{4}[q,k_z]\ket{f_j^\ell}\right), \\
    T_{l,l^\prime}^\h{} &= -\frac{3i}{2}\bra{f_l^h}[q,k_z]\ket{f_{l^\prime}^h}\\
    T_{j,j^\prime}^\e{} &= -\frac{3i}{\sqrt{2}}\left(\braket{f_j^\circ|u_+|f_{j^\prime}^\perp} - \braket{f_j^\perp|u_-|f_{j^\prime}^\circ}\right) - 5i\bra{f_j^\ell}[q,k_z]\ket{f_{j^\prime}^\ell}, \\
    \mu_{l,j} &= \frac{3}{2}\braket{f_l^h|\gamma_2 + \gamma_3|f_j^\circ}, \\
    \delta_{l,j} &= \frac{3}{2}\braket{f_l^h|\gamma_2 - \gamma_3|f_j^\circ},
  \end{align}
\end{subequations}

\noindent where $u_\pm = \{\gamma_3,k_z\} \pm [\kappa,k_z]$ and

\begin{align}
  \ket{f_j^\perp} &\equiv \frac{1}{\sqrt{3}}\left(\sqrt{2}\ket{f_j^\ell} - \ket{f_j^s}\right),
  &\ket{f_j^\circ} &\equiv \frac{1}{\sqrt{3}}\left(\ket{f_j^\ell} + \sqrt{2}\ket{f_j^s}\right).
\end{align}

The $\mathbf{N}$ matrices are

\begin{align}
    \mathbf{N}_\gamma &= 4\begin{bmatrix}
        \boldsymbol{\Gamma}_\perp^\h{} & 0 & 0 & 0 \\
        0 & \boldsymbol{\Gamma}_\perp^\e{} & 0 & 0 \\
        0 & 0 & \boldsymbol{\Gamma}_\perp^\e{} & 0 \\
        0 & 0 & 0 & \boldsymbol{\Gamma}_\perp^\h{}
    \end{bmatrix}, & \tilde{\mathbf{N}}_\gamma &= 4\begin{bmatrix}
        \tilde{\boldsymbol{\Gamma}}_\perp^\h{} & 0 & 0 & 0 \\
        0 & \tilde{\boldsymbol{\Gamma}}_\perp^\e{} & 0 & 0 \\
        0 & 0 & \tilde{\boldsymbol{\Gamma}}_\perp^\e{} & 0 \\
        0 & 0 & 0 & \tilde{\boldsymbol{\Gamma}}_\perp^\h{}
    \end{bmatrix}, \\
    \mathbf{N}_g &= \begin{bmatrix}
        0 & \mathbf{G}_\parallel^\times & 0 & 0 \\
        0 & 0 & \mathbf{G}_\parallel^\e{} & 0 \\
        0 & 0 & 0 & \mathbf{G}_\parallel^{\times\dagger} \\
        \mathbf{G}_\parallel^\h{} & 0 & 0 & 0
    \end{bmatrix}, & \mathbf{N}_1^\pm=\mathbf{N}_1 &= \begin{bmatrix}
        0 & \mathbf{R}^\times & 0 & 0 \\
        0 & 0 & \mathbf{R}^\e{} & 0 \\
        0 & 0 & 0 & \mathbf{R}^{\times\dagger} \\
        0 & 0 & 0 & 0
    \end{bmatrix},
\end{align}

\noindent where

\begin{subequations}
    \begin{align}
    \Gamma_{\perp l,l^\prime}^\h{} &= -\bra{f_l^h}\gamma_-\ket{f_{l^\prime}^h}, \\
    \tilde{\Gamma}_{\perp l,l^\prime}^\h{} &= -\frac{1}{2}\bra{f_l^h}\{\gamma_-,k_z\}\ket{f_{l^\prime}^h}, \\
    \Gamma_{\perp j,j^\prime}^\e{} &= -\bra{f_j^\circ}\gamma_-\ket{f_{j^\prime}^\circ} - \bra{f_j^\perp}\gamma_+\ket{f_{j^\prime}^\perp}, \\
    \tilde{\Gamma}_{\perp j,j^\prime}^\e{} &= -\frac{1}{2}\bra{f_j^\circ}\{\gamma_-,k_z\}\ket{f_{j^\prime}^\circ} - \frac{1}{2}\bra{f_j^\perp}\{\gamma_+,k_z\}\ket{f_{j^\prime}^\perp}, \\
    G_{\parallel l,j}^\times &= -\sqrt{2}\bra{f_l^h}\left(3\kappa\ket{f_j^\perp}+\frac{7\sqrt{6}q}{4}\ket{f_j^\ell} - \sqrt{3}\ket{f_j^s}\right), \\
    G_{\parallel l,l^\prime}^\h{} &= -3\bra{f_l^h}q\ket{f_{l^\prime}^h} \\
    G_{\parallel j,j^\prime}^\e{} &= -\sqrt{2}\bra{f_j^\circ}3\kappa+1\ket{f_{j^\prime}^\perp} - \sqrt{2}\bra{f_j^\perp}3\kappa+1\ket{f_{j^\prime}^\circ} + 2\braket{f_j^\perp|f_{j^\prime}^\perp} - 10\bra{f_j^\ell}q\ket{f_{j^\prime}^\ell}, \\
    R_{l,j}^\times &= 3\sqrt{2}i\bra{f_l^h}\gamma_3\ket{f_j^\perp}, \\
    R_{j,j^\prime}^\e{} &= 3\sqrt{2}i\left(\bra{f_j^\circ}\gamma_3\ket{f_{j^\prime}^\perp} - \bra{f_j^\perp}\gamma_3\ket{f_{j^\prime}^\circ}\right).
\end{align}
\end{subequations}

\section{\texorpdfstring{Perturbation theory}{}}\label{sec:pert}
\subsection{\texorpdfstring{Band dispersion}{}}

Away from $\mathbf{K}_\parallel = 0$, Hamiltonian \eqref{Hk} becomes non-diagonal, and folding it down onto any given set of subbands, labeled as the $\mathcal{A}$ set, can be performed by means of a Schrieffer-Wolff transformation (SWT)~\cite{Bravyi2011,Winkler2003} with $\mathbf{K}_\parallel$ as the perturbation parameter. The contributions from all remote subbands, the $\mathcal{B}$ set, are eliminated up to the desired perturbation order. More explicitly, the unperturbed Hamiltonian is $\mathbf{E}_0^\mathrm{qw}$ ($\mathbf{K_\parallel}$-independent), while the perturbation $\mathbf{W}(\mathbf{K}_\parallel) = \mathbf{W}_1(\mathbf{K}_\parallel) + \mathbf{W}_2(\mathbf{K}_\parallel)$ is at least linear in $\mathbf{K}_\parallel$:

\begin{align}
    \mathbf{W}_1(\mathbf{K}_\parallel) &=  - \frac{\alpha_0\sin\theta}{2l_B^2}\tilde{\mathbf{N}}_\gamma r_\phi + \left(i\alpha_0\mathbf{M}_1 K_- + \text{h.c.}\right), \\
    \mathbf{W}_2(\mathbf{K}_\parallel) &= \alpha_0\left\{\mathbf{M}_\gamma K_\parallel^2 + \frac{\cos\theta}{2l_B^2}\mathbf{M}_g + \frac{\sin^2\theta}{4l_B^4}\mathbf{N}_\gamma r_\phi^2 + \left[\mathbf{M}_2K_-^2 + \frac{\sin\theta}{2l_B^2}\left(e^{-i\phi}\mathbf{N}_g + i\mathbf{N}_1^+r_\phi K_- + i\mathbf{N}_1^-K_-r_\phi\right) + \text{h.c.}\right]\right\}.
\end{align}
    
We use the following convention:

\begin{align*}
    i,j,k,\dots: & &\text{indices that run through the whole Hilbert space} \\
    \alpha,\beta,\gamma,\delta,\dots: & &\text{indices that run through the }\mathcal{A}\text{ set} \\
    \lambda,\mu,\nu,\dots: & &\text{indices that run through the }\mathcal{B}\text{ set}
\end{align*}

\noindent We also define the following projection operators:

\begin{align}
    P &= \sum_{\alpha\in\mathcal{A}}{\ket{\varphi_\alpha}\bra{\varphi_\alpha}}, \\
    Q &= \sum_{\lambda\in\mathcal{B}}{\ket{\varphi_\lambda}\bra{\varphi_\lambda}},
\end{align}

\noindent and superoperators:

\begin{align}
    \mathcal{P}[X] &= PXP, \\
    \mathcal{Q}[X] &= QXQ, \\
    \mathcal{D}[X] &= PXP + QXQ, \\
    \mathcal{O}[X] &= PXQ + QXP, \\
    \mathcal{L}[X] &= \sum_{i,j}{\ket{\varphi_i}\frac{\bra{\varphi_i}\mathcal{O}[X]\ket{\varphi_j}}{E_i-E_j}\bra{\varphi_j}}.
\end{align}

\noindent where $E_i$ is the energy of subband $i$ of the unperturbed Hamiltonian (or at $\mathbf{K}_\parallel = 0$). It is clear from the definition of $\mathcal{L}[X]$ that $\mathcal{O}[\mathcal{L}[X]] = \mathcal{L}[X]$ and $\mathcal{D}[\mathcal{L}[X]] = 0$. Up to $3$\textsuperscript{rd} order (and only keeping terms up to $k^3$), the effective Hamiltonian for the $\mathcal{A}$ set is given by

\begin{equation}
\begin{split}
    \mathcal{H}_\mathrm{eff}(\mathbf{K}_\parallel) &= \mathcal{P}[\mathbf{E}_0^\mathrm{qw}] + \mathcal{P}[\mathbf{W}_1] + \mathcal{P}[\mathbf{W}_2] \\
    &+ \mathcal{P}\left[\mathrm{sw}_2(\mathbf{W}_1,\mathbf{W}_1)\right] + \mathcal{P}\left[\mathrm{sw}_2(\mathbf{W}_1,\mathbf{W}_2)\right] + \mathcal{P}\left[\mathrm{sw}_2(\mathbf{W}_2,\mathbf{W}_1)\right] \\
    &+ \mathcal{P}\left[\mathrm{sw}_3(\mathbf{W}_1,\mathbf{W}_1,\mathbf{W}_1)\right],
\end{split}
\end{equation}

\noindent where

\begin{subequations}
\begin{align}
    \mathrm{sw}_2(X,Y) &= \frac{1}{2}\left(\mathcal{L}[X]Y - X\mathcal{L}[Y]\right), \\
    \mathrm{sw}_3(X,Y,Z) &= \frac{1}{2}\left(\mathcal{L}[\mathcal{L}[X]Y]Z + X\mathcal{L}[Y\mathcal{L}[Z]] + \mathcal{L}[XP\mathcal{L}[Y]]Z + X\mathcal{L}[\mathcal{L}[Y]PZ]\right).
\end{align}
\end{subequations}

In the special case where $B=0$ and the $\mathcal{A}$ set consists of the $2$-fold HH ground state, $\mathcal{A}=\{\ket{\h{1};\pm}\}$, one obtains the following effective Hamiltonian (assuming $E_1^\h{}=0$):

\begin{equation}\label{HHeff_}
    \mathcal{H}_\mathrm{eff} = \alpha_0\boldsymbol{\upgamma} k_\parallel^2+\alpha_0\left[i\left(\boldsymbol{\upbeta}_2k_-^3 +\boldsymbol{\upbeta}_3k_-k_+k_- \right)+ \text{h.c.}\right],
\end{equation}

\noindent where the bold symbols are $2\times 2$ matrices directly proportional to the corresponding effective parameter. In terms of the $\mathbf{M}$-matrices:

\begin{align*}
\boldsymbol{\upgamma} &= \mathcal{P}\left[\mathbf{M}_\gamma\right] + \alpha_0\mathcal{P}\left[\mathrm{sw}_2\left(\mathbf{M}_1,\mathbf{M}_1^\dagger\right) + \mathrm{sw}_2\left(\mathbf{M}_1^\dagger,\mathbf{M}_1\right)\right] \\
&= \gamma\mathbb{1}_{2\times 2}, \\
\boldsymbol{\upbeta}_2 &= \alpha_0\mathcal{P}\left[\mathrm{sw}_2\left(\mathbf{M}_1,\mathbf{M}_2\right) + \mathrm{sw}_2\left(\mathbf{M}_2,\mathbf{M}_1\right)\right] - \alpha_0^2\mathcal{P}\left[\mathrm{sw}_3\left(\mathbf{M}_1,\mathbf{M}_1,\mathbf{M}_1\right)\right] \\
&= \beta_2\sigma_+, \\
\boldsymbol{\upbeta}_3 &= \alpha_0\mathcal{P}\left[\mathrm{sw}_2\left(\mathbf{M}_\gamma,\mathbf{M}_1\right) + \mathrm{sw}_2\left(\mathbf{M}_1,\mathbf{M}_\gamma\right) -\mathrm{sw}_2\left(\mathbf{M}_1^\dagger,\mathbf{M}_2\right) - \mathrm{sw}_2\left(\mathbf{M}_2,\mathbf{M}_1^\dagger\right)\right] +\alpha_0^2\mathcal{P}\left[\mathrm{sw}_3\left(\mathbf{M}_1,\mathbf{M}_1,\mathbf{M}_1^\dagger\right) + \text{c.p.}\right] \\
&= \beta_3\sigma_-,
\end{align*}

\noindent where c.p. means cyclic permutations. Inserting this back into \eqref{HHeff_} gives

\begin{equation}\label{HHeff}
    \mathcal{H}_\mathrm{eff} = \alpha_0\gamma k_\parallel^2+\alpha_0\left[-i\left(\beta_2k_+^3 -\beta_3k_-k_+k_- \right)\sigma_-+ \text{h.c.}\right].
\end{equation}

We point out that to $1$\textsuperscript{st} order, there would also be a term $i\alpha_0\boldsymbol{\upbeta}_1k_-$, where $\boldsymbol{\upbeta}_1 = \mathcal{P}[\mathbf{M}_1] = \beta_1\sigma_-$ and $\beta_1 = -(3i/2)\braket{[q,k_z]}$. However we find this term to be negligibly small and is therefore neglected in our calculations.

\subsection{\texorpdfstring{Quantum Dot}{}}
Another application of perturbation theory is to construct a reduced quantum well basis set on which quantum dot calculations can be performed efficiently. For instance, the SOI parameters converge ($3$\textsuperscript{rd} order SWT, $2\times 2$ eff. Hamiltonian) once the quantum well Hilbert space is $\sim 200$-dimensional. This, times the number of in-plane basis states of the quantum dot problem results in large matrices. Here, we employ $2$\textsuperscript{nd} order perturbation theory to find {\it multi-subband} effective Hamiltonians that describe  well the ground state without requiring $\sim 200$ levels.

A general $2$\textsuperscript{nd} order effective Hamiltonian is given by

\begin{equation}
    \mathcal{H}_\mathrm{eff}(\mathbf{K}_\parallel) = \mathcal{P}[\mathbf{E}_0^\mathrm{qw}] + \mathcal{P}[\mathbf{W}_1] + \mathcal{P}[\mathbf{W}_2] + \mathcal{P}[\mathrm{sw}_2(\mathbf{W}_1,\mathbf{W}_1)].
\end{equation}

The second order term involves products of $\mathbf{W}_1$ with itself, and will ultimately result in terms that can be collected back into the $\mathbf{M}$ and $\mathbf{N}$ matrices that define the original Hamiltonian. Explicitly:

\begin{align*}
    \frac{\mathbf{W}_1\mathbf{W}_1}{\alpha_0^2} &= \left(-\frac{\sin\theta}{2l_B^2}\tilde{\mathbf{N}}_\gamma r_\phi + i\mathbf{M}_1K_- - i\mathbf{M}_1^\dagger K_+\right)\left(-\frac{\sin\theta}{2l_B^2}\tilde{\mathbf{N}}_\gamma r_\phi + i\mathbf{M}_1K_- - i\mathbf{M}_1^\dagger K_+\right) \\
    &= \left(\mathbf{M}_1\mathbf{M}_1^\dagger + \mathbf{M}_1^\dagger\mathbf{M}_1\right)K_\parallel^2 + \frac{\cos\theta}{2l_B^2}\left(2\mathbf{M}_1\mathbf{M}_1^\dagger - 2\mathbf{M}_1^\dagger\mathbf{M}_1\right) + \frac{\sin^2\theta}{4l_B^4}\tilde{\mathbf{N}}_\gamma\tilde{\mathbf{N}}_\gamma r_\phi^2 \\
    &- \left[\mathbf{M}_1\mathbf{M}_1K_-^2 + \frac{\sin\theta}{2l_B^2}\left(i\tilde{\mathbf{N}}_\gamma\mathbf{M}_1 r_\phi K_- + i\mathbf{M}_1\tilde{\mathbf{N}}_\gamma K_-r_\phi\right) + \mathrm{h.c.}\right],
\end{align*}

\noindent where we used $K_\pm K_\mp = K_\parallel^2\mp\cos\theta/l_B^2$. By identification with the original Hamiltonian \eqref{Hk} we define the $\mathcal{A}$-set $\mathcal{M}$-matrices

\begin{subequations}
    \begin{align}
        \mathcal{M}_\gamma &= \mathcal{P}[\mathbf{M}_\gamma] + \alpha_0\mathcal{P}\left[\mathrm{sw}_2(\mathbf{M}_1,\mathbf{M}_1^\dagger) + \mathrm{sw}_2(\mathbf{M}_1^\dagger,\mathbf{M}_1)\right], \\
        \mathcal{M}_g &= \mathcal{P}[\mathbf{M}_g] + 2\alpha_0\mathcal{P}\left[\mathrm{sw}_2(\mathbf{M}_1,\mathbf{M}_1^\dagger) - \mathrm{sw}_2(\mathbf{M}_1^\dagger,\mathbf{M}_1)\right], \\
        \mathcal{M}_2 &= \mathcal{P}[\mathbf{M}_2] - \alpha_0\mathcal{P}\left[\mathrm{sw}_2(\mathbf{M}_1,\mathbf{M}_1)\right], \\
        \mathcal{N}_\gamma &= \mathcal{P}[\mathbf{N}_\gamma] + \alpha_0\mathcal{P}\left[\mathrm{sw}_2(\tilde{\mathbf{N}}_\gamma,\tilde{\mathbf{N}}_\gamma)\right], \\
        \mathcal{N}_1^+ &= \mathcal{P}[\mathbf{N}_1] - \alpha_0\mathcal{P}\left[\mathrm{sw}_2(\tilde{\mathbf{N}}_\gamma,\mathbf{M}_1)\right], \\
        \mathcal{N}_1^- &= \mathcal{P}[\mathbf{N}_1] - \alpha_0\mathcal{P}\left[\mathrm{sw}_2(\mathbf{M}_1,\tilde{\mathbf{N}}_\gamma)\right],
    \end{align}
\end{subequations}

\noindent and for completeness:

\begin{subequations}
    \begin{align}
        \mathcal{E}_0^\mathrm{qw} &= \mathcal{P}[\mathbf{E}_0^\mathrm{qw}], \\
        \mathcal{M}_1 &= \mathcal{P}[\mathbf{M}_1], \\
        \tilde{\mathcal{N}}_\gamma &= \mathcal{P}[\tilde{\mathbf{N}}_\gamma], \\
        \mathcal{N}_g &= \mathcal{P}[\mathbf{N}_g].
    \end{align}
\end{subequations}

The matrix $\mathbf{N}_g$ does not acquire corrections from $2$\textsuperscript{nd} order SWT. These matrices of reduced dimension can then replace those in the original Hamiltonian \eqref{Hk} to get

\begin{equation}\label{Hk_}
\begin{split}
    \mathcal{H}_{k\cdot p}(\mathbf{K}_\parallel) = \mathcal{E}_0^\mathrm{qw} &+ \alpha_0\left[\mathcal{M}_\gamma K_\parallel^2 + \frac{\cos\theta}{2l_B^2}\mathcal{M}_g + \frac{\sin\theta}{2l_B^2}\left(\frac{\sin\theta}{2l_B^2}\mathcal{N}_\gamma r_\phi - \tilde{\mathcal{N}}_\gamma\right)r_\phi\right] \\
    &+ \alpha_0\left[i\mathcal{M}_1 K_- + \mathcal{M}_2K_-^2 + \frac{\sin\theta}{2l_B^2}\left(e^{-i\phi}\mathcal{N}_g + i\mathcal{N}_1^+r_\phi K_- + i\mathcal{N}_1^-K_-r_\phi\right) + \text{h.c.}\right] \\
    &+ V_\parallel(x,y).
\end{split}
\end{equation}

\begin{figure}
    \centering
    \includegraphics[width=0.9\linewidth]{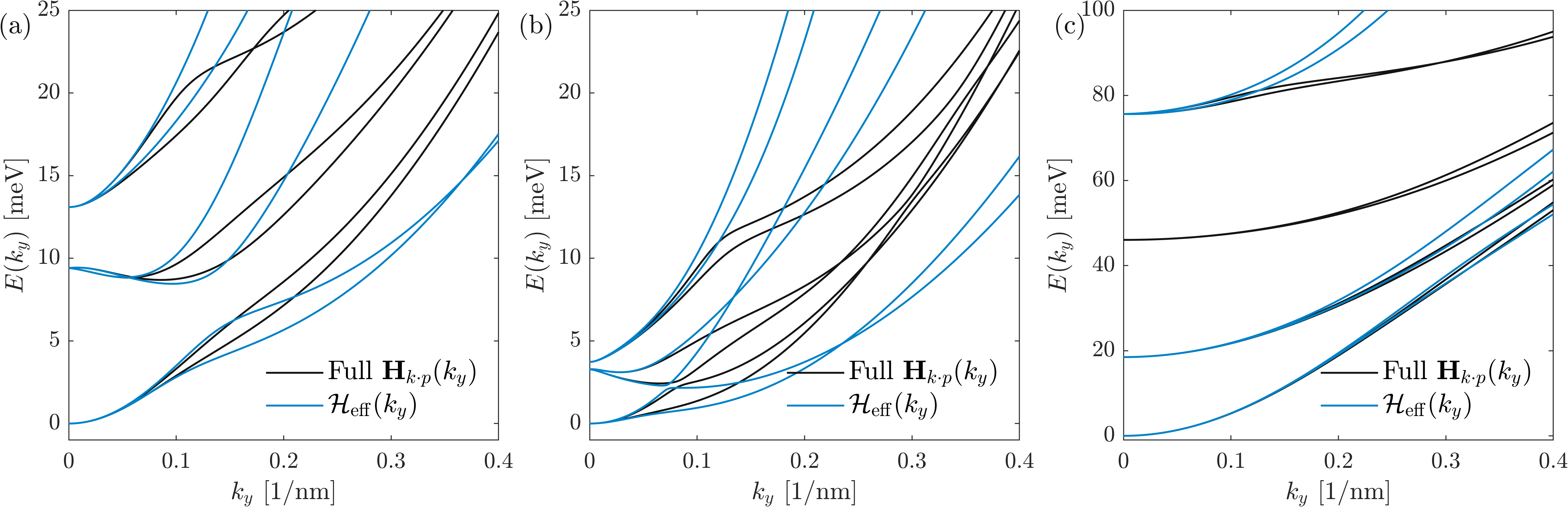}
    \caption{$E(k_y)$ dispersion from the full Hamiltonian (black) and from a $2$\textsuperscript{nd} order effective Hamiltonian. (a) unstrained Ge. (b) Ge+ with optimized spikes. The $\mathcal{A}$ set consists of $\mathcal{A} = \{\h{1},\h{2},\e{1}\}$. (c) $\varepsilon$-Ge with $\mathcal{A} = \{\h{1},\h{2},\e{1}, \e{2}\}$. $\e{2}$ lies beyond the plot window and is therefore not visible. The lonely black level is $\h{3}$.}
    \label{fig:si_Heff}
\end{figure}

Fig. \ref{fig:si_Heff} shows the energy dispersion ($V_\parallel=0$) for unspiked Ge, optimized spikes and $\varepsilon$-Ge computed from the diagonalization of the full $k\cdot p$ matrix $\mathbf{H}_{k\cdot p}$ compared to the dispersion provided by a $2$\textsuperscript{nd} order SWT effective Hamiltonian $\mathcal{H}_\mathrm{eff}$. The dimension of the full Hilbert space is $2\cdot 250 = 500$ and includes the lowest $104$ ($103$) $\h{}$ spin doublets and the lowest $146$ ($147$) $\e{}$ spin doublets in both Ge  and $\varepsilon$-Ge.

\subsection{\texorpdfstring{Derivation of the $\beta_2$ expression}{}}

Keeping only the 2nd order contribution to $\beta_2$ for simplicity, we had from the previous section

\begin{equation}
    \boldsymbol{\upbeta}_2 \approx \alpha_0\mathcal{P}\left[\mathrm{sw}_2\left(\mathbf{M}_1,\mathbf{M}_2\right) + \text{c.p.}\right] = \beta_2\sigma_+.
\end{equation}

Moreover, when the $\mathcal{A}$ set is degenerate ($E_\alpha = E$ for all $\alpha\in\mathcal{A}$), the 2nd order $\mathrm{sw}_2(X,Y)$ function simplifies to

\begin{align}
    \mathrm{sw}_2(X,Y)_{\alpha\beta} &= \frac{1}{2}\left(\mathcal{L}[X]Y - X\mathcal{L}[Y]\right)_{\alpha\beta} \\
    &= \frac{1}{2}\sum_\lambda{\left(\frac{X_{\alpha\lambda}Y_{\lambda\beta}}{E_\alpha-E_\lambda} - \frac{X_{\alpha\lambda}Y_{\lambda\beta}}{E_\lambda-E_\beta}\right)} \\
    &= \frac{1}{2}\sum_\lambda{\left(\frac{X_{\alpha\lambda}Y_{\lambda\beta}}{E_\alpha-E_\lambda} + \frac{X_{\alpha\lambda}Y_{\lambda\beta}}{E_\alpha-E_\lambda}\right)} \\
    &= \sum_\lambda{\frac{X_{\alpha\lambda}Y_{\lambda\beta}}{E_\alpha-E_\lambda}} \\
    &= \left(\mathcal{L}[X]Y\right)_{\alpha\beta}.
\end{align}

Assuming $q=0$ and neglecting the SO band for simplicity, the matrix elements of the $\mathbf{T}^\times$ block and of the $\boldsymbol{\upmu}$ block are

\begin{align}
    T^\times_{l,j} &= -i\sqrt{3}\bra{f_l^h}\{\gamma_3,k_z\}\ket{f_j^\ell}, \\
    \mu_{l,j} &= \frac{\sqrt{3}}{2}\bra{f_l^h}\gamma_2+\gamma_3\ket{f_j^\ell}.
\end{align}

Then the $\beta_2$ parameter is

\begin{align}
    \boldsymbol{\upbeta}_2 &= \alpha_0\mathcal{P}\left[\mathcal{L}[\mathbf{M}_1]\mathbf{M}_2 + \mathcal{L}[\mathbf{M}_2]\mathbf{M}_1\right] \\
    \rightarrow \beta_2 &= \alpha_0\sum_\lambda{\frac{T^\times_{1\lambda}\mu_{\lambda 1}^\dagger + \mu_{1\lambda}T_{\lambda 1}^{\times\dagger}}{E_1^\h{} - E_\lambda^\mathrm{L}}} \\
    &= -2\sqrt{3}\alpha_0\sum_\lambda{\frac{\mu_\lambda}{\Delta_\lambda}\Im\left(\bra{f_1^h}\{\gamma_3,k_z\}\ket{f_\lambda^\ell} + \bra{f_1^h}[\kappa,k_z]\ket{f_\lambda^\ell}\right)} \\
    &= -2\sqrt{3}\alpha_0\sum_\lambda{\frac{\mu_\lambda}{\Delta_\lambda}\int{dz\,f_1^h(z)\left[2\gamma_3(z)f_\lambda^{\ell\prime}(z) + \left(\gamma_3^\prime(z) - \kappa^\prime(z)\right)f_\lambda^\ell(z)\right]}},
\end{align}

\noindent which is the result provided in the main text, Eq.~(3). Here, the global minus sign comes from the electron-energy convention, which is flipped in the main text.

\newpage
\section{\texorpdfstring{SOI spin splitting in machine learning-optimized Si bumps}{}}\label{sec:bump}

In this section we show the SOI spin splitting $E_\mathrm{so}(\mathbf{k})$ for $k_x = 0$ against $k_y$ for the machine learning-optimized bumped Ge+.

\begin{figure}[bh]
    \centering
    \includegraphics[width=0.3\linewidth]{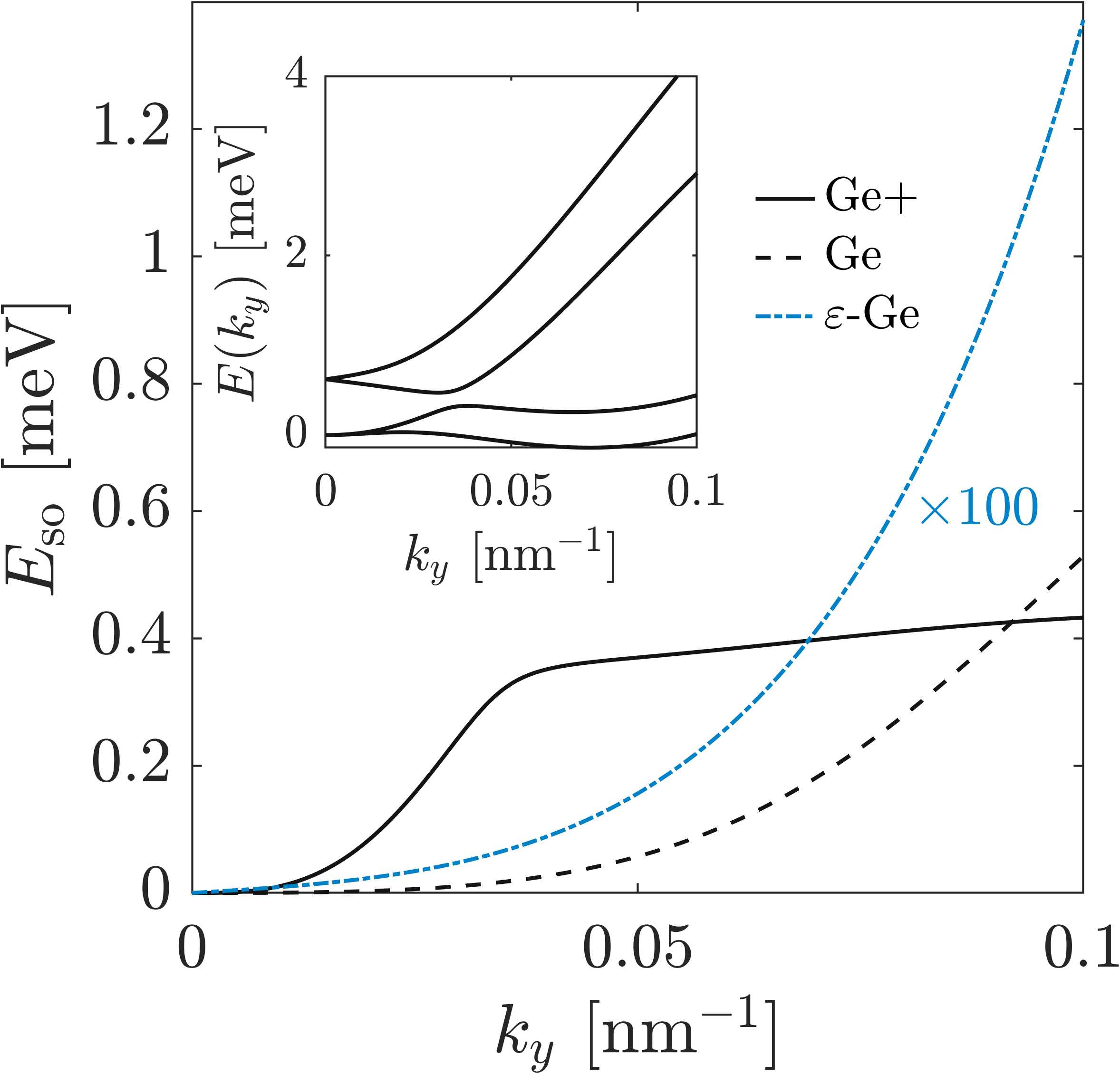}
    \caption{SOI energy $E_\mathrm{so}$ of the HH ground state at $k_x=0$ against $k_y$ (black solid lines) for the machine learning-optimized bumped Ge+ (see Table 1 in main text for the list of parameters). For comparison, we show $E_\mathrm{so}$ in unstrained Ge with black dashed lines and in a $16\,\mathrm{nm}$-wide $\varepsilon$-Ge quantum well with blue dashed-dotted lines (rescaled by 2 orders of magnitude). Insets show the energy dispersion $E(k_y)$ of the optimal Ge+ heterostructure.}
    \label{fig:kbump}
\end{figure}

\newpage
\section{\texorpdfstring{Additional data on Si-spikes Ge+}{}}\label{sec:spikes}
\subsection{Manual optimization}
\begin{figure}[h]
    \centering
    \includegraphics[width=0.8\linewidth]{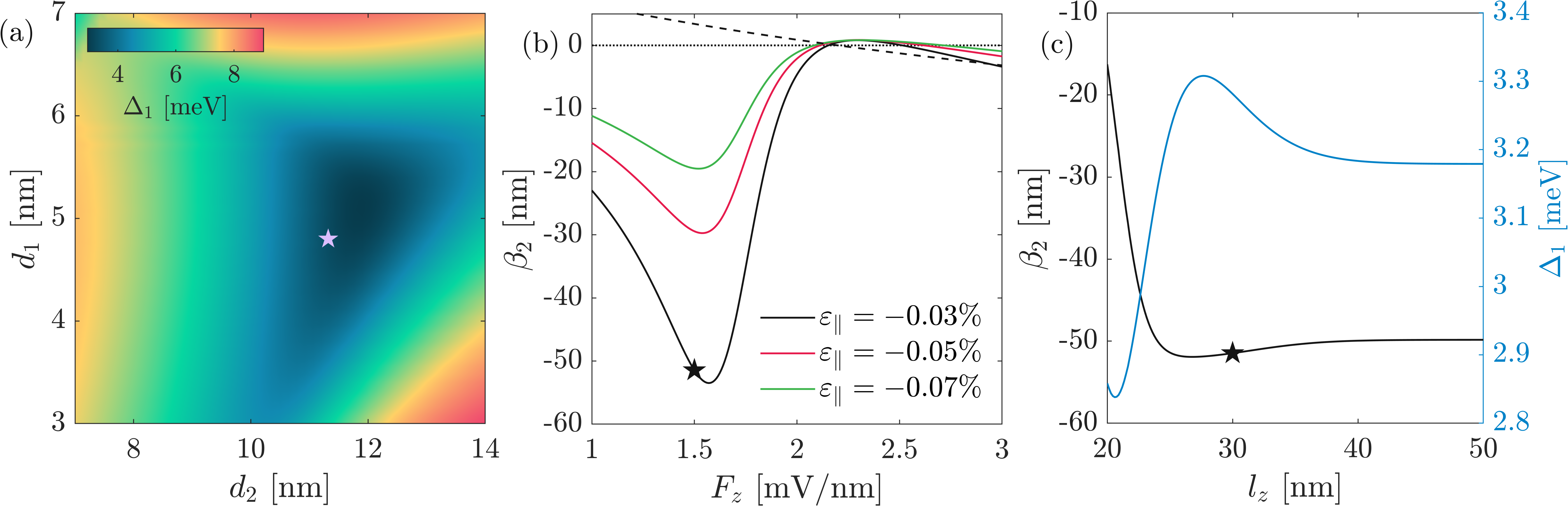}
    \caption{(a) $\Delta_1$ against $d_{1,2}$ for $F_z = 1.5\,\mathrm{mV}/\mathrm{nm}$, $\varepsilon_\parallel = -0.03\%$, and $l_z = 30\,\mathrm{nm}$. The star marks the maximal $|\beta_2|$ at $d_1 = 4.8\,\mathrm{nm}$ and $d_2 = 11.3\,\mathrm{nm}$. (b) $\beta_2$ against $F_z$ for increasing residual strain in Ge. The dashed line corresponds to no Si spikes. (c) $\beta_2$ (left axis) and $\Delta_1$ (right axis) against $l_z$.}
    \label{fig:opt}
\end{figure}

In this section we report additional simulation results on the manually optimized Si-spikes. In Fig. \ref{fig:opt} we show the variation of $\Delta_1$ against the spikes distances $d_{1,2}$ at gate field $F_z=1.5\,\mathrm{mV}/\mathrm{nm}$. We also show in panel (b) and (c) the SOI strength $\beta_2$ against $F_z$ and $l_z$ respectively.

\begin{figure}[h]
    \centering
    \includegraphics[width=\linewidth]{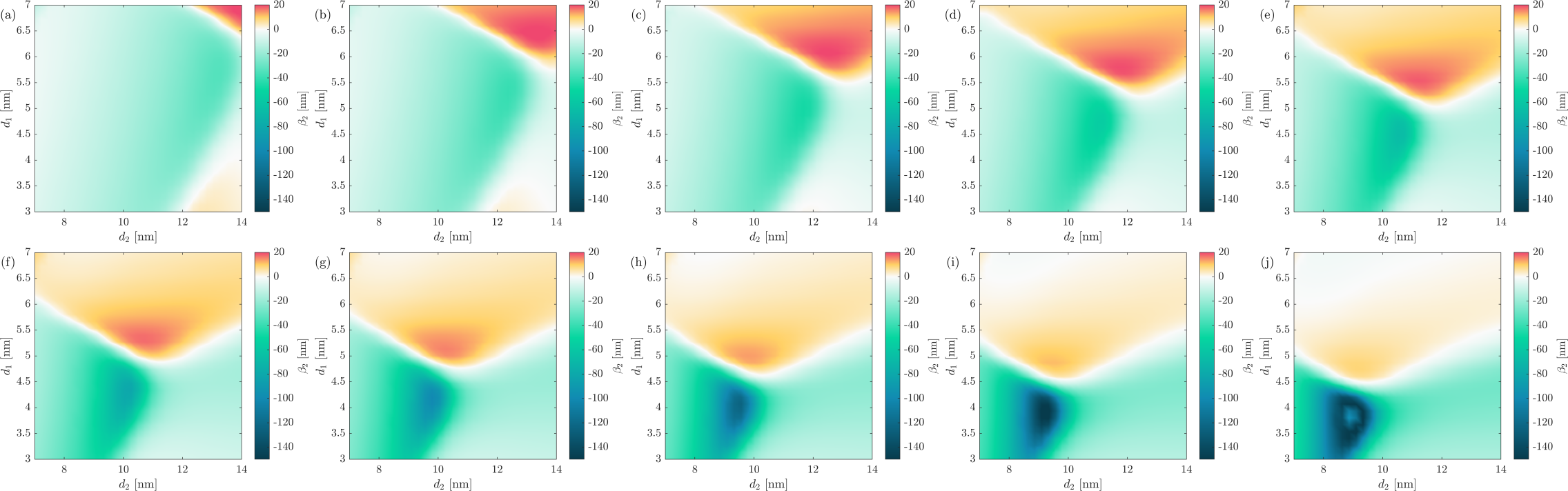}
    \caption{Calculated $\beta_2$ RSOI parameter as a function of the spike distances $d_{1,2}$ for gate field strengths from (a) $F_z=1.0\,\mathrm{mV}/\mathrm{nm}$ to (j) $F_z=2.8\,\mathrm{mV}/\mathrm{nm}$ in steps of $0.2\,\mathrm{mV}/\mathrm{nm}$. All plots share the same $x$-$y$-$z$ scale.}
    \label{fig:b2-Fz}
\end{figure}

\begin{figure}[h]
    \centering
    \includegraphics[width=\linewidth]{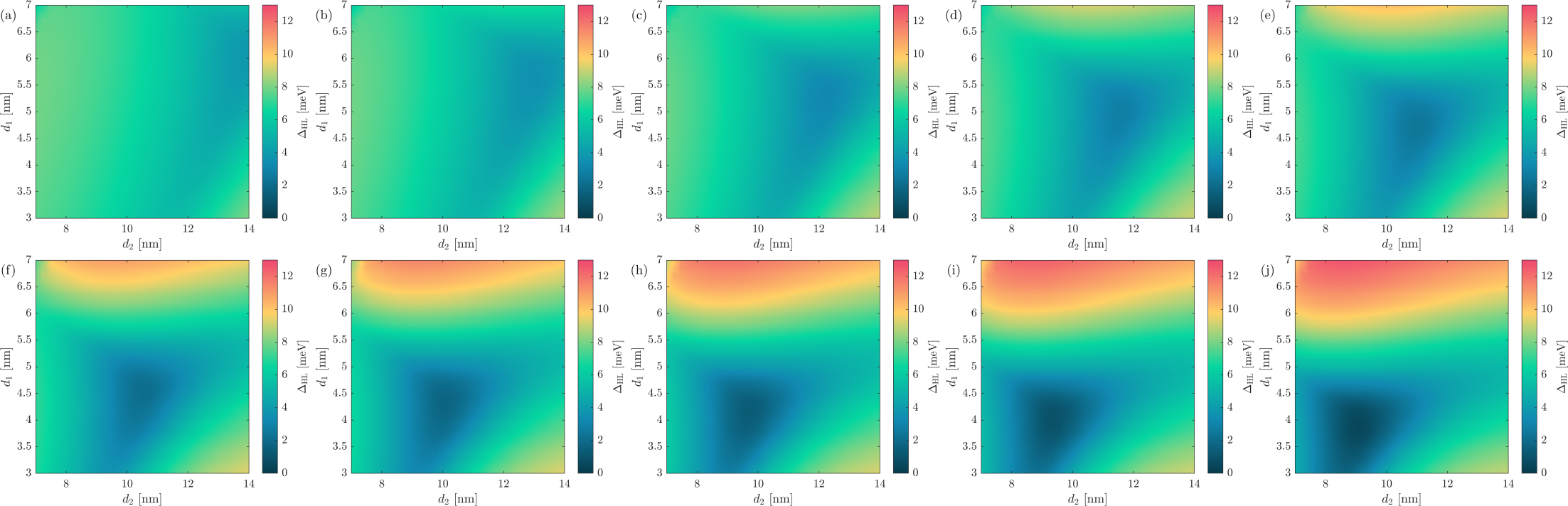}
    \caption{Calculated HH-LH splitting energy as a function of the spike distances $d_{1,2}$ for gate field strengths from (a) $F_z=1.0\,\mathrm{mV}/\mathrm{nm}$ to (j) $F_z=2.8\,\mathrm{mV}/\mathrm{nm}$ in steps of $0.2\,\mathrm{mV}/\mathrm{nm}$. All plots share the same $x$-$y$-$z$ scale.}
    \label{fig:hl-Fz}
\end{figure}

We also report the variation of $\beta_2$ (Fig. \ref{fig:b2-Fz}) and the LH-HH splitting energy (Fig. \ref{fig:hl-Fz}) with respect to $d_1$ and $d_2$ for various gate field strengths, other than the one presented in the main article.

\subsection{Machine learning optimization}

\begin{figure}[h]
    \centering
    \includegraphics[width=0.4\linewidth]{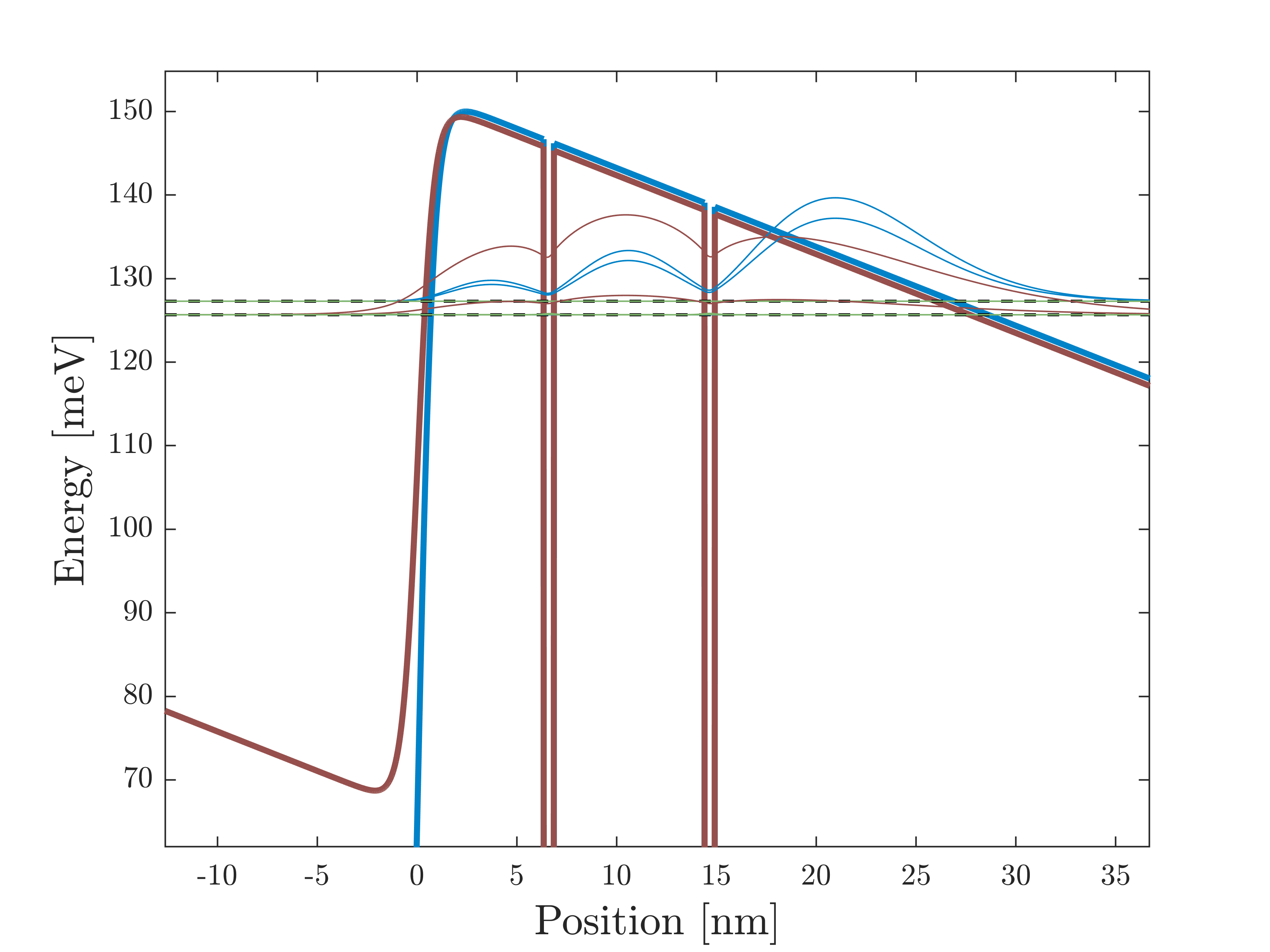}
    \caption{HH envelope function $|\braket{z|f_1^h}|^2$ (blue curves) of the ground state and LH envelope function $|\braket{z|f_1^\ell}|^2$ (red curves) of the first excited state in the machine learning-optimized spiked Ge$+$. The green lines are the small SO components $|\braket{z|f_1^s}|^2$ belonging to the LH ground state. Thicker red and blue lines (the latter mostly hidden behind the red lines because of small strains) indicate the bulk band edge energies of the LH and HH bands, respectively. Increasing hole energy points downwards.}
    \label{fig:psi}
\end{figure}

 In Fig. \ref{fig:psi} we show the ground state and first excited state (which is LH-like) wave function of the machine learning-optimized Ge+, corresponding to the system in the 3\textsuperscript{rd} column of Table 1 in the main text.

\newpage
\section{\texorpdfstring{Andreev spin qubits}{}}\label{sec:asq}
\subsection{\texorpdfstring{Linearized model of Andreev spin qubits}{}}

We study the non-interacting model for an Andreev Spin Qubit (ASQ), looking at a general equation for the spin-splitting in terms of Fermi velocities of up and down spins.
We consider a single ASQ first in a 1D SNS junction with 3 regions for $z<-L/2$, $|z|<L/2$, and $z>L/2$ described by the three Hamiltonians, respectively

\begin{align}
    H_\mathrm{L}&= (\hbar v_\mathrm{L} k_z -\mu) \tau_z +\Delta_\mathrm{L} (e^{-i\varphi_\mathrm{L}}\tau_++e^{i\varphi_\mathrm{L}}\tau_-), \\
    H_\mathrm{N}&= \left[\hbar\left(v+\sigma_z\delta v/2\right)k_z - \mu\right]\tau_z \\
    H_\mathrm{R}&= (\hbar v_\mathrm{R} k_z -\mu) \tau_z +\Delta_\mathrm{R} (e^{-i\varphi_\mathrm{R}}\tau_++e^{i\varphi_\mathrm{R}}\tau_-).
\end{align}

Here we consider a linearized Hamiltonian close to the chemical potential $\mu\gg\Delta_{m=\mathrm{L},\mathrm{R}}$ of the different regions with $\bra{z}k_z=-i\partial_z\bra{z}$ and $\tau_i$ acting on the Nambu space. Each Hamiltonian above has its respective copy at opposite momentum with negative velocities. We do not include Zeeman fields, but we include a possible difference in Fermi velocities from the up and down spins $\delta v$ due to SOI. The velocity $v$ is the average velocity at the given $\mu$, and we anticipate that it can differ from the velocity without SOI.

The Hamiltonians $H_\mathrm{L}$ and $H_\mathrm{R}$ are independent of the spin $\sigma$ and can be diagonalized only within a given spin subspace without loss of generality. The eigenvalue problem then reduces to a $2\times 2$ problem in Nambu space ($m=\mathrm{L},\mathrm{R}$):

\begin{equation}
    \begin{bmatrix}
        \hbar v_mk_z-\mu & \Delta_me^{-i\varphi_m} \\
        \Delta_me^{i\varphi_m} & -\hbar v_mk_z+\mu
    \end{bmatrix}\begin{bmatrix}
        \ket{\psi_m^\uparrow} \\ \ket{\psi_m^\downarrow}
    \end{bmatrix} = \epsilon
    \begin{bmatrix}
        \ket{\psi_m^\uparrow} \\ \ket{\psi_m^\downarrow}
        \end{bmatrix},
\end{equation}

\noindent where $\uparrow,\downarrow$ refer to the up and down Nambu component, respectively. This Hamiltonian has the general solution:

\begin{subequations}
    \begin{align}
    \psi_m^\uparrow(z) &= \Delta_me^{-i\varphi_m/2}e^{iz/\xi_m}\left(\frac{Ce^{-z/l_m}}{\epsilon-i\Gamma_m}+\frac{De^{z/l_m}}{\epsilon+i\Gamma_m}\right), \\
    \psi_m^\downarrow(z) &= e^{i\varphi_m/2}e^{iz/\xi_m}\left(Ce^{-z/l_m} + De^{z/l_m}\right),
\end{align}
\end{subequations}

\noindent where $\Gamma_m= \sqrt{\Delta_m^2-\epsilon^2}$, $\xi_m = \hbar v_m/\mu$ and $l_m = \hbar v_m/\Gamma_m$. The coefficients $C$ and $D$ are arbitrary constants. We restrict ourselves to bound states, and thus we require the wavefunctions to vanish at infinity,  and in the left (right) region we impose $C=0$ ($D=0$). Reintroducing the spin degree of freedom, we have the total solutions for the left and right regions

\begin{align}
    \ket{\psi_\mathrm{L}} &= \int{dz\,\ket{z}\left[\ket{\uparrow}\psi_\mathrm{L}^\uparrow(z) + \ket{\downarrow}\psi_\mathrm{L}^\downarrow(z)\right]\left(\alpha\ket{+}+\beta\ket{-}\right)}, \\
    \ket{\psi_\mathrm{R}} &= \int{dz\,\ket{z}\left[\ket{\uparrow}\psi_\mathrm{R}^\uparrow(z) + \ket{\downarrow}\psi_\mathrm{R}^\downarrow(z)\right]\left(\gamma\ket{+}+\delta\ket{-}\right)}, \\
\end{align}

\noindent where $\alpha$, $\beta$, $\gamma$ and $\delta$ are arbitrary constants and

\begin{align}
    \psi_\mathrm{L}^\uparrow(z) &= \frac{\Delta_\mathrm{L}}{\epsilon+i\Gamma_\mathrm{L}}e^{-i\varphi_\mathrm{L}/2}e^{iz/\xi_\mathrm{L}}e^{z/l_\mathrm{L}}, \\
    \psi_\mathrm{R}^\uparrow(z) &= \frac{\Delta_\mathrm{R}}{\epsilon-i\Gamma_\mathrm{R}}e^{-i\varphi_\mathrm{R}/2}e^{iz/\xi_\mathrm{R}}e^{-z/l_\mathrm{R}}, \\
    \psi_\mathrm{L}^\downarrow(z) &= e^{i\varphi_\mathrm{L}/2}e^{iz/\xi_\mathrm{L}}e^{z/l_\mathrm{L}}, \\
    \psi_\mathrm{R}^\downarrow(z) &= e^{i\varphi_\mathrm{R}/2}e^{iz/\xi_\mathrm{R}}e^{-z/l_\mathrm{R}}.
\end{align}

\noindent Here we absorbed the $C$ and $D$ constants into $\alpha\dots\delta$.

In the semiconductor, the Hamiltonian is diagonal in both Nambu and spin space. The general solution for Nambu component $\tau=\uparrow,\downarrow$ and spin component $\sigma=+,-$ is

\begin{gather}
    \left[\hbar\left(v+\sigma\delta v/2\right)k_z - \mu\right]\tau\ket{\psi_\mathrm{N}^{\tau\sigma}} = \epsilon\ket{\psi_\mathrm{N}^{\tau\sigma}} \\
    \Leftrightarrow\psi_\mathrm{N}^{\tau\sigma}(z) = \braket{z|\psi_\mathrm{N}^{\tau\sigma}} = e^{iz/\xi_\mathrm{N}^\sigma}e^{i\tau z/l_\mathrm{N}^\sigma},
\end{gather}

\noindent where $\xi_\mathrm{N}^\sigma = \hbar v_\sigma/\mu$, $l_\mathrm{N}^\sigma = \hbar v_\sigma/\epsilon$ and $v_\sigma = v + \sigma\delta v/2$. A total general state for the semiconducting region is therefore

\begin{equation}
    \ket{\psi_\mathrm{N}} = a\ket{\uparrow}\ket{+}\ket{\psi_\mathrm{N}^{\uparrow +}} + b\ket{\uparrow}\ket{-}\ket{\psi_\mathrm{N}^{\uparrow -}} + c\ket{\downarrow}\ket{+}\ket{\psi_\mathrm{N}^{\downarrow +}} + d\ket{\downarrow}\ket{-}\ket{\psi_\mathrm{N}^{\downarrow -}},
\end{equation}

\noindent with $a$, $b$, $c$ and $d$ arbitrary constants. If we neglect any flipping of the spin upon reflection at the superconducting-semiconducting interfaces, then we can discard the spin-down components of the total wavefunction, and solve the simultaneous boundary conditions only for the $\ket{\uparrow}\ket{+}$ and $\ket{\downarrow}\ket{+}$ components. This is equivalent to setting $\beta =\delta = 0$ and $b=d=0$ and gives the following four continuity conditions at $|z|=L/2$:

\begin{align}
    \gamma\psi_\mathrm{R}^\uparrow\left(\frac{L}{2}\right) &= a\psi_\mathrm{N}^{\uparrow+}\left(\frac{L}{2}\right),
    &\gamma\psi_\mathrm{R}^\downarrow\left(\frac{L}{2}\right) &= c\psi_\mathrm{N}^{\downarrow+}\left(\frac{L}{2}\right),\label{pL2} \\
    \alpha\psi_\mathrm{L}^\uparrow\left(-\frac{L}{2}\right) &= a\psi_\mathrm{N}^{\uparrow+}\left(-\frac{L}{2}\right),
    &\alpha\psi_\mathrm{L}^\downarrow\left(-\frac{L}{2}\right) &= c\psi_\mathrm{N}^{\downarrow+}\left(-\frac{L}{2}\right).\label{mL2}
\end{align}

From Eqs.~\eqref{pL2} and \eqref{mL2} we obtain, respectively

\begin{align}
    \frac{a}{c} &= \frac{\Delta_\mathrm{R}}{\epsilon-i\Gamma_\mathrm{R}}e^{-i\varphi_\mathrm{R}}e^{-iL/l_\mathrm{N}^+}, \\
    \frac{a}{c} &= \frac{\Delta_\mathrm{L}}{\epsilon+i\Gamma_\mathrm{L}}e^{-i\varphi_\mathrm{L}}e^{iL/l_\mathrm{N}^+},
\end{align}

\noindent which leads to the general solution for $\epsilon$, assuming $\Delta_\mathrm{R}=\Delta_\mathrm{L}=\Delta$:

\begin{equation}\label{e}
    \epsilon\sin\left(\epsilon/E_L - \varphi/2\right) = \sqrt{\Delta^2-\epsilon^2}\cos\left(\epsilon/E_L - \varphi/2\right),
\end{equation}

\noindent where $E_L = \hbar v_+/L$ and $\varphi = \varphi_\mathrm{L}-\varphi_\mathrm{R}$. The same equation also holds for the spin-down level with $v_+$ replaced by $v_-$, and agrees with Reference \onlinecite{Sauls2018} in the short junction limit at perfect transparency. Taking the square followed by an $\arccos$ function on both sides of \eqref{e} leads to the alternative formulation~\cite{Coppini2026}:

\begin{equation}
    \arccos\left(\frac{\epsilon}{\Delta}\right) - \frac{\epsilon}{E_L} = -\frac{\varphi}{2} + n\pi,
\end{equation}

\noindent where $n$ is an integer.

\subsection{\texorpdfstring{Spin splitting}{}}
We now analyze the solutions to the transcendental equation \eqref{e} and their variations with respect to small changes in Fermi velocities due to SOI. We first re-write \eqref{e} in terms of the dimensionless constants $x=\epsilon/\Delta\in[-1,1]$ and $\alpha = L\Delta/(\hbar v)$ where $v$ could be any Fermi velocity, and we define the function $f_\alpha(x)$:

\begin{equation}
    f_\alpha(x)=x\tan(\alpha x-\varphi/2)-\sqrt{1-x^2}.
\end{equation}

Solving $f_\alpha(x_0) = 0$ for $x_0$ also solves \eqref{e} with $\epsilon = x_0\Delta$. A small variation in the Fermi velocity will result in a small variation $\delta\alpha$ of the parameter $\alpha$: $\alpha\to\alpha+\delta\alpha$. Under such a change the function $f_\alpha(x)$ becomes

\begin{align}
    f_{\alpha+\delta\alpha}(x) &= f_\alpha(x) + \frac{d}{d\alpha}f_\alpha(x)\delta\alpha + \mathcal{O}(\delta\alpha^2) \\
    &= x\tan\left(\alpha x-\varphi/2\right)\left[1 + \delta\alpha x\tan\left(\alpha x-\varphi/2\right)\right] - \sqrt{1-x^2} + \delta\alpha x^2 + \mathcal{O}(\delta\alpha^2).
\end{align}

\noindent Evaluating $f_{\alpha+\delta\alpha}(x)$ at $x=x_0$ gives $f_{\alpha+\delta\alpha}(x_0) = \delta\alpha+\mathcal{O}(\delta\alpha^2)$. The new zero $x_1$ of the function $f_{\alpha+\delta\alpha}(x)$ can be estimated by one iteration of the Newton method:

\begin{align}
    x_1&\approx x_0-\frac{f_{\alpha+\delta\alpha}(x_0)}{\left.\frac{d}{dx}f_{\alpha+\delta\alpha}(x)\right|_{x=x_0}} \\
    &= x_0\left[1-\frac{\delta\alpha\sqrt{1-x_0^2}}{\left(1+\alpha\sqrt{1-x_0^2}\right)\left(1+2\delta\alpha\sqrt{1-x_0^2}\right)}\right].
\end{align}

If the shift in $\alpha$ for the spin-up branch is $\alpha\to\alpha-\delta\alpha_+$ and the shift in the spin-down branch is $\alpha\to\alpha+\delta\alpha_-$, then the difference of the two shifted zeros $x_1^--x_1^+$ is proportional to the ASQ spin splitting due to SOI:

\begin{equation}\label{ASQx}
    x_1^- - x_1^+ = -\frac{x_0\sqrt{1-x_0^2}}{1 + \alpha\sqrt{1-x_0^2}}\left[\frac{\delta\alpha_+ + \delta\alpha_-}{\left(1 - 2\delta\alpha_+\sqrt{1-x_0^2}\right)\left(1 + 2\delta\alpha_-\sqrt{1-x_0^2}\right)}\right].
\end{equation}

If $v_0$ is the velocity without SOI, then we generally expect the following for the velocities $v_\pm$:

\begin{align}
    v_+ &= v_0 + v_d + \delta v/2, \\
    v_- &= v_0 + v_d - \delta v/2,
\end{align}

\noindent where $v_d$ is a SOI-induced velocity drift that is identical for both spins. If $f_\alpha(x_0)=0$ where $\alpha$ and $x_0$ pertain to the {\it drifted velocity} $v\equiv v_0+v_d$, then the changes in velocity $\pm\delta v/2$ are symmetrical with respect to $v$ and the variations in $\alpha$ become

\begin{align}
    \delta\alpha_+ &= \frac{L\Delta}{\hbar}\left(\frac{1}{v} - \frac{1}{v+\delta v/2}\right) = \alpha\left(\frac{\delta v/2}{v+\delta v/2}\right), \\
    \delta\alpha_- &= \frac{L\Delta}{\hbar}\left(\frac{1}{v-\delta v/2}-\frac{1}{v}\right) = \alpha\left(\frac{\delta v/2}{v-\delta v/2}\right).
\end{align}

Inserting $\delta\alpha_\pm$ into the bracket of Eq. \eqref{ASQx} and expanding with respect to $\delta v$ gives the ASQ spin splitting formula for arbitrary junction length $L$:

\begin{equation}
    \delta\epsilon \equiv \epsilon_--\epsilon_+ = \left(\frac{-\epsilon/v}{1+E_L/\Gamma}\right)\delta v + \mathcal{O}\left(\delta v^3\right),
\end{equation}

\noindent where $\Gamma = \sqrt{\Delta^2-\epsilon^2}$ and $E_L=\hbar v/L$. Here $\epsilon$ is the bound state energy of the system computed for  the {\it average} velocity between those with SOI, i.e. $v = (v_++v_-)/2$. This equation is equivalent to Eq. (4) in the main text.

In the short junction limit $\alpha\ll 1$, we can assume that the solution $x_0$ of $f_\alpha(x_0)=0$ takes the general form

\begin{equation}
    x_0 = c_0+c_1\alpha+c_2\alpha^2+\dots
\end{equation}

\noindent Inserting this in $f_\alpha(x)$, expanding in terms of $\alpha$ and setting the condition $f_\alpha(x_0)=0$ results in the following $c_i$ coefficients:

\begin{subequations}
    \begin{align}
    c_0 &= -\cos(\varphi/2), \\
    c_1 &= \sin(\varphi)/2, \\
    c_2 &= \frac{1}{8}\left[\cos(\varphi/2) + 3\cos(3\varphi/2)\right], \\
    c_3 &= -\frac{1}{6}\left[\sin(\varphi) + 2\sin(2\varphi)\right], \\
    &\dots\nonumber
\end{align}
\end{subequations}

\noindent Neglecting $\alpha^3$ terms and higher, the difference of zeros $x_0^\pm$ becomes 

\begin{equation}
    x_0^- - x_0^+ \approx \frac{1}{2}\left(\alpha_--\alpha_+\right)\sin\varphi + \frac{1}{8}\left(\alpha_-^2-\alpha_+^2\right)\left[\cos(\varphi/2) + 3\cos(3\varphi/2)\right],
\end{equation}

\noindent which to lowest order in $\delta v$ leads to the spin splitting  provided in the main text [see Eq.~(5)]:

\begin{equation}\label{deS}
    \delta\epsilon \approx \frac{L\Delta^2}{2\hbar}\left\{\sin(\varphi) + \frac{L\Delta}{2\hbar v}\left[\cos\left(\frac{\varphi}{2}\right) + 3\cos\left(\frac{3\varphi}{2}\right)\right]\right\}\frac{\delta v}{v^2} + \mathcal{O}(\delta v^3).
\end{equation}







\noindent The approximation sign in \eqref{deS} comes from neglecting $\alpha^3$ and higher terms.

\subsection{\texorpdfstring{Expansion of velocities in terms of subband parameters}{}}

We consider here a wide junction, such that confinement effects along the $x$-direction can be neglected. In Ge, the ASQ still presents a difference in Fermi velocity arising from the cubic SOI. To model this situation, we integrate out the $x$-direction by taking the $k_x=0$ limit of the dispersion relation.

The effective low-energy Hamiltonian of the HH ground state at $k_x=0$ and arbitrary $k_y$ can be generally written as a sum of powers of $k\equiv k_y$, with the linear-in-$k$ term omitted due to its small magnitude:

\begin{equation}\label{Hseries}
    \mathcal{H}_\mathrm{eff}^\sigma(k)=\sum_{i=2}{c_ik^i\sigma^i},
\end{equation}

\noindent where $\sigma=\pm 1$ is the pseudo-spin with $\sigma^2=1$. Fixing the energy at a given chemical potential $\mu$ and inverting the series \eqref{Hseries} for the two pseudo-spins, we get an expansion for $k$ in terms of powers of $\mu$:

\begin{equation}\label{kvsmu}
    k_\sigma(\mu) = \sum_{j=1}{a_j^\sigma\mu^{j/2}},
\end{equation}

\noindent where the $a_j^\sigma$ are expansion coefficients that depend on the $c_i$'s. The first four terms are

\begin{equation}
    k_\sigma(\mu) = \frac{1}{c_2^{1/2}}\mu^{1/2} + \frac{\sigma}{2c_2^2}c_3\mu + \frac{1}{8c_2^{7/2}}\left(5c_3^2-4c_2c_4\right)\mu^{3/2} - \frac{\sigma}{2c_2^5}\left(2c_3^3-3c_2c_3c_4+c_2^2c_5\right)\mu^2 + \mathcal{O}\left(\mu^{5/2}\right).
\end{equation}

On the other hand, the velocity $v_\sigma$ is given by

\begin{equation}\label{vsigma}
    v_\sigma \equiv \frac{1}{\hbar}\frac{d\mathcal{H}_\mathrm{eff}^\sigma(k)}{dk} = \frac{1}{\hbar}\sum_{i=2}{ic^ik^{i-1}\sigma^i}.
\end{equation}

\noindent We recall the following relations pertaining to the spin-dependent velocities:

\begin{subequations}
    \begin{align}
    v_+ &= v_0+v_d+\delta v/2, \\
    v_- &= v_0+v_d-\delta v/2,
\end{align}
\end{subequations}

\noindent where $v_0$ is the velocity without SOI (corresponding to all odd-$k$ powers in $\mathcal{H}_\mathrm{eff}^\sigma(k)$ set to zero), $v_d$ is a velocity drift that is the same for both spins and $\delta v$ is the difference of Fermi velocities. Inserting $k_\sigma(\mu)$ into the velocities gives their expansions in powers of $\mu$:

\begin{subequations}
\begin{align}
    \hbar v_0 &= 2c_2^{1/2}\mu^{1/2} + \frac{3c_4}{c_2^{3/2}}\mu^{3/2} + \mathcal{O}\left(\mu^{5/2}\right), \\
    \hbar\delta v &= \frac{4c_3}{c_2}\mu + \frac{1}{c_2^4}\left(5c_3^3-12c_2c_3c_4+8c_2^2c_5\right)\mu^2 + \mathcal{O}\left(\mu^3\right), \\
    \hbar v_d &= -\frac{7c_3^2}{4c_2^{5/2}}\mu^{3/2} + \mathcal{O}\left(\mu^{5/2}\right), \\
    \frac{\delta v}{v} &= \frac{2c_3}{c_2^{3/2}}\mu^{1/2} + \frac{1}{4c_2^{9/2}}\left(17c_3^3-36c_2c_3c_4+16c_2^2c_5\right)\mu^{3/2} + \mathcal{O}\left(\mu^{5/2}\right), \\
    \frac{\delta v}{\hbar v^2} &= \frac{c_3}{c_2^2} + \frac{1}{c_2^5}\left(4c_3^3-6c_2c_3c_4+2c_2^2c_5\right)\mu \\
    &+ \frac{3}{c_2^8}\left(7c_3^5 -20c_2c_3^3c_4 + 10c_2^2c_3^2c_5 + 10c_2^2c_3c_4^2 - 4c_2^3c_3c_6 - 4c_2^3c_4c_5 + c_2^4c_7\right)\mu^2 + \mathcal{O}\left(\mu^3\right). \nonumber
\end{align}
\end{subequations}

One can replace the $c_i$ coefficients with the corresponding ones from \eqref{HHeff}: $c_2=\alpha_0\gamma$ and $c_3=\alpha_0\tilde{\beta}$, where $\tilde{\beta}=-\beta_2+\beta_3$. We point out that \eqref{HHeff} is only cubic in $k$ and thus cannot capture the contributions from $c_4$ and above. An exact description of velocities up to $c_i$ requires a $i$\textsuperscript{th} order SWT on \eqref{Hk}.

\section{\texorpdfstring{Quantum dot Spin qubits}{}}\label{sec:qd}
\subsection{\texorpdfstring{Model}{}}
The Hamiltonian of the quantum dot (QD) system is given by \eqref{Hk} or \eqref{Hk_} with $V_\parallel(x,y)$ given by an effective parabolic potential with effective lengths $l_x$ and $l_y$:

\begin{equation}
    V_\parallel(x,y) = -\alpha_0\left(\frac{x^2}{l_x^4} + \frac{y^2}{l_y^4}\right).
\end{equation}

\noindent where the minus sign comes from our electron-energy convention. We first write the Hamiltonian at $B=0$ and diagonalize it by the usual ladder operator approach. At $B=0$ the Hamiltonian is

\begin{equation}\label{Hqd}
    \mathcal{H}_0^\mathrm{qd} = \mathcal{E}_0^\mathrm{qw} + \alpha_0\left[\mathcal{M}_\gamma k_\parallel^2 + \left(i\mathcal{M}_1 k_- + \mathcal{M}_2k_-^2 + \text{h.c.}\right)\right] - \alpha_0\left(\frac{x^2}{l_x^4} + \frac{y^2}{l_y^4}\right).
\end{equation}

The $\mathcal{M}$-matrices are computed following the method described in Section \ref{sec:pert}. It is a $2$\textsuperscript{nd} order SWT Hamiltonian for the first $3$ quantum well spin doublets (first 2 $\h{}$ levels and first $\e{}$ level, $\mathcal{A} = \{\h{1},\h{2},\e{1}\}$) computed from a total space of $250$ quantum well subbands. This effective Hamiltonian is the only one that includes an $\e{}$ level in the $\mathcal{A}$-set whose band structure does not bend down at large $k_\parallel$ for both unspiked and spiked Ge. For $\varepsilon$-Ge we use the same approach but with $\mathcal{A} = \{\h{1},\h{2},\e{1},\e{2}\}$, also computed from a total space of $250$ quantum well subbands. We write the position and momentum operators with the following ladder operators:

\begin{align}
    a_1 &= \frac{\rho_-}{l} + \frac{ilk_-}{2}, & a_2 &= \frac{\rho_+}{l} + \frac{ilk_+}{2}, \\
    \Leftrightarrow\rho_- &= \frac{l}{2}\left(a_1 + a_2^\dagger\right), &k_- &= \frac{1}{il}\left(a_1-a_2^\dagger\right),
\end{align}

\noindent where $\rho_\pm = (x\pm iy)/2$ and 

\begin{equation}
    \frac{1}{l^4} = \frac{1}{2}\left(\frac{1}{l_x^4} + \frac{1}{l_y^4}\right).
\end{equation}

The $a_i$, $\rho_\pm$ and $k_\pm$ satisfy the following commutation rules:

\begin{align}
    [a_i,a_j^\dagger] &= \delta_{i,j}, &[\rho_\pm,k_\pm] &= 0, &[\rho_\pm,k_\mp] &= i.
\end{align}

In terms of $\rho_\pm$, the potential becomes

\begin{equation}
    V_\parallel(x,y) = -\frac{\alpha_0}{l^4}\left[4\rho_-\rho_+ + 2\left(\rho_-^2+\rho_+^2\right)\cos\xi\right],
\end{equation}

\noindent with $\xi = 2\arctan(l_x^2/l_y^2)\in [0,\pi]$. Finally, mutual eigenstates $\ket{n_1,n_2}$ of $a_1^\dagger a_1$ and $a_2^\dagger a_2$ (with integers $n_i\geq 0$) are also eigenstates of the angular momentum $L_z = xk_y-yk_x$. This is summarized as follow:

\begin{align}
    a_1\ket{n_1,n_2} &= \sqrt{n_1}\ket{n_1-1,n_2}, & a_2\ket{n_1,n_2} &= \sqrt{n_2}\ket{n_1,n_2-1}, \\
    a_1^\dagger\ket{n_1,n_2} &= \sqrt{n_1+1}\ket{n_1+1,n_2}, & a_2^\dagger\ket{n_1,n_2} &= \sqrt{n_2+1}\ket{n_1,n_2+1}, \\
    a_1^\dagger a_1\ket{n_1,n_2} &= n_1\ket{n_1,n_2}, & a_2^\dagger a_2\ket{n_1,n_2} &= n_2\ket{n_1,n_2},
\end{align}

\begin{equation}
    L_z\ket{n_1,n_2} = (n_1-n_2)\ket{n_1,n_2}.
\end{equation}

\noindent One can also define the quantum numbers $j\equiv n_1+n_2$ and $m\equiv n_1-n_2$ and write $L_z\ket{j,m}=m\ket{j,m}$, where $j=0,1,\dots$ and $m = -j,-j+2,\dots,j-2,j$.

To diagonalize $\mathcal{H}_0^\mathrm{qd}$ we project it down onto a basis consisting of all the shells from $j=0$ up to $j=20$ (total $231$ spin doublets). The uncertainty on the in-plane momentum $\sqrt{\braket{k_\parallel^2}} = \frac{1}{l}\sqrt{1+j}$ with $j=20$ and $l=11.2\,\mathrm{nm}$ samples regions in $k$-space as far as $k_\parallel\sim 0.41\,\mathrm{nm}^{-1}$. We define the $B=0$ QD pseudo-spin index $\varsigma$ from the in-plane symmetry of the wavefunctions. For example, the $(\h{+},\h{+})$ sector of the QD Hamiltonian is quadratic in $k$ and $\rho$ and therefore does not change the parity of the wavefunction in the $\h{+}$ sector. On the other hand, the $(\h{+},\e{+})$ sector is linear in $k$ and therefore changes the parity of the wavefunction in the sector $\e{+}$. The two pseudo-spins are therefore chosen to be (with sector ordering $\{\h{+},\e{+},\e{-},\h{-}\}$)

\begin{align}
    \varsigma=+:\ket{\varphi}&=\begin{bmatrix}
        \mathrm{A} \\ \mathrm{S} \\ \mathrm{A} \\ \mathrm{S}
    \end{bmatrix},
    &\varsigma=-:\ket{\varphi}&=\begin{bmatrix}
        \mathrm{S} \\ \mathrm{A} \\ \mathrm{S} \\ \mathrm{A}
    \end{bmatrix},
\end{align}

\noindent where the symmetric (S) and antisymmetric (A) parts involve only even $j$ or odd $j$ orbital shells, respectively. More explicitly, the $\varsigma=+1$ eigenstates can be expanded in terms of the quantum well subbands $\ket{\tau\sigma;l}$ and of the Fock orbitals $\ket{j,m}$:

\begin{equation}
\begin{split}
    \ket{\varphi_{i,J,M}^+} &= \sum_{l,\mathrm{odd}\,j,m}{\ket{\h{+};l}\ket{j,m}A_{i,J,M}^{\h{};l,j,m}} + \sum_{l,\mathrm{even}\,j,m}{\ket{\e{+};l}\ket{j,m}S_{i,J,M}^{\e{};l,j,m}} \\
    &+ \sum_{l,\mathrm{odd}\,j,m}{\ket{\e{-};l}\ket{j,m}A_{i,J,M}^{\e{};l,j,m}} + \sum_{l,\mathrm{even}\,j,m}{\ket{\h{-};l}\ket{j,m}S_{i,J,M}^{\h{};l,j,m}},
\end{split}
\end{equation}

\noindent where $i$ is a general $z$-subband index, and $(J,M)$ are general in-plane orbital indices, and where the $A$ and $S$ coefficients vanish for even $j$ or odd $j$ respectively. The $\varsigma = -1$ pseudo-spin levels are constructed from time reversal rules:

\begin{equation}
\begin{split}
    \ket{\varphi_{i,J,M}^-} &= -\sum_{l,\mathrm{odd}\,j,m}{\ket{\h{-};l}\ket{j,-m}A_{i,J,M}^{\h{};l,j,m}} + \sum_{l,\mathrm{even}\,j,m}{\ket{\e{-};l}\ket{j,-m}S_{i,J,M}^{\e{};l,j,m}} \\
    &- \sum_{l,\mathrm{odd}\,j,m}{\ket{\e{+};l}\ket{j,-m}A_{i,J,M}^{\e{};l,j,m}} + \sum_{l,\mathrm{even}\,j,m}{\ket{\h{+};l}\ket{j,-m}S_{i,J,M}^{\h{};l,j,m}}.
\end{split}
\end{equation}

\noindent We note that the $A$ and $S$ coefficients are all real. The quantum dot basis at $B=0$ is then used to project the full Hamiltonian \eqref{Hk} or \eqref{Hk_} onto it. Using the following gauge for the out-of-plane component of the $B$-field:

\begin{equation}
    \mathbf{A}_\parallel(\mathbf{r}) = \frac{B\cos\theta}{2}\left(-y\mathbf{e}_x+x\mathbf{e}_y\right),
\end{equation}

\noindent the full quantum dot Hamiltonian is

\begin{equation}
    \mathbf{H}^\mathrm{qd}(B) = \mathbf{E}_0^\mathrm{qd}+\frac{\alpha_0}{2l_B^2}\left[\cos\theta\mathbf{L}_{2\perp} + \frac{\cos^2\theta}{2l_B^2}\mathbf{L}_{4\perp} + \frac{\sin^2\theta}{2l_B^2}\mathbf{L}_{4\parallel}+\sin\theta\left(e^{-i\phi}\mathbf{L}_{2\parallel}+\frac{\sin\theta}{2l_B^2}e^{-2i\phi}\tilde{\mathbf{L}}_{4\parallel}+\frac{\cos\theta}{2l_B^2}e^{-i\phi}\mathbf{L}_{4\times}+\mathrm{h.c.}\right)\right],
\end{equation}

\noindent where $\mathbf{E}_0^\mathrm{qd}$ are the eigenvalues of $\mathcal{H}_0^\mathrm{qd}$ and the $\mathbf{L}$-matrices are

\begin{subequations}
    \begin{align}
    \mathbf{L}_{2\perp} &= \mathbf{V}^\dagger\left[\mathcal{M}_g+2\mathcal{M}_\gamma L_z + \left(2\mathcal{M}_1\rho_--4i\mathcal{M}_2\rho_-k_-+\mathrm{h.c.}\right)\right]\mathbf{V}, \\
    \mathbf{L}_{2\parallel} &= \mathbf{V}^\dagger\left[\mathcal{N}_g - i\tilde{\mathcal{N}}_\gamma\rho_+ - \mathcal{N}_1^+\rho_+k_- - \mathcal{N}_1^-k_-\rho_+ + \left(\mathcal{N}_1^++\mathcal{N}_1^-\right)^\dagger\rho_+k_+\right]\mathbf{V}, \\
    \mathbf{L}_{4\perp} &= \mathbf{V}^\dagger\left[4\mathcal{M}_\gamma\rho_-\rho_+ - 4\left(\mathcal{M}_2\rho_-^2+\mathrm{h.c.}\right)\right]\mathbf{V}, \\
    \mathbf{L}_{4\parallel} &= \mathbf{V}^\dagger\left[2\mathcal{N}_\gamma\rho_-\rho_+\right]\mathbf{V}, \\
    \tilde{\mathbf{L}}_{4\parallel} &= \mathbf{V}^\dagger\left[-\mathcal{N}_\gamma\rho_+^2\right]\mathbf{V}, \\
    \mathbf{L}_{4\times} &= \mathbf{V}^\dagger\left[2i\left(\mathcal{N}_1^++\mathcal{N}_1^-\right)\rho_-\rho_++2i\left(\mathcal{N}_1^++\mathcal{N}_1^-\right)^\dagger\rho_+^2\right]\mathbf{V}, 
    \end{align}
\end{subequations}

\noindent with $\mathbf{V}$ the matrix of (column) eigenvectors of $\mathcal{H}_0^\mathrm{qd}$, such that $\mathbf{V}^\dagger\mathbf{V} = \mathbb{1}$.

\begin{figure}[t]
    \centering
    \includegraphics[scale=0.6]{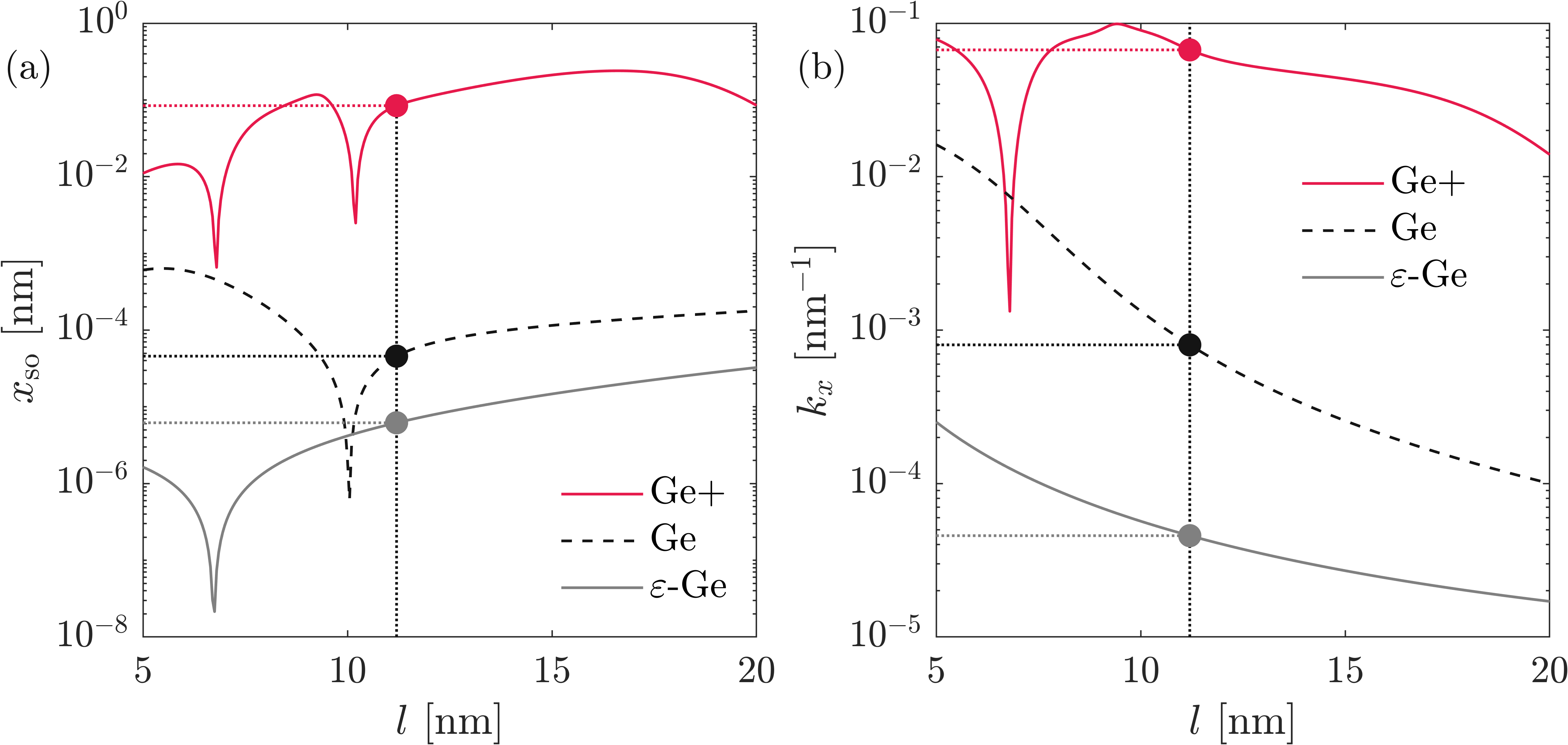}
    \caption{(a) $|\braket{\varphi_{1,0,0}^+|x|\varphi_{1,0,0}^-}|$ and (b) $|\braket{\varphi_{1,0,0}^+|k_x|\varphi_{1,0,0}^-}|$ matrix element for the optimized spikes, unspiked and $\varepsilon$-Ge heterostructure. $\mathbf{B} = (0.1\,\mathrm{T})\mathbf{e}_x$. Dotted lines are guidelines indicating the values for $l = 11.2\,\mathrm{nm}$.}
    \label{fig:li}
\end{figure}

Figure \ref{fig:li} shows the off-diagonal matrix element $x_\mathrm{so} = |\braket{\varphi_{1,0,0}^+|x|\varphi_{1,0,0}^-}|$ and $k_\mathrm{so} = |\braket{\varphi_{1,0,0}^+|k_x|\varphi_{1,0,0}^-}|$ for the quantum dot ground state orbital with an in-plane magnetic field $\mathbf{B} = (0.1\,\mathrm{T})\mathbf{e}_x$, and a gate field $F_z = 1.5\,\mathrm{mV}/\mathrm{nm}$.

\begin{figure}[ht]
    \centering
    \includegraphics[scale=0.6]{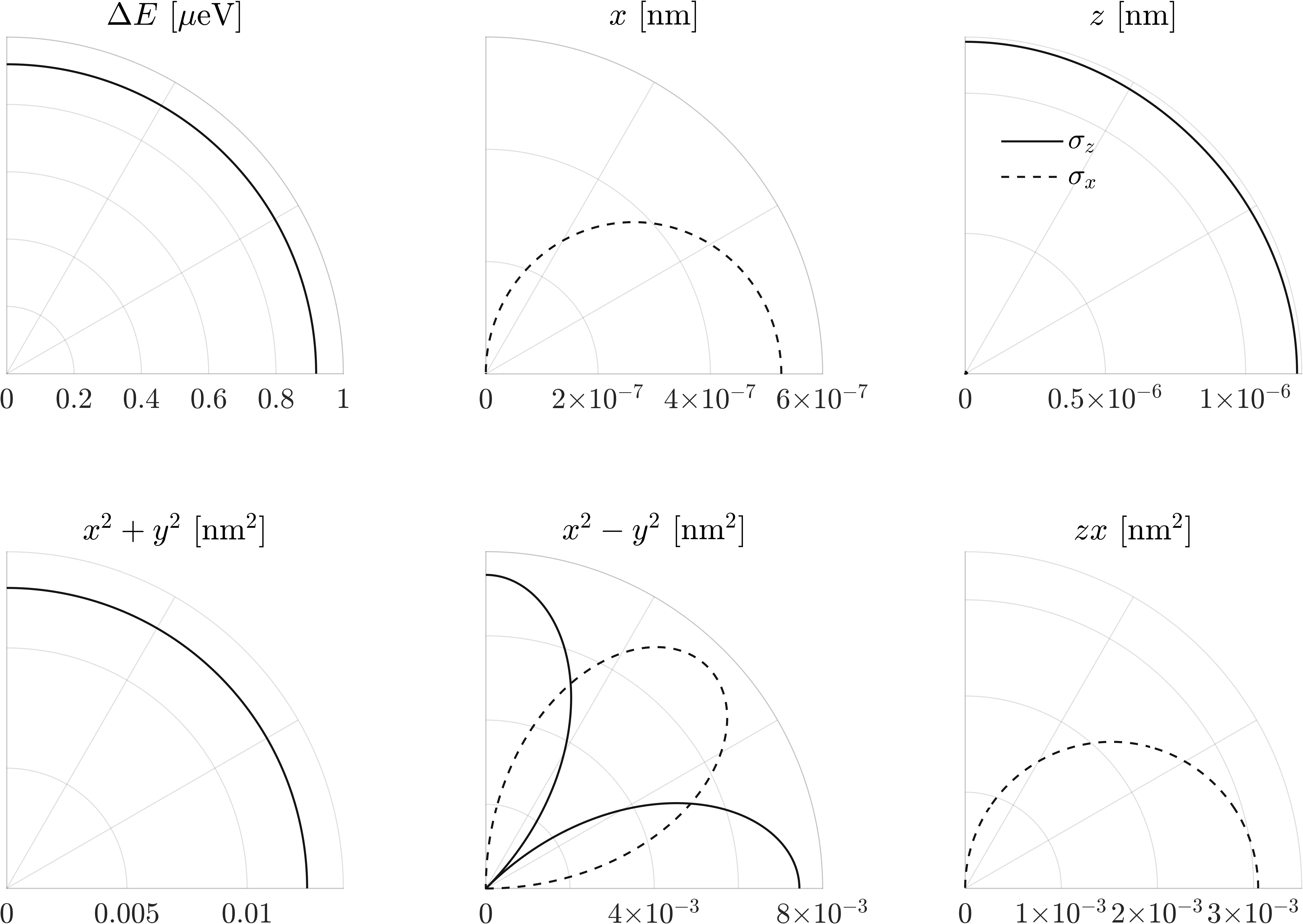}
    \caption{$\varepsilon$-Ge: Longitudinal (solid lines) and transverse (dashed lines) components of linear and quadratic position operators as a function of the magnetic field in-plane angle $\phi$ for $B = 0.1\,\mathrm{T}$ and $l = 11.2\,\mathrm{nm}$. Also shown is the Zeeman splitting energy $\Delta E$. Computed with an $\mathcal{A}$-set that includes the first two $\h{}$ spin doublets and the first two $\e{}$ spin doublets.}
    \label{fig:opse}
\end{figure}

\begin{figure}[ht]
    \centering
    \includegraphics[scale=0.6]{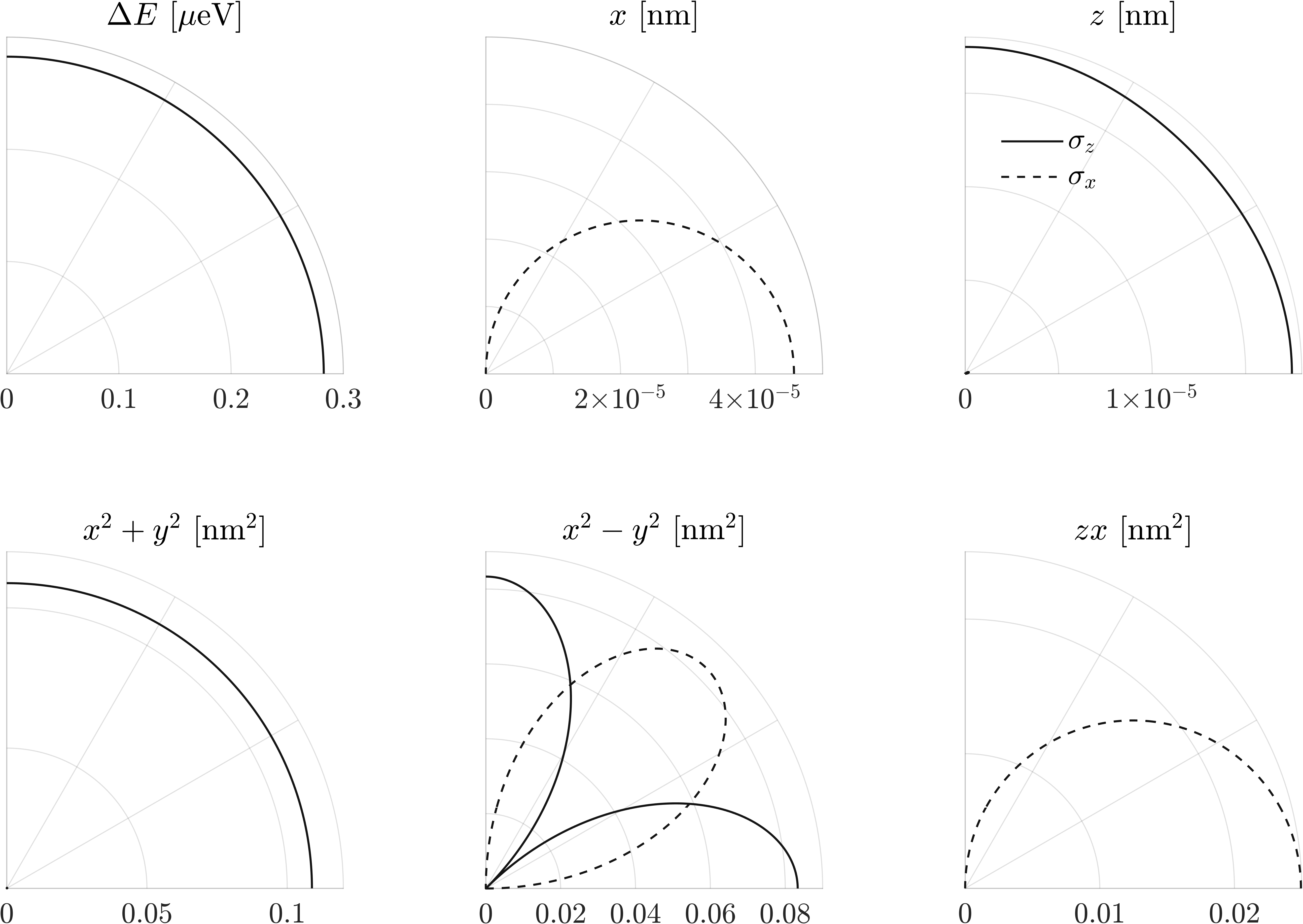}
    \caption{Unspiked Ge: Longitudinal (solid lines) and transverse (dashed lines) components of linear and quadratic position operators as a function of the magnetic field in-plane angle $\phi$ for $B = 0.1\,\mathrm{T}$ and $l = 11.2\,\mathrm{nm}$. Also shown is the Zeeman splitting energy $\Delta E$.}
    \label{fig:opsu}
\end{figure}

\begin{figure}[ht]
    \centering
    \includegraphics[scale=0.6]{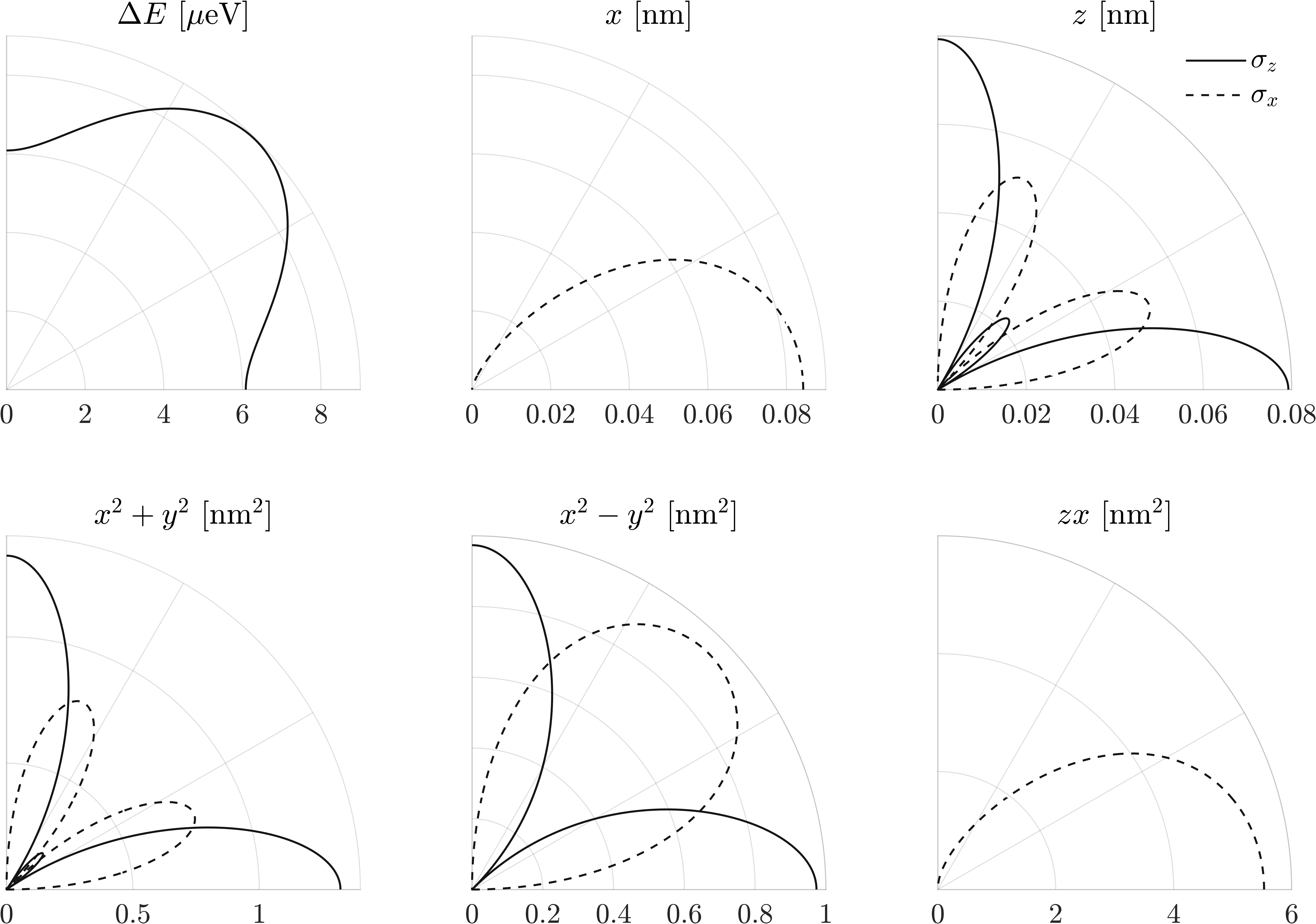}
    \caption{Spiked Ge: Longitudinal (solid lines) and transverse (dashed lines) components of linear and quadratic position operators as a function of the magnetic field in-plane angle $\phi$ for $B = 0.1\,\mathrm{T}$ and $l = 11.2\,\mathrm{nm}$. Also shown is the Zeeman splitting energy $\Delta E$.}
    \label{fig:opss}
\end{figure}

Similarly, we also plot in Fig. \ref{fig:opse}, \ref{fig:opsu} and \ref{fig:opss} the longitudinal ($\sigma_z$) and the transverse ($\sigma_x$) components, in the QD ground state subspace, of the operators that appear in the quadratic expansion of the electrostatic potential underneath the gates (as discussed in the next section) as a function of the magnetic field in-plane angle $\phi$.

\newpage

\subsection{\texorpdfstring{Dephasing time calculations}{}}

We consider the dephasing in a quantum dot system where charge traps cause small fluctuations of the gate voltages. The gate layout consists of one square plunger gate of dimension $170\,\mathrm{nm}\times 170\,\mathrm{nm}$ with a voltage of $V_p^0=-170\,\mathrm{mV}$ at the DC operation point. The four identical barrier gates have dimensions $170\,\mathrm{nm}\times 50\,\mathrm{nm}$, are separated from the plunger by a $5\,\mathrm{nm}$ spacer, and operate at a voltage of $V_b^0 = -70\,\mathrm{mV}$. The simulation domain has dimensions $300\times 300\times400\,\mathrm{nm}$, with the center of the plunger gate at $x,y$ coordinates $(0,0)$ and with the Ge layer starting $20\,\mathrm{nm}$ below the surface. The electrostatic potential $\phi$ is found by solving the Laplace equation

\begin{align}
    \nabla\cdot\left[\epsilon(\mathbf{r})\nabla \phi\right] &= 0, \\
    \Leftrightarrow\left[\nabla^2 + \frac{\epsilon^\prime(z)}{\epsilon(z)}\frac{\partial}{\partial z}\right]\phi &= 0,
\end{align}

\noindent with a mesh size $\delta x = \delta y = \delta z = 5\,\mathrm{nm}$. We take Dirichlet boundary conditions of $0\,\mathrm{mV}$ everywhere except at the surface where the value of $\phi$ is either $0\,\mathrm{mV}$ or set by the corresponding gate voltage. The dielectric constant only depends on $z$ and is equal to $16$ in Ge and $15.18$ in Si$_{0.2}$Ge$_{0.8}$ by linear interpolation~\cite{Madelung1991}. The voltages and dimensions mentioned above are such that at $(x,y,z) = (0,0,z_0)$ (where $z_0\equiv 0\,\mathrm{nm}$ refers to the $z$-coordinate in Ge $5\,\mathrm{nm}$ below the top Si$_{0.2}$Ge$_{0.8}$ barrier), the gate field $F_z\approx1.487\,\mathrm{mV/\mathrm{nm}}$. The corresponding quantum dot effective length is $l\approx 11.2\,\mathrm{nm}$. We write $\vec{V} = (V_p,V_1,V_2,V_3,V_4)$ being the voltages of the plunger and of the four barrier gates at $+x$, $+y$, $-x$ and $-y$, respectively. In particular, we write for gate $i$: $V_i = V_i^0 + \delta_i$, where $V_i^0$ is the DC operation point and $\delta_i$ is the small fluctuation from noise. We take $\delta_i = 1\,\mathrm{mV}$ for each gate, such that $T_2^*\sim1\,\upmu\mathrm{s}$ for $\varepsilon$-Ge which is the right order of magnitude for hole spin qubits in Ge~\cite{StanoArXiv}.

When barrier noise is zero, the general solution of the Poisson equation can be expressed as a linear combination of polynomials that transform according to the trivial irreducible representation of the $C_{4v}$ point group. Each individual barrier noise further breaks the symmetry of the solution to a linear combination of functions that transform according to the trivial irrep of $C_{1v}$. The Poisson equation is linear and therefore the total solution is the sum of the potential generated by each gate:

\begin{equation}\label{phiT}
\begin{split}
    \phi(\vec{V},\mathbf{r}) &= \sum_{ij}{\left[f_{ij}(\vec{V}^0) + h_{ij}(\delta_p)\right]I_i(x,y)z^j} \\
    &+ \sum_k{z^k\sum_{ij}{x^iy^j}\left\{\left[g_{ijk}(\delta_1) + (-1)^ig_{ijk}(\delta_3)\right]y^j + \left[g_{jik}(\delta_2) + (-1)^jg_{jik}(\delta_4)\right]x^i\right\}},
\end{split}
\end{equation}

\noindent where $I_i(x,y)$ are invariant functions of the trivial irrep of $C_{4v}$ $\left(1,x^2+y^2,(x^2-y^2)^2,x^2y^2,\dots\right)$, and $f_{ij}$, $h_{ij}$ and $g_{ijk}$ are expansion coefficients. We Taylor expand the full solution \eqref{phiT} close to the point $(0,0,z_0)$ to quadratic order (neglecting $z^2$ terms):

\begin{equation}
\begin{split}
    \phi(\vec{V},\mathbf{r})&\approx\left[g_{100}(\delta_1) - g_{100}(\delta_3)\right]x + \left[g_{100}(\delta_2) - g_{100}(\delta_4)\right]y \\
    &+ \left[f_{01}(\vec{V}^0) + h_{01}(\delta_p) + g_{001}(\delta_1) + g_{001}(\delta_2) + g_{001}(\delta_3) + g_{001}(\delta_4)\right]z \\
    &+ \left[f_{10}(\vec{V}^0) + h_{10}(\delta_p) + g_{200}(\delta_1) + g_{010}(\delta_2) + g_{200}(\delta_3) + g_{010}(\delta_4)\right]x^2 \\
    &+ \left[f_{10}(\vec{V}^0) + h_{10}(\delta_p) + g_{010}(\delta_1) + g_{200}(\delta_2) + g_{010}(\delta_3) + g_{200}(\delta_4)\right]y^2 \\
    &+ \left[g_{101}(\delta_1) - g_{101}(\delta_3)\right]zx + \left[g_{101}(\delta_2) - g_{101}(\delta_4)\right]yz.
\end{split}
\end{equation}

\noindent Note that $xy$-terms are forbidden by symmetry. When the fluctuations $\vec{\delta}$ are small, the full solution can be expressed in terms of the fluctuations $\delta_i$ multiplied by the derivative of the expansion coefficients close to the DC operation point, i.e. $g_{ijk}(\delta)\approx g_{ijk}^\prime(0)\delta$. Applying this and casting the full solution close to the point $(0,0,z_0)$ in terms of harmonic oscillator parameters we obtain

\begin{equation}
\begin{split}
    \phi(\mathbf{r}) &\approx F_zz + \frac{1}{e}\frac{\alpha_0}{l^4}\left(x^2+y^2\right) \\
    &+ g_{100}^\prime(0)(\delta_1 - \delta_3)x + g_{100}^\prime(0)(\delta_2 - \delta_4)y + \left[h_{01}^\prime(0)\delta_p + g_{001}^\prime(0)\left(\delta_1+\delta_2+\delta_3+\delta_4\right)\right]z \\
    &+ \left[h_{10}^\prime(0)\delta_p + \frac{1}{2}\left(g_{200}^\prime(0) + g_{010}^\prime(0)\right)\left(\delta_1+\delta_2+\delta_3+\delta_4\right)\right]\left(x^2+y^2\right) \\
    &+ \frac{1}{2}\left(g_{200}^\prime(0) - g_{010}^\prime(0)\right)\left(\delta_1-\delta_2+\delta_3-\delta_4\right)\left(x^2-y^2\right) \\
    &+ g_{101}^\prime(0)(\delta_1 - \delta_3)zx + g_{101}^\prime(0)(\delta_2 - \delta_4)yz,
\end{split}
\end{equation}

\noindent where $F_z = f_{01}(\vec{V}^0)$ and $\alpha_0/l^4 = ef_{10}(\vec{V}^0)$. The Larmor frequency shifts $\varepsilon_i$ of the qubit for gate $i$ can then be computed directly from the seven constants $h_{01}^\prime(0)$, $h_{10}^\prime(0)$, $g_{100}^\prime(0)$, $g_{001}^\prime(0)$, $g_{200}^\prime(0)$, $g_{010}^\prime(0)$ and $g_{101}^\prime(0)$, and the $\sigma_z$-component in the qubit subspace of the position operators $x^2+y^2$, $x^2-y^2$, $x$, $y$, $z$, $zx$ and $yz$ (see Fig. \ref{fig:opse}, \ref{fig:opsu} and \ref{fig:opss}). At the DC operation point, we find the following constants:

\begin{subequations}
    \begin{align}
    h_{01}^\prime(0) &= -9.4340\cdot10^{-3}\,\mathrm{nm}^{-1}, \\
    h_{10}^\prime(0) &= -1.8569\cdot10^{-5}\,\mathrm{nm}^{-2}, \\
    g_{100}^\prime(0) &= 3.9668\cdot10^{-4}\,\mathrm{nm}^{-1}, \\
    g_{001}^\prime(0) &= 4.1757\cdot10^{-4}\,\mathrm{nm}^{-1}, \\
    g_{200}^\prime(0) &= 6.1099\cdot10^{-6}\,\mathrm{nm}^{-2}, \\
    g_{010}^\prime(0) &= -8.9887\cdot10^{-7}\,\mathrm{nm}^{-2}, \\
    g_{101}^\prime(0) &= 9.6812\cdot10^{-6}\,\mathrm{nm}^{-2}.
\end{align}
\end{subequations}

\begin{figure}[h!]
    \centering
    \includegraphics[width=0.6\linewidth]{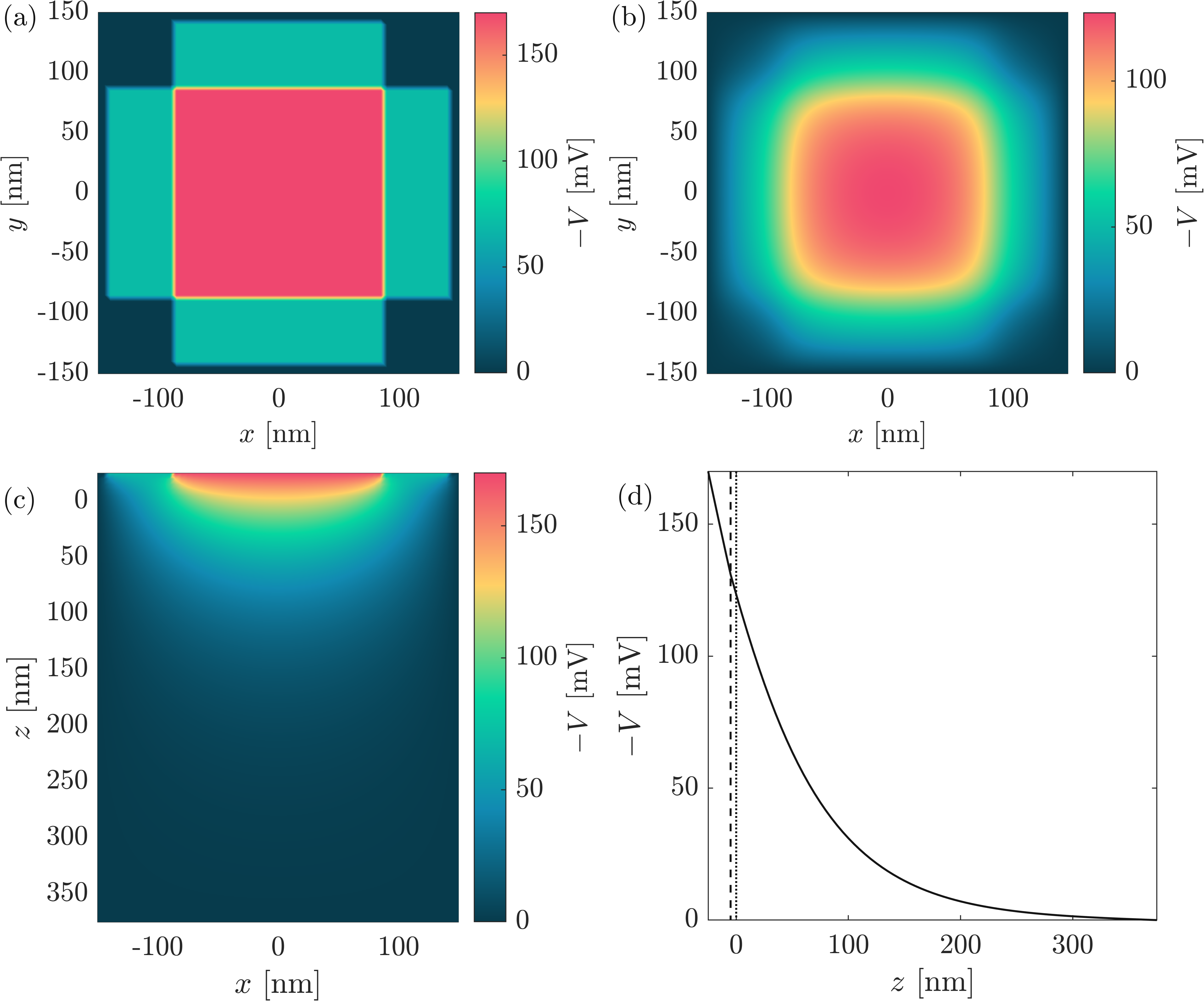}
    \caption{(a)-(b) $x$-$y$ slices of the potential $\phi(\mathbf{r})$ (a) just below the gates (b) in the Ge layer at $z = z_0$. (c) $z$-$x$ slice of the potential $\phi(\mathbf{r})$ at $y = 0$. (d) Graph of $\phi(0,0,z)$. The dotted line indicates $z = z_0 \equiv 0$ and the dashed line is the Ge/SiGe interface.}
    \label{fig:3D}
\end{figure}

Fig. \ref{fig:3D} shows the solution of the Poisson equation on the simulation domain described at the beginning of this section.

\newpage

Assuming the plunger gate and the four barriers act as uncorrelated sources of $1/f$ charge noise, the dephasing time $T_2^*$ is equal to

\begin{equation}
    T_2^* = \frac{\hbar}{\sqrt{\sum_i\varepsilon_i^2}}\sqrt{\frac{\pi}{\int_{\omega_ct/2}^\infty{dx\,\frac{\sin^2(x)}{x^3}}}},
\end{equation}

\noindent where the sum is made on all $5$ sources of noise, $t$ is the measurement time and $\omega_c$ is a cut-off frequency. If $\omega_c t\approx 0.10879$ the integral is close to $\pi$ and the square-root term $\approx1$.

\begin{figure}[h!]
    \centering
    \includegraphics[width=0.7\linewidth]{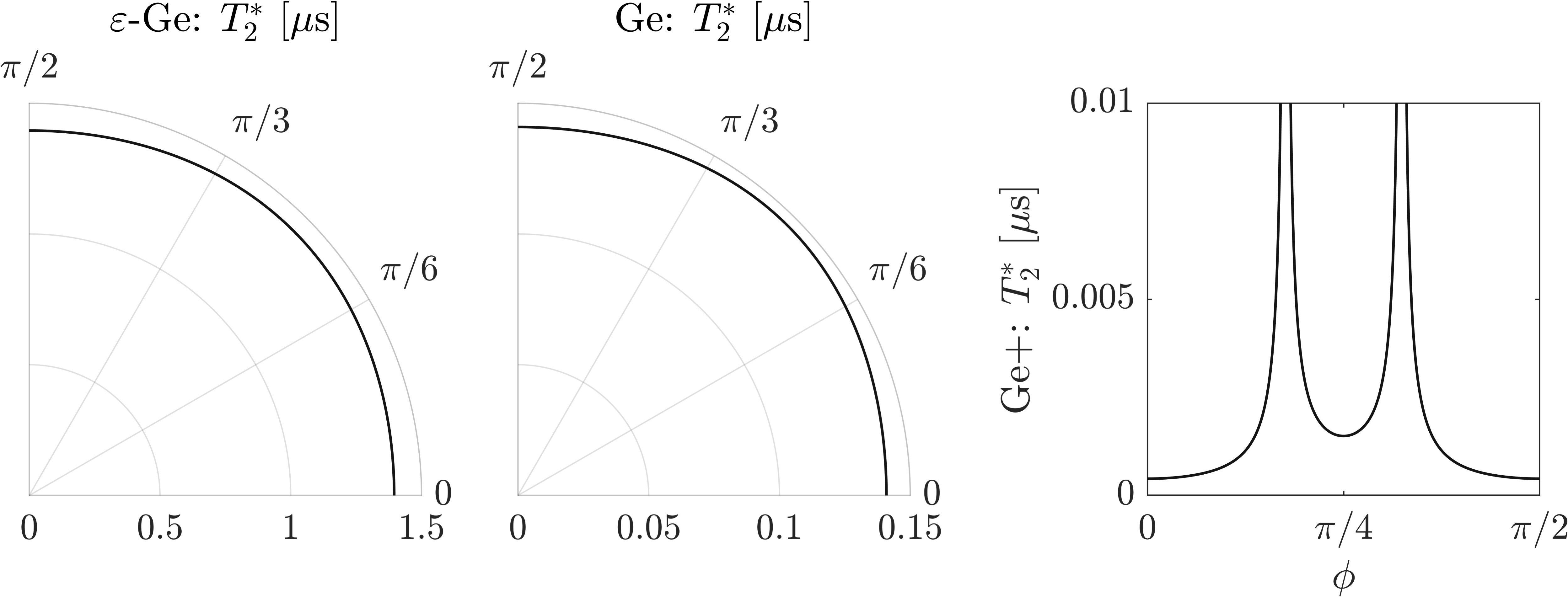}
    \caption{$T_2^*$ as a function of the $B$-field in-plane angle $\phi$ for the three systems at $B=0.1\,\mathrm{T}$ and voltage fluctuations $\delta_i = 1\,\mathrm{mV}$ for all gates.}
    \label{fig:T2}
\end{figure}

Plots of the $T_2^*$ for the three systems as a function of the $B$-field in-plane angle $\phi$ are shown in Fig. \ref{fig:T2}. The two sharp peaks in the spiked Ge appear due to the vanishing of the longitudinal component of $z$ and $x^2+y^2$, see Fig. \ref{fig:opss}.

\hspace{0.5cm}
\newpage

\newpage

\section{\texorpdfstring{Multi-Objective Bayesian Optimisation Methodology}{}}\label{sec:ml-method}


The optimisation problem considered here involves simultaneously maximising two competing objectives over a multi-dimensional continuous input space. 
A \emph{Bayesian optimisation} (BO) framework is employed. BO maintains a cheap-to-evaluate probabilistic surrogate model that is updated after every batch of simulation calls, and uses an acquisition function to decide which candidate points to evaluate next. The specific variant used here is a \emph{multi-objective Bayesian optimisation} loop based on a \textbf{Multitask Gaussian Process} (MTGP) surrogate and a \textbf{Pareto Upper Confidence Bound} (UCB) acquisition strategy.

\begin{table}[h!]
\label{alg:bo}
\centering
\caption{Algorithm protocol overview.}
\begin{tabular}{lp{10cm}}
\toprule
Step & Description \\
\midrule
1 & Initialise surrogate on $n_0$ seed evaluations \\
2 & Normalise inputs and outputs \\
3a & Fit MTGP by maximising exact MLL \\
3b & Compute UCB per task at all candidates \\
3c & Extract Pareto-optimal set from UCB vectors \\
3d & Evaluate MATLAB oracle at Pareto-selected candidates \\
3e & Augment dataset; remove evaluated points from pool \\
4 & Repeat steps 3a--3e for $N_{\mathrm{iter}}$ iterations \\
\bottomrule
\end{tabular}
\end{table}

\begin{table}[h!]
\centering
\caption{Bayesian optimisation hyperparameter summary.}
\label{tab:hyperparams}
\begin{tabular}{lll}
\toprule
Symbol & Description & Value \\
\midrule
$n_0$ & Initial training set size & 1000 \\
$N_{\mathrm{iter}}$ & Number of BO iterations & 40 \\
$N_{\mathrm{cand}}$ & Candidate pool size & 50{,}000 \\
$N_{\mathrm{GP}}$ & GP training steps per iteration & 200 \\
$\alpha$ & Adam learning rate & 0.05 \\
$\beta$ & UCB exploration parameter & 1.0 \\
$R$ & Multitask kernel rank & 2 \\
Jitter & Cholesky diagonal jitter & $10^{-3}$ \\
\bottomrule
\end{tabular}
\end{table}

\subsection{\texorpdfstring{Optimisation Objectives}{}}

Let $\mathbf{x} \in \mathcal{X} \subset \mathbb{R}^{N}$ be the input vector.
The two scalar objectives evaluated by the MATLAB oracle $f_{\mathrm{sb}}$ are

\begin{equation}
    f_1(\mathbf{x}) = \bigl|b_2(\mathbf{x})\bigr|, \qquad
    f_2(\mathbf{x}) = \log\left(\frac{1.}{\bigl|\partial b_2^{(2)}(\mathbf{x})\bigr|}\right),
    \label{eq:objectives}
\end{equation}

\noindent where $b_2$ and $\partial b_2^{(2)}$ are two output quantities returned by the simulation. 
For the bump system we consider $\partial b_2^{(2)}$, for the spike system we consider $\partial b_2$ 
Both objectives are to be \emph{maximised} simultaneously, forming a Pareto front in the objective space $\mathbb{R}^2$.

\subsection{\texorpdfstring{Input Space and Candidate Pool}{}}
The $N$ input variables together with their admissible ranges are listed in Table~\ref{tab:inputs} and ~\ref{tab:inputs2} for the two systems under scrutiny (Si bump and Si spike). All variables are treated as continuous.
At the start of each optimisation run a \emph{candidate pool} of $N_{\mathrm{cand}} = 50{,}000$ points is sampled independently and uniformly at random from the Cartesian product of the intervals in Table~\ref{tab:inputs}. Evaluated points are removed from the pool after each BO iteration so that no point is selected and evaluated twice.

\begin{table}[h!]
\centering
\caption{Input parameters and search bounds for Si bump structure optimization}
\label{tab:inputs}
\begin{tabular}{ccc}
\toprule
 & Si bump & \\
\midrule
Index & Variable & Search range \\
\midrule
1 & $d$                              & $[1,\;30]\,\mathrm{nm}$ \\
2 & $w$                               & $[0.5,\;5]\,\mathrm{nm}$ \\
3 & $A$                             & $[0,\;40]\%$ \\
4 & $\varepsilon_\parallel$               & $[-0.0002,\;0]\%$ \\ 
5 & $4\tau_+$                         & $[0,\;4]\,\mathrm{nm}$ \\
6 & $4\tau_-$                         & $[0,\;4]\,\mathrm{nm}$ \\
7 & $4\tau_i$                        & $[0,\;4]\,\mathrm{nm}$ \\
8 & $x$                             & $[10,\;40]\%$ \\
9 & $r$                        & $[0,\;1]$ \\
\bottomrule
\end{tabular}
\end{table}

\begin{table}[h!]
\centering
\caption{Input parameters and search bounds for Si spike structure optimization}
\label{tab:inputs2}
\begin{tabular}{ccc}
\toprule
 & Si spike & \\
\midrule
Index & Variable              & Search range \\
\midrule
1 & $d_1$                             & $[1,\;15]\,\mathrm{nm}$ \\
2 & $d_2$                              & $[1,\;15]\,\mathrm{nm}$ \\
3 & $d_3=d_4=d_2$                & $[1,\;15]\,\mathrm{nm}$ \\
4 & $\varepsilon_\parallel$               & $[-0.0003,\;0]\%$ \\ 
5 & $4\tau_i$                          & $[0,\;4]\,\mathrm{nm}$ \\
6 & $r$                             & $[0,\;1]$ \\
\bottomrule
\end{tabular}
\end{table}

The range for the gate electric field $F_z$ is chosen such that the ground state is located in Ge, not in the triangular potential formed at the surface in the \SiGe{} barrier.  We first define the mapping $r\in[0,1]\mapsto F_z(r)\in[F_z^-,F_z^+]$, where

\begin{equation}
    F_z(r) = 10^{(1-r)f_\mathrm{min}+rf_\mathrm{max}}\,\mathrm{mV}/\mathrm{nm}.
\end{equation}

For the lower bound we take $f_\mathrm{min}=-1$ such that $F_z^-=F_z(r=0)\equiv 0.1\,\mathrm{mV}/\mathrm{nm}$, which always returns a ground state in Ge for our parameter space. The upper bound $F_z^+(d_i,\varepsilon_\parallel,A,\dots)$ generally depends on the heterostructure parameters. A sufficient condition for a ground state in Ge is to ensure that the energy of the LH band at the surface remains higher than that of the HH band just underneath the bump or below the second spike (assuming the \SiGe{} barrier is tensile strained and Ge is compressive strained). If $L$ is the thickness of the \SiGe{} barrier, then $F_z$ must be smaller than

\begin{equation}
    F_z < F_z^+ = \frac{1}{e}\frac{E_\mathrm{SiGe} - E_\mathrm{Ge}}{L+\ell},
\end{equation}

\noindent where $\ell = d+w+4\tau_-$ for bumped Ge$+$ and $\ell = d_2 + 0.5\,\mathrm{nm}$ for spiked Ge$+$. The band edge energies are

\begin{align}
    E_\mathrm{Ge} &= \mathcal{E}_{\Gamma_5^+}^\mathrm{Ge} + \frac{\Delta_0^\mathrm{Ge}}{3} + a_v^\mathrm{Ge}\text{Tr}\,\varepsilon^\mathrm{Ge} + b^\mathrm{Ge}\delta\varepsilon^\mathrm{Ge}, \\
    E_\mathrm{SiGe} &= \mathcal{E}_{\Gamma_5^+}^\mathrm{SiGe} + \frac{\Delta_0^\mathrm{SiGe}}{3} + a_v^\mathrm{SiGe}\text{Tr}\,\varepsilon^\mathrm{SiGe} + \frac{\Delta_0^\mathrm{SiGe}}{2}\left(\sqrt{1-2\rho+9\rho^2} - \rho-1\right),
\end{align}

\noindent where $\rho = b^\mathrm{SiGe}\delta\varepsilon^\mathrm{SiGe}/\Delta_0^\mathrm{SiGe}$. It is straightforward to verify that the last term in $E_\mathrm{SiGe}$ tends to $-b^\mathrm{SiGe}\delta\varepsilon^\mathrm{SiGe}$ in the limit $\Delta_0^\mathrm{SiGe}\to\infty$, as expected for a LH band in the limit of infinite bulk SOI. The strain components in the SiGe layer are determined from the in-plane lattice constant of the strained Ge layer, assuming a coherent growth:

\begin{equation}
    \varepsilon_\parallel^\mathrm{SiGe} = \frac{(1+\varepsilon_\parallel^\mathrm{Ge})a_0^\mathrm{Ge}}{a_0^\mathrm{SiGe}} -1,
\end{equation}

\noindent with $a_0$ being the lattice constant of the respective material without strain. Then we take the logarithm of $F_z^+$ to obtain $f_\mathrm{max} = \log_{10}[F_z^+/(\mathrm{mV}/\mathrm{nm})]$.



\newpage

\subsection{\texorpdfstring{Model specification}{}}

The surrogate is an \textbf{Exact Multitask Gaussian Process} (MTGP) implemented in GPyTorch~\cite{gardner2018gpytorch}. An MTGP models the two objectives jointly, allowing the model to exploit inter-task correlations. Formally, the joint prior
over outputs $\mathbf{f}(\mathbf{x}) = [f_1(\mathbf{x}),\, f_2(\mathbf{x})]^\top$ is a Multitask Multivariate Normal distribution:
\begin{equation}
    p\!\left(\mathbf{f}(\mathbf{x})\right) = \mathcal{N}\!\left(\bm{\mu}(\mathbf{x}),\;
    \mathbf{K}(\mathbf{x}, \mathbf{x}')\right),
\end{equation}
\noindent where the mean and covariance are structured as described below.
\\

\paragraph{Mean function.}
A \texttt{MultitaskMean} composed of independent \texttt{ConstantMean} functions is used per task:
\begin{equation}
    \mu_t(\mathbf{x}) = c_t, \qquad t \in \{1, 2\},
\end{equation}
\noindent with learned constants $c_1, c_2 \in \mathbb{R}$.
\\

\paragraph{Covariance (kernel) function.}
The joint covariance is constructed via the \texttt{MultitaskKernel} of rank $R=2$:
\begin{equation}
    k\bigl((\mathbf{x},t),\,(\mathbf{x}',t')\bigr)
    = k_{\mathrm{RBF}}(\mathbf{x}, \mathbf{x}') \cdot B_{tt'},
\end{equation}
\noindent where $k_{\mathrm{RBF}}$ is the Radial Basis Function (squared-exponential) kernel
\begin{equation}
    k_{\mathrm{RBF}}(\mathbf{x}, \mathbf{x}') =
    \sigma_f^2\exp\!\left(-\frac{1}{2}\sum_{j=1}^{9}
    \frac{(x_j - x'_j)^2}{\ell_j^2}\right),
\end{equation}
\noindent and $\mathbf{B} \in \mathbb{R}^{2 \times 2}$ is a positive semi-definite task covariance matrix parameterised as $\mathbf{B} = \mathbf{v}\mathbf{v}^\top + \mathrm{diag}(\bm{\kappa})$ for a rank-2 vector $\mathbf{v} \in \mathbb{R}^2$ and noise variances $\bm{\kappa}$. 
The learnable parameters of the kernel are the output scale $\sigma_f^2$, the nine length-scales $\ell_1,\ldots,\ell_9$, and the task covariance parameters $\mathbf{v}$ and $\bm{\kappa}$.
\\

\paragraph{Likelihood.}
A \texttt{MultitaskGaussianLikelihood} with per-task homoscedastic noise $\sigma_{\mathrm{noise},t}^2$ is assumed:
\begin{equation}
    y_t(\mathbf{x}) = f_t(\mathbf{x}) + \varepsilon_t, \qquad
    \varepsilon_t \sim \mathcal{N}(0, \sigma_{\mathrm{noise},t}^2).
\end{equation}
A Cholesky jitter of $10^{-3}$ is added to the diagonal of the kernel matrix to ensure numerical stability during factorisation.

\subsection{Training}
At every BO iteration the hyperparameters are re-estimated by maximising the \textbf{Exact Marginal Log-Likelihood} (MLL):
\begin{equation}
    \log p(\mathbf{Y} \mid \mathbf{X}, \bm{\theta}) =
    -\tfrac{1}{2}\mathbf{y}^\top \mathbf{K}_{\bm{\theta}}^{-1} \mathbf{y}
    -\tfrac{1}{2}\log\det \mathbf{K}_{\bm{\theta}}
    - \tfrac{n}{2}\log 2\pi,
\end{equation}
\noindent where $\mathbf{y}$ is the stacked vector of all observed outputs, $\mathbf{X}$ is the matrix of training inputs, and $\bm{\theta}$ collects all kernel and noise hyperparameters. Optimisation is performed with the \textbf{Adam optimiser} \cite{kingma2015adam} using
\begin{itemize}
    \item learning rate $\alpha = 0.05$,
    \item $N_{\mathrm{GP}} = 200$ gradient steps per BO iteration,
\end{itemize}

\subsection{Data Normalisation}

\paragraph{Input normalisation.}
All nine input features are normalised using \texttt{StandardScaler} (zero mean, unit variance), fitted on the full training set available at each iteration and re-fitted each time the training data is extended:
\begin{equation}
    \tilde{x}_j = \frac{x_j - \hat{\mu}_{x_j}}{\hat{\sigma}_{x_j}}.
\end{equation}
\\

\paragraph{Output normalisation.}
Each of the two output columns is standardised independently using its own \texttt{StandardScaler}, again re-fitted at every iteration:
\begin{equation}
    \tilde{y}_t = \frac{y_t - \hat{\mu}_{y_t}}{\hat{\sigma}_{y_t}}, \qquad t \in \{1,2\}.
\end{equation}
The GP is trained entirely in the normalised space. Predictions are mapped back to the original scale before computing acquisition values and before recording results.

\subsection{Acquisition Function: Pareto UCB}

For each candidate point $\mathbf{x}$ and each task $t$, the \textbf{Upper Confidence Bound} (UCB) is computed from the GP posterior as
\begin{equation}
    \mathrm{UCB}_t(\mathbf{x}) = \mu_t(\mathbf{x}) + \beta\,\sigma_t(\mathbf{x}),
    \label{eq:ucb}
\end{equation}
\noindent where $\mu_t(\mathbf{x})$ and $\sigma_t(\mathbf{x})$ are the posterior mean and standard deviation (in the \emph{original} output scale, after inverse transform), and $\beta \geq 0$ is an \emph{exploration--exploitation} trade-off parameter, here set $= 1.0$.


The UCB values are then assembled into a two-dimensional acquisition vector:
\begin{equation}
    \mathbf{u}(\mathbf{x}) = \bigl[\mathrm{UCB}_1(\mathbf{x}),\;
\mathrm{UCB}_2(\mathbf{x})\bigr]^\top.
\end{equation}
A point $\mathbf{x}$ is said to \emph{Pareto-dominate} $\mathbf{x}'$ with respect to $\mathbf{u}$ if $u_t(\mathbf{x}) \geq u_t(\mathbf{x}')$ for all $t$ and $u_t(\mathbf{x}) > u_t(\mathbf{x}')$ for at least one $t$. The \textbf{Pareto front} of the candidate pool is the set of non-dominated points:
\begin{equation}
    \mathcal{P} = \bigl\{\mathbf{x} \in \mathcal{X}_{\mathrm{cand}} :
    \nexists\, \mathbf{x}' \in \mathcal{X}_{\mathrm{cand}}\text{ s.t.\ }
    \mathbf{u}(\mathbf{x}') \text{ dominates } \mathbf{u}(\mathbf{x})\bigr\}.
\end{equation}

All points on the UCB Pareto front $\mathcal{P}$ are selected as the next batch to evaluate.  The selected points are removed from the candidate pool before the next
iteration begins.

\subsection{Computational performance}
Model training and inference is performed on a MacBook Pro, M3 Chip, 18GB Memory. 
In light of the feasibility of the study with this computational equipment, detailed performance analysis and optimization were deemed beyond the scope of the work. 
In its current implementation,  training time at each Bayesian Optimization iteration is in the  $10^{1}$ seconds order of magnitude, inference is $10^{2}$ faster w.r.t. the physics based ground-truth approach.

\newpage

\section{Multi-Objective Bayesian Optimisation Results}

\begin{figure}[h!]
    \centering
    \includegraphics[width=0.45\linewidth]{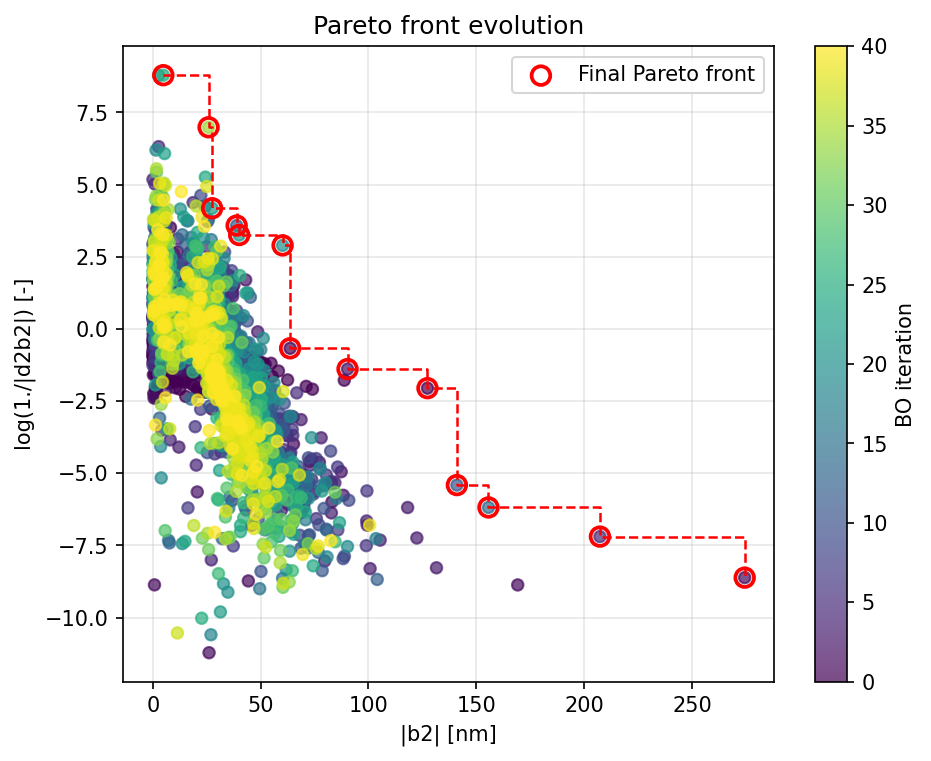}
    \includegraphics[width=0.9\linewidth]{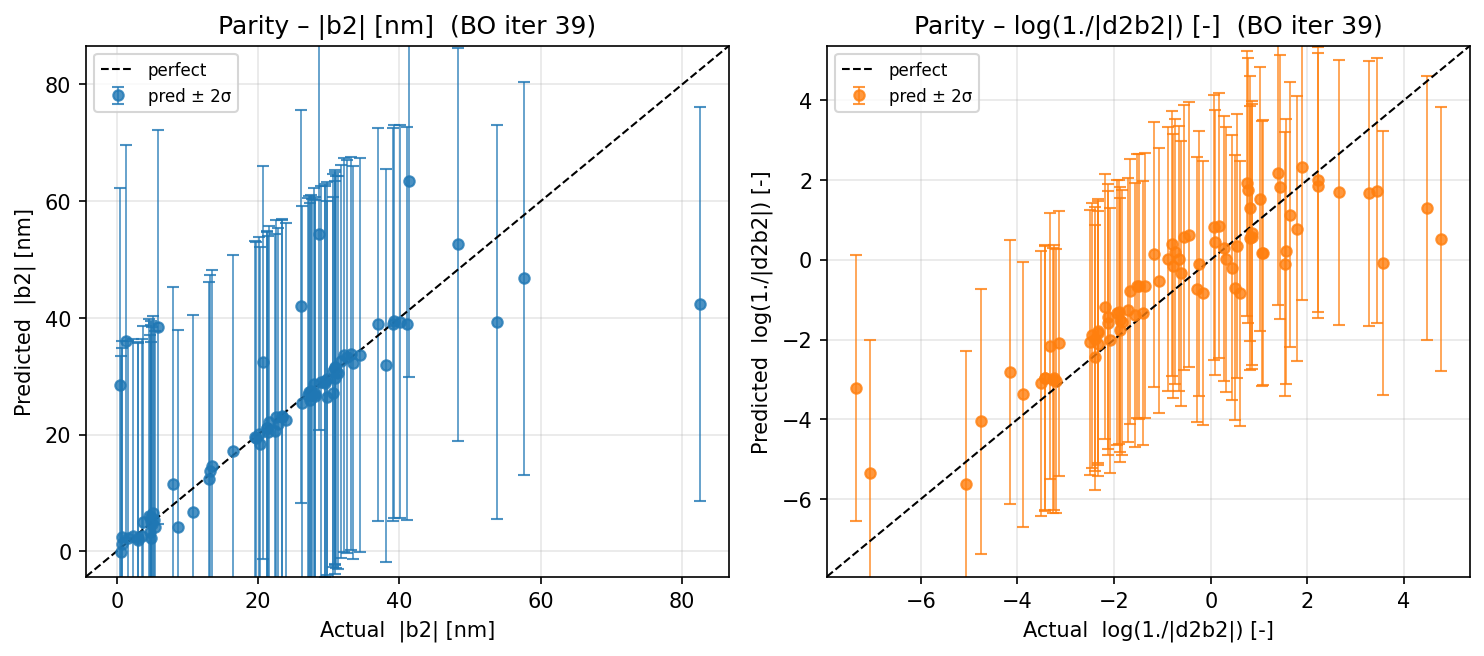}
    \caption{Top panel illustrates the Pareto Front identified by the multi-objective Bayesian optimization of GeSi Bumps. Lower panel illustrates the model prediction accuracy for the final selected candidates, corroborating the trust in the candidate selection approach.}
    \label{fig:bo-bump}
\end{figure}

Among the sample structures, we consider only the ones with largest $\beta_2$ , and fine-tune parameters to ensure $ (|\partial\beta_2/\partial F_z|)$= 0.

\newpage

\begin{figure}[h!]
    \centering
    \includegraphics[width=0.45\linewidth]{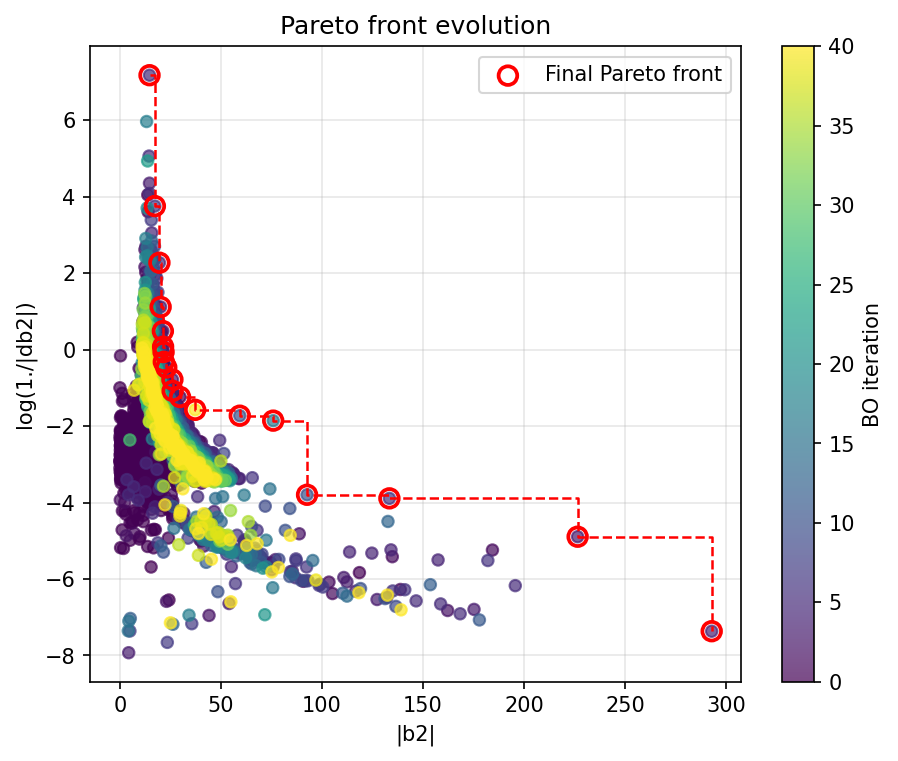}
    \includegraphics[width=0.9\linewidth]{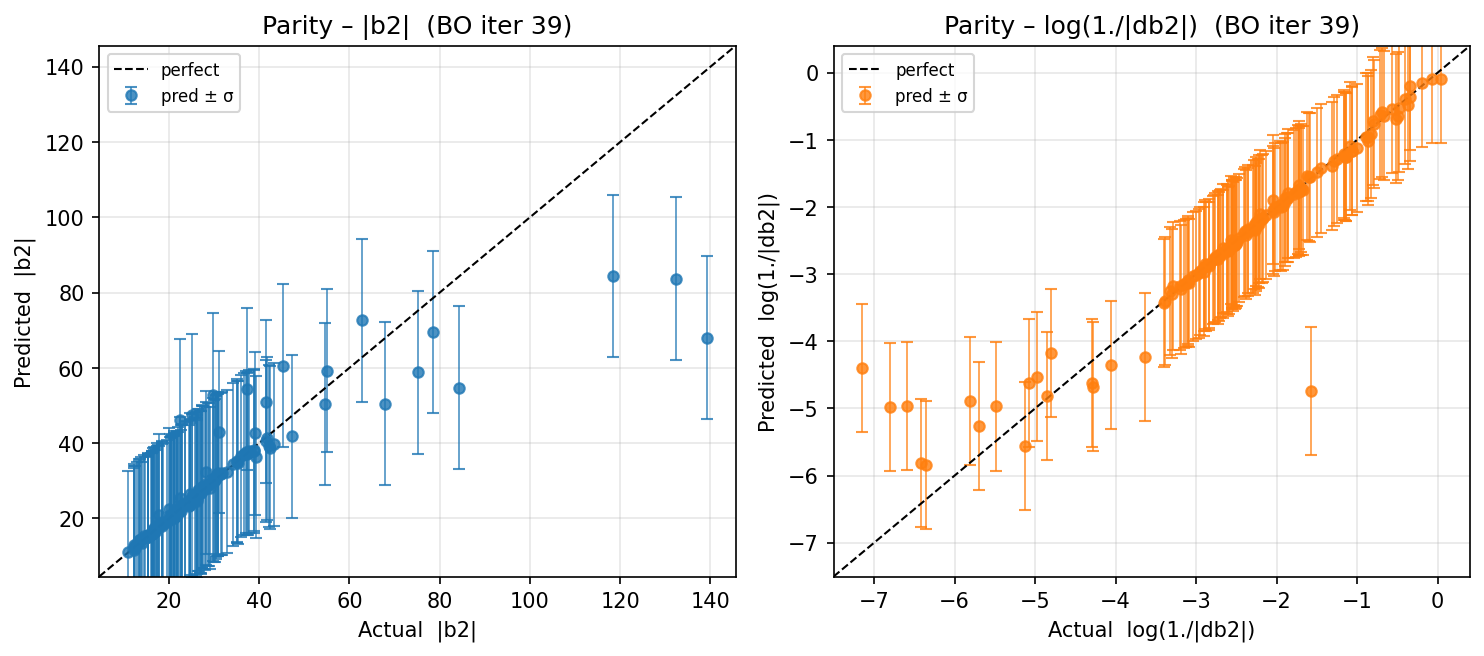}
    \caption{Top panel illustrates the Pareto Front identified by the multi-objective Bayesian optimization of GeSi spikes. Lower panel illustrates the model prediction accuracy for the final selected candidates, corroborating the trust in the candidate selection approach.}
    \label{fig:bo-spike}
\end{figure}

Among the sample structures, we consider only the ones with $\beta_2$ > 30  and $\log (|\partial\beta_2/\partial F_z|^{-1})$ > -2 .
\newpage

%
